\documentclass[prd,aps,preprint,showpacs,nofootinbib,superscriptaddress,tightenlines]{revtex4-1}
\usepackage{dcolumn}
\usepackage{bm}
\usepackage[english]{babel}
\usepackage{hyperref}
\usepackage{epsf}
\usepackage{amsmath}
\usepackage{amssymb}
\usepackage{amsfonts}
\usepackage{bbold}
\usepackage{graphicx}
\usepackage{epstopdf}
\usepackage{url}
\usepackage{xcolor}

\newcommand{\nn}{\nonumber} 
\newcommand{\beq}{\begin{equation}}
\newcommand{\eeq}{\end{equation}} 
\newcommand{\beqa}{\begin{eqnarray}} 
\newcommand{\eeqa}{\end{eqnarray}}

\def\bc{\begin{center}}
\def\ec{\end{center}}
\def\bi{\begin{itemize}}
\def\ei{\end{itemize}}
\def\be{\begin{equation}}
 \def\ee{\end{equation}}
\def\ben{\begin{equation*}}
 \def\een{\end{equation*}}
 \def\bea{\begin{eqnarray}}
 \def\eea{\end{eqnarray}}
 \def\bean{\begin{eqnarray*}}
 \def\eean{\end{eqnarray*}}
\newcommand{\ie}{{\em i.e.}}  \newcommand{\eg}{{\em e.g.}}

\newcommand{\morder}[1]{{\cal O}\left(#1 \right)}
\newcommand{\eq}[1]{(\ref{#1})}
\newcommand{\ed}{\end{document}}

\newcommand{\ave}[1]{\langle{#1}\rangle}
\newcommand{\Ave}[1]{\left\langle{#1}\right\rangle}
\newcommand{\com}[2]{\left[{#1},{#2}\right]}

\newcommand{\zerovec}{{\boldsymbol 0}}
\newcommand{\avec}{{\boldsymbol a}}
\newcommand{\bvec}{{\boldsymbol b}}
\newcommand{\fvec}{{\boldsymbol f}}
\newcommand{\fvect}{{\boldsymbol{\tilde f}}}
\newcommand{\Vt}{{\tilde V}}
\newcommand{\Cvec}{{\boldsymbol C}}
\newcommand{\pvec}{{\boldsymbol p}}
\newcommand{\ellvec}{{\boldsymbol \ell}}
\newcommand{\kvec}{{\boldsymbol k}}
\newcommand{\qvec}{{\boldsymbol q}}
\newcommand{\xvec}{{\boldsymbol x}}
\newcommand{\epsvec}{{\boldsymbol \varepsilon}}

\newcommand{\tf}{t_{\mathrm{f}}}

\newcommand{\pt}{p_{_\perp}}

\newcommand{\pp}{p--p}
\newcommand{\pA}{p--A}

\newcommand{\dd}{{\rm d}}
  \newcommand{\halft}{{\textstyle \frac{1}{2}}}
\newcommand{\lsim}{\lesssim} \newcommand{\gsim}{\gtrsim}

\def\bm#1{\mbox{\boldmath$#1$}}
 
 \def\esim{\,\mathrel{\rlap{\lower0.2em\hbox{$-$}}\raise0.15em\hbox{\footnotesize $\hskip0.04em\sim$}}\,}
 \def\gsim{\mathrel{\rlap{\lower0.2em\hbox{$\sim$}}\raise0.2em\hbox{$>$}}}
 \def\ksim{\mathrel{\rlap{\lower0.2em\hbox{$\sim$}}\raise0.2em\hbox{$<$}}}

\def\pt{p_{_\perp}}

\begin{document}
\title{Coherent medium-induced gluon radiation \\ in hard forward $1\to 1$ partonic processes}

\author{Fran\c{c}ois Arleo}
\affiliation{Laboratoire Leprince-Ringuet (LLR), \'Ecole polytechnique, CNRS/IN2P3 91128 Palaiseau, France}
\affiliation{Laboratoire d'Annecy-le-Vieux de Physique Th\'eorique (LAPTh), UMR5108, Universit\'e de Savoie \& CNRS,  BP 110, 74941 Annecy-le-Vieux cedex, France}
\author{Rodion Kolevatov}
\affiliation{SUBATECH, UMR 6457, Universit\'e de Nantes, Ecole des
Mines de Nantes, IN2P3/CNRS \\ 4 rue Alfred Kastler, 44307 Nantes cedex 3, France}
\affiliation{Department of High Energy Physics, Saint-Petersburg State University\\ Ulyanovskaya 1, 198504, Saint-Petersburg, Russia}
\author{St\'ephane Peign\'e}
\affiliation{SUBATECH, UMR 6457, Universit\'e de Nantes, Ecole des
Mines de Nantes, CNRS/IN2P3 \\ 4 rue Alfred Kastler, 44307 Nantes cedex 3, France}


\begin{abstract}
We revisit the medium-induced coherent radiation associated to {\it hard} and {\it forward} (small angle) scattering of an energetic parton through a nuclear medium. We consider all $1\to 1$ hard forward processes ($g \to g$, $q \to q$, $q \to g$ and $g \to q$), and derive the energy spectrum of induced coherent radiation rigorously to all orders in the opacity expansion and for the specific case of a Coulomb scattering potential. We obtain a simple general formula for the induced coherent spectrum, which encompasses the results corresponding to previously known special cases.  
\end{abstract}

\maketitle

\setcounter{footnote}{0}
\renewcommand{\thefootnote}{\arabic{footnote}} 	


 \section{Introduction and summary} 
 \label{sec:intro}

The theory of medium-induced gluon radiation has received a lot of attention over the last twenty years, as it became clear that parton energy loss is most likely responsible for the quenching of large $\pt$ hadrons and jets in heavy-ion collisions, first observed at RHIC~\cite{Adler:2003qi,Adams:2003kv} and then at the LHC~\cite{Aad:2012vca,Aamodt:2010jd,CMS:2012aa}. 

Multiple scatterings of a high-energy parton propagating in a medium\footnote{Here the medium refers either to cold nuclear matter or to a quark-gluon plasma, even though the theoretical setup discussed in Section~\ref{sec:hard-smallangle} applies more naturally to the former case.} of length $L$ induce the emission of gluons, a process occurring on a typical time $\tf \sim k^+/\kvec^2$, where $k^+$ is the light-cone momentum of the radiated gluon, and $\kvec \equiv \vec{k}_\perp$. It is useful to recall the three regimes identified in~\cite{Baier:1996kr} depending on the value of the gluon formation time $\tf$ (for a more detailed discussion in QED and in QCD, see~\cite{Peigne:2008wu}):
\begin{itemize}
\item[(i)] {\bf Bethe--Heitler, $\bm{\tf \ll \lambda}$}, where $\lambda$ is the parton mean free path in the medium. In this regime, \emph{each} scattering center acts as an independent source of radiation. The induced gluon spectrum is thus proportional to the (Bethe-Heitler) gluon spectrum induced by a \emph{single} scattering. This regime is at work for very soft gluons, $k^+ \ll \mu^2 \lambda$, where $\mu$ is the typical transverse momentum exchange in a single scattering. 
\item[(ii)] {\bf Landau--Pomeranchuk--Migdal (LPM), $\bm{\lambda \ll \tf \ll L}$}, for which a \emph{group} of $(\tf / \lambda)$ scattering centers acts as a single radiator, leading to a relative suppression of the gluon radiation intensity with respect to that in the Bethe-Heitler regime. The above condition on $\tf$ translates into $\mu^2 \lambda \ll k^+ \ll \ave{\kvec^2} L \sim \hat{q} L^2$, where $\hat{q}=\mu^2/\lambda$ is the so-called medium transport coefficient. (In a large medium, $\ave{\kvec^2}$ turns out to be on the order of $\hat{q} L$, the transverse momentum  exchanged over the length $L$.)
\item[(iii)] {\bf Long formation time, $\bm{t_f \gg L}$}. In this regime (also known as the `factorization regime'), \emph{all} scattering centers in the medium act coherently as a source of radiation. It occurs  at large gluon momenta,  $k^+ \gg \hat{q}L^2$.
\end{itemize}

The region of large formation time leads to `large' mean parton energy loss, $\Delta p^+ \propto p^+$ (with $p^+$ the parton light-cone momentum), coming from the broad time interval for gluon radiation. However, because of the weak dependence of $\Delta p^+$ on the medium length, it is often considered that radiation of gluons with long formation times does not contribute to the \emph{medium-induced} energy loss. As a consequence, gluon radiation in the LPM regime would dominate the induced spectrum, leading to an average parton energy loss $\Delta p^+ \sim \hat{q} L^2$, independent of $p^+$.

However, it was shown in Ref.~\cite{Arleo:2010rb} that the latter holds when the energetic parton is suddenly created (or annihilated) in the medium, but {\it does not hold} in the case of an `asymptotic' parton prepared long before the medium, undergoing small angle scattering through the medium, and hadronizing long after. In this case, the explicit calculation shows that the medium-induced spectrum is dominated by gluons with large formation time $\tf \gg L$, being thus {\it fully coherent} over the medium. This leads to an induced parton energy loss proportional to the parton energy, $\Delta p^+ \propto p^+$. (In particular, this demonstrates that the induced energy loss in a finite length medium is not bounded at high energy). In light-cone gauge, the {\it medium-induced} coherent radiation spectrum arises from the interference between gluon emission in the `initial state' (\ie, before the medium) and in the `final state' (after the medium)~\cite{Arleo:2010rb}.  As discussed in Ref.~\cite{Arleo:2010rb}, coherent parton energy loss is expected to play an important role in the phenomenology of hadron production in proton-nucleus collisions. The effects of coherent energy loss on quarkonium nuclear suppression were studied in Refs.~\cite{Arleo:2012hn,Arleo:2012rs,Arleo:2013zua}.

The kinematical setup used in~\cite{Arleo:2010rb} where fully coherent radiation occurs is as follows: a parton of high energy (large $p^+$) and with $\pvec \equiv \vec{p}_\perp = \vec{0}_\perp$ experiences a single hard scattering (with transverse momentum exchange $\qvec \equiv \vec{q}_\perp$) off a nuclear target and is tagged with $\pvec^\prime \simeq \qvec$ at small angle $|\pvec^\prime|/p^+\ll1$. In addition to such $1 \to 1$
{\it hard} and {\it forward} scattering, the parton undergoes multiple \emph{soft} scattering in the finite-size target, and thus receives a transverse momentum kick $\Delta q_\perp \ll q_\perp$. 

In this setup, Ref.~\cite{Arleo:2010rb} (see also~\cite{Arleo:2012rs}) considered the $g \to Q\bar{Q}$ hard process (mediated by single gluon exchange in the $t$-channel), with the final $Q\bar{Q}$ pair being massive (of mass $M$) and in a compact color octet state. The associated induced coherent spectrum was derived in a Feynman diagram calculation and at first order in the opacity expansion, \ie, with the transverse momentum broadening $\Delta q_\perp$ across the medium modelled by a {\it single} rescattering $\ell_\perp$. The result at all orders in the opacity expansion was obtained by identifying $\ell_\perp^2 = \Delta q_\perp^2(L) = \hat{q} L$ (where $\hat{q} = \mu^2/\lambda_g$, with $\lambda_g$ the gluon mean free path in the medium). The induced spectrum found in this semi-heuristic way reads~\cite{Arleo:2010rb,Arleo:2012rs} 
\be
\label{spec-APS-0}
x \frac{\dd I}{\dd x} =  N_c \, \frac{\alpha_s}{\pi} \, \log{\left(1 + \frac{\Delta q_\perp^2(L)}{x^2 M_\perp^2} \right)} \, ,
\ee
when defined with respect to a zero-size target, or equivalently
\be
\label{spec-APS}
\left. x \frac{\dd I}{\dd x}  \right|_{{\rm pA}-{\rm pp}} =  N_c \, \frac{\alpha_s}{\pi} \, \log{\left(\frac{x^2 M_\perp^2 + \Delta q_\perp^2(L_{\rm A})}{x^2 M_\perp^2 + \Delta q_\perp^2(L_{\rm p})} \right)} \, ,
\ee
when defined in \pA\ vs. \pp \ collisions. We denote $x \equiv k^+/p^+$ and $M_\perp \equiv \sqrt{M^2 + q_\perp^2}$.

Medium-induced coherent radiation was also studied using a semi-classical method in Refs.~\cite{Armesto:2012qa,Armesto:2013fca}, at first order~\cite{Armesto:2012qa} and all orders~\cite{Armesto:2013fca} in opacity and in a similar kinematical setup as that of Ref.~\cite{Arleo:2010rb}, however for the $q \to q$ case of a massless quark experiencing a hard scattering mediated by a {\it color singlet} exchange in the $t$-channel. Although no compact form of the coherent energy spectrum is given in Refs.~\cite{Armesto:2012qa,Armesto:2013fca}, the integral over $\vec{k}_\perp$ of the $\vec{k}_\perp$-differential spectrum found in~\cite{Armesto:2013fca} can be worked out analytically and reads (with $Q_{s {\rm p}} = \hat{q} L_{\rm p}$, $Q_{s {\rm A}} = \hat{q} L_{\rm A}$)
\be
\label{spec-armesto2}
\left. x \frac{\dd I}{\dd x} \right|_{{\rm pA}-{\rm pp}} =  2C_F \, \frac{\alpha_s}{\pi} \int_{{x^2 q_\perp^2}/{Q_{s {\rm A}}^2}}^{{x^2 q_\perp^2}/{Q_{s {\rm p}}^2}} \frac{\dd u}{u} \, e^{-u} = 2C_F \, \frac{\alpha_s}{\pi} \left[ {\rm Ei} \left( - \frac{x^2 q_\perp^2}{Q_{s {\rm p}}^2} \right) - {\rm Ei} \left( - \frac{x^2 q_\perp^2}{Q_{s {\rm A}}^2} \right)  \right]\, .
\ee
This can be checked to have the same parametric limits as the spectrum \eq{spec-APS}, up to the replacements $M \to 0$ and $N_c \to 2 C_F$ in Eq.~\eq{spec-APS}. 

In the present paper, we revisit induced coherent radiation and consider all $1\to 1$ hard forward  processes, namely $g \to g$, $q \to q$, $q \to g$ and $g \to q$, assuming the hard exchange in the $t$-channel to be color octet (for $g \to g$ and $q \to q$) or color triplet (for $q \to g$ and $g \to q$). We work in the same setup as Ref.~\cite{Arleo:2010rb}, but derive the induced coherent spectrum in full rigor, to all orders in the opacity expansion and for the specific case of a Coulomb rescattering potential.  

As a main result, we find that the induced coherent spectrum associated to $1 \to 1$ hard forward processes is given by the general (approximate) expression \eq{spec-general} (or equivalently \eq{spec-general-pAvspp}), with $C_R$, $C_{R'}$ the color charges of the incoming and outgoing partons, $C_t$ the color charge of the $t$-channel exchange, and $\Delta q_\perp(L)$ the {\it typical} transverse momentum broadening induced by multiple Coulomb rescatterings. The results obtained previously for the spectra associated to $g \to Q\bar{Q}$~\cite{Arleo:2010rb,Arleo:2012rs} and $q \to q$~\cite{Armesto:2012qa,Armesto:2013fca} are simply interpreted as particular cases of Eq.~\eq{spec-general}, corresponding respectively to $\{C_R =C_{R'} =N_c; \, C_t =N_c\}$ and $\{C_R =C_{R'} =C_F;\, C_t=0\}$. 

Our paper is organized as follows. 

First, the case of an asymptotic parton of color charge $C_R$ scattering at small angle off a \emph{pointlike} target through single gluon exchange is studied in Section \ref{sec:pointlike}. The radiation spectrum associated to this process is determined in the soft radiation ($x \equiv k^+/p^+ \ll 1$) and high-energy limit, see Eq.~(\ref{GBspec-exact}). This allows one to recover the Gunion-Bertsch spectrum, Eq.~(\ref{GBspec}), in the eikonal limit (scattering angle $\theta_s \to 0$) as well as the spectrum in the hard scattering limit ($|\qvec| \gg |\kvec|$), Eq.~(\ref{GBspec-hard}). The $\kvec$-integrated spectrum $x \dd{I}/\dd x$ is discussed in section~\ref{sec:qdep}, where we note that the {\it variation} of the spectrum with respect to the hard exchange $\qvec$ arises from gluon emission in the `soft' ($|\kvec| \sim  x |\qvec|$) and `hard' ($|\kvec| \sim |\qvec|$) transverse momentum regions. We argue that the color factor of the soft contribution heralds that of the induced spectrum in a finite size target.

The gluon spectrum associated to the scattering of an asymptotic parton off a {\it finite-size} target is investigated in detail in Section \ref{sec:hard-smallangle}, for the case of a {\it Coulomb} rescattering potential. The medium-induced gluon spectrum is determined rigorously in the soft ($x \ll 1$) and large formation time ($\tf\gg L$) limits. Calculations are carried out at first order in the opacity expansion,  $r\equiv L/\lambda_g \ll 1$, Eq.~(\ref{specn1}), and to all orders, $r \gg 1$, Eqs.~\eq{spec-alln}--\eq{sar}. We observe that the radiation spectrum is given by the overlap between the initial parton--gluon wavefunction (`evolved' by soft rescatterings, and thus medium dependent) and the wavefunction of the outgoing parton--gluon system. When $r \gg 1$ the induced spectrum can be approximated by a simple expression, Eq.~\eq{spec-alln-appr}, which in the case $C_R = N_c$ coincides with Eq.~\eq{spec-APS-0} derived semi-heuristically in~\cite{Arleo:2010rb,Arleo:2012rs}, provided $\Delta q_\perp^2(L)$ in \eq{spec-APS-0} is understood as the {\it typical} exchange in multiple Coulomb scattering, $\Delta q_\perp^2(L) = \mu^2 r \log{r}$. 
 
The results obtained in Section~\ref{sec:hard-smallangle} for the scattering of a fast asymptotic parton (\ie\ $q\to{q}$ or $g\to{g}$) are generalized to other processes in Section~\ref{sec:other-processes}. The case of an incoming fast quark scattered to an outgoing fast gluon ($q\to{g}$) is first examined. We find that the associated medium-induced spectrum is identical to that obtained in the case of an asymptotic gluon ($g\to{g}$), despite the different color charge in the initial state. The case of a fast gluon scattered to a pointlike color octet of mass $M$ (like for instance a compact heavy $Q\bar{Q}$ pair) is also treated. As for $g\to{g}$, the medium-induced gluon spectrum associated to $g \to Q\bar{Q}$ can be accurately approximated by a simple expression (Eq.~\eq{massive-spec-appr} to all orders in the opacity expansion), which has been used for phenomenological applications in~\cite{Arleo:2012hn,Arleo:2013zua}.  

We also discuss in Section~\ref{sec:IS-FS} the purely initial-state (respectively, final-state) radiation, associated to hard processes where the color charge of the outgoing particle (respectively, incoming particle) vanishes and coherent radiation between initial and final states is thus absent. In those cases, we explicitly verify that radiation with $\tf \gg L$ does not contribute to the induced gluon spectrum. The purely initial-state (or final-state) induced spectrum $x \dd{I}/\dd x$ arises from radiation with $\tf \sim \morder{ L}$, leading to the known result $\Delta p^+ \sim \hat{q} L^2$, independent (up to logarithms) of the parton energy.

Finally, we summarize our main results in Section~\ref{sec:discussion}, where we state and discuss the general (approximate) formulae \eq{spec-general} (or equivalently \eq{spec-general-pAvspp}) for the induced coherent spectrum associated to hard forward $1\to 1$ processes, at all orders in the opacity expansion. In this final Section we also discuss the connection of our study with a previous study of gluon production in the saturation formalism~\cite{Kovchegov:1998bi}.

\section{Small angle scattering of an `asymptotic parton' off a pointlike target and associated gluon radiation}
\label{sec:pointlike}

\subsection{Soft radiation amplitude and spectrum}
\label{sec:exact-pointlike}

\begin{figure}[t]
\centering
\includegraphics[width=12cm]{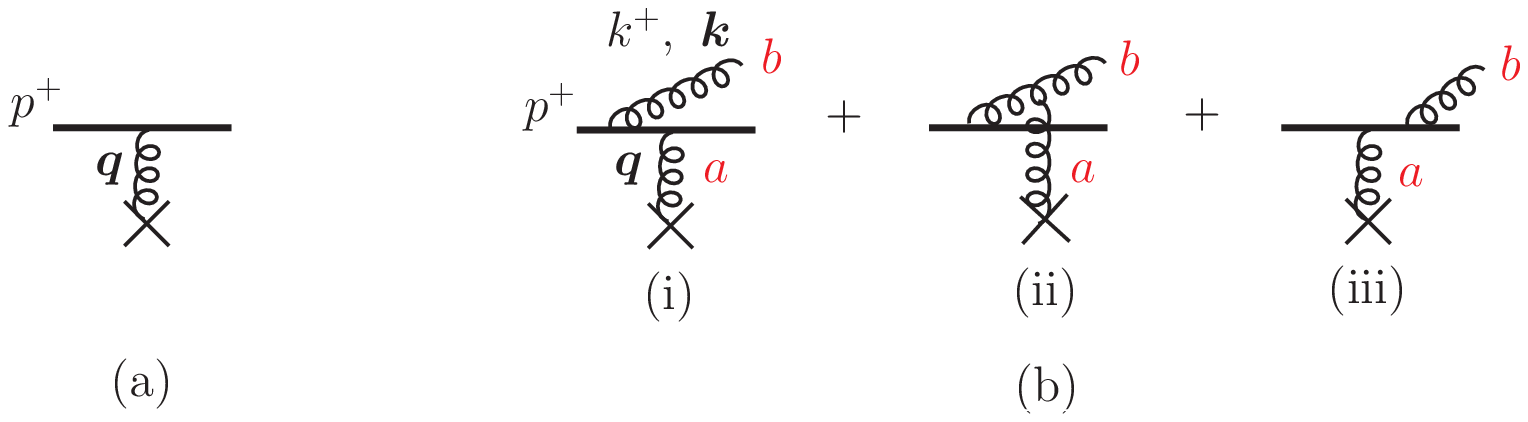}
\caption{(a) Elastic scattering amplitude ${\cal M}_{\rm el}$ of an energetic parton off a pointlike target. (b) Induced gluon radiation amplitude ${\cal M}_{\rm rad}$.}
\label{fig:GBamplitude}
\end{figure}

Here we briefly review the radiation spectrum associated to small angle scattering of an asymptotic parton off a pointlike target. Consider a massless quark of large momentum $p=(p^+,0,\vec{0}_\perp)$\footnote{We use light-cone variables, $p=(p^+,p^-,\pvec)$, with $p^{\pm} = p^0 \pm p^z$ and $\pvec \equiv \vec{p}_\perp$.} acquiring the transverse momentum $\vec{q}_\perp \equiv \qvec$ via single gluon exchange, see Fig.~\ref{fig:GBamplitude}a. We focus on small angle scattering, $|\qvec| \ll p^+$. The gluon emission amplitude induced by the latter scattering is given by the diagrams of Fig.~\ref{fig:GBamplitude}b, and was first derived by Gunion and Bertsch~\cite{Gunion:1981qs}. Denoting the radiated gluon momentum $k=(k^+,\kvec^2/k^+,\kvec)$ and focussing on soft ($x \equiv k^+/p^+ \ll 1$) and small angle ($k_\perp \ll k^+$) radiation, the gauge invariant radiation amplitude reads~\cite{Gunion:1981qs}
\be
\label{GB}
{\cal M}_{\rm rad} =  \hat{\cal M}_{\rm el} \cdot (2g) \cdot \left[  C_1 \, \frac{\kvec}{\kvec^{2}} + C_2 \, \frac{\kvec - \qvec}{(\kvec - \qvec)^{2}} - C_3 \, \frac{\kvec - x\qvec}{(\kvec - x\qvec)^{2}} \right] \cdot \epsvec \, .
\ee
In this expression $g$ is the QCD coupling, $\epsvec \equiv \vec{\varepsilon}_\perp$ stands for the (real) {\em physical} polarization of the radiated gluon,\footnote{Since $\epsvec$ formally disappears after squaring the amplitude and summing over the two physical polarization states, it will be implicit in the following.} and $C_1$, $C_2$, $C_3$ are the {\em full} color factors associated to the diagrams (i), (ii), (iii) of Fig.~\ref{fig:GBamplitude}b. Hence $\hat{\cal M}_{\rm el}$ corresponds to the Lorentz part of the elastic amplitude  ${\cal M}_{\rm el}$ of Fig.~\ref{fig:GBamplitude}a, \ie, ${\cal M}_{\rm el} = C_{\rm el} \, \hat{\cal M}_{\rm el}$ with $C_{\rm el}$ the `elastic' color factor.

In light-cone perturbation theory and in light-cone gauge $A^+=0$, it is easy to check that the first two terms 
in the r.h.s. of \eq{GB} correspond to initial state radiation (diagrams (i) and (ii) of Fig.~\ref{fig:GBamplitude}b), and the last term to final state radiation (diagram (iii)). The Gunion-Bertsch amplitude is then very simply interpreted, since $\psi(x,\kvec) \propto g \, \kvec \cdot \epsvec / \kvec^2$ is nothing but the light-cone wavefunction (in light-cone gauge) of the quark-gluon fluctuation in the incoming quark state \cite{Mueller:2001fv}. The radiated gluon with $k^+ = x p^+$ and $\vec{k}_\perp = \kvec$ can thus arise directly from the incoming quark-gluon fluctuation $\psi(x,\kvec)$ (diagram (i)), from the incoming fluctuation $\psi(x,\kvec-\qvec)$ followed by gluon scattering (diagram (ii)), or from the quark-gluon fluctuation in the quark {\em after} the scattering (diagram (iii)). The latter fluctuation has a total transverse momentum $\qvec$ and thus a wavefunction $\sim \psi(x,\kvec- x\qvec)$, as dictated by Lorentz 
transformation laws of light-cone wavefunctions~\cite{Lepage:1980fj}.

The general structure of the Gunion-Bertsch amplitude \eq{GB} is actually independent of the incoming parton type, as well as of the target type. In the following we consider an energetic parton of color charge $C_R$, with $C_R = C_A = N_c$ for a gluon and $C_R = C_F = (N_c^2-1)/(2N_c)$ for a quark. 

\begin{figure}[t]
\centering
\includegraphics[width=12cm]{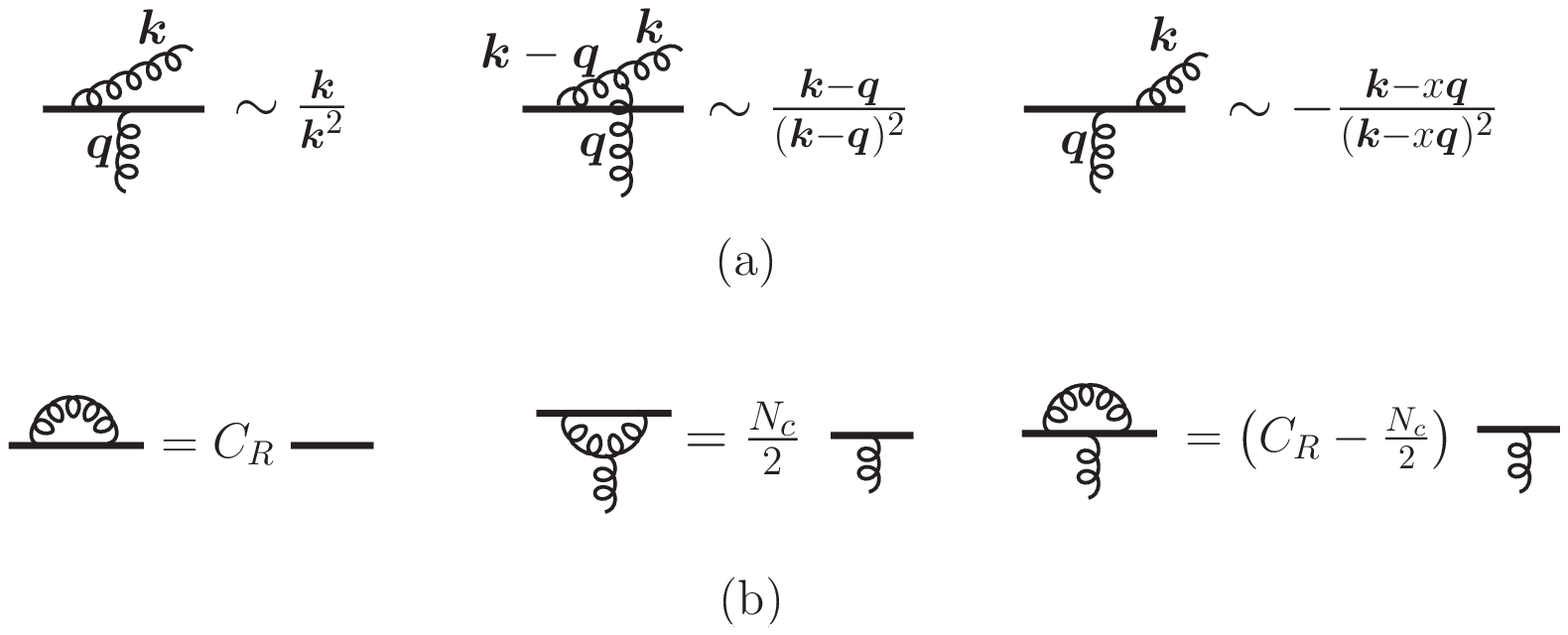}
\caption{Pictorial rules for (a) emission vertices and (b) color factors. The energetic parton of color charge $C_R$ is denoted by the solid line.}
\label{fig:pictorial-rules}
\end{figure}

The gluon radiation intensity $\dd I$ is obtained from \eq{GB} as 
\be
\dd I = \frac{\dd \sigma_{\rm rad}}{\sigma_{\rm el}} = \left( \frac{\sum |{\cal M}_{\rm rad}|^2 }{\sum |{\cal M}_{\rm el}|^2} \right) \frac{\dd k^+ \dd^2 \kvec}{2 k^+  (2\pi)^3} \, ,
\label{phase-space-factor}
\ee
where $|{\cal M}_{\rm rad}|^2$ and $|{\cal M}_{\rm el}|^2$ are summed over initial and final color indices. This gives the radiation spectrum 
\be
\label{qcd-spec-pict}
x \frac{\dd{I}}{\dd x \dd^2 \kvec} = \frac{\alpha_s}{\pi^2} \, \frac{\mbox{\raisebox{-6mm}{\hbox{\epsfxsize=10cm\epsfbox{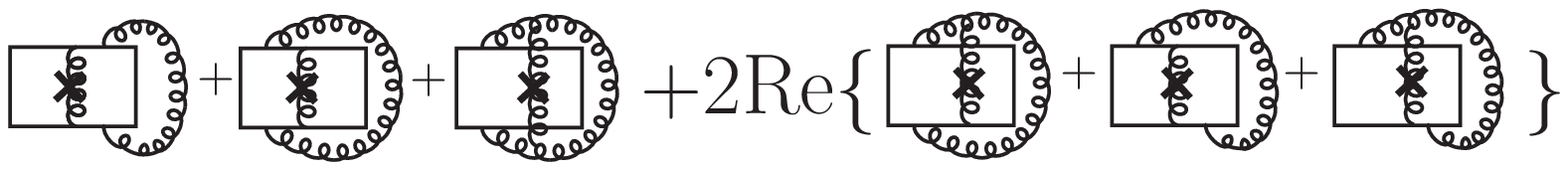}}}}}{\mbox{\raisebox{0mm}{\hbox{\epsfxsize=1cm\epsfbox{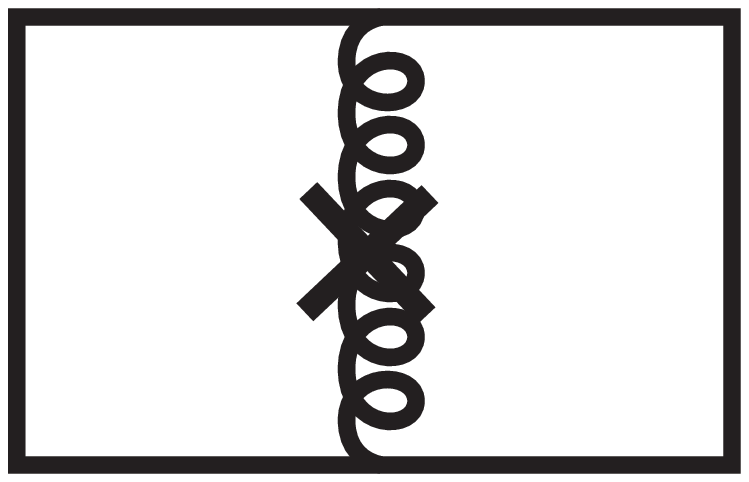}}}}} \, ,
\ee
where we use a pictorial representation defined as follows. For each diagram appearing in the numerator of \eq{qcd-spec-pict}, the upper (lower) part arises from one of the three contributions to the emission amplitude (conjugate amplitude) given in \eq{GB} or in Fig.~\ref{fig:GBamplitude}b. The Lorentz part of each diagram follows from the rules depicted in Fig.~\ref{fig:pictorial-rules}a, and the color factor from the rules of Fig.~\ref{fig:pictorial-rules}b.\footnote{For the pictorial representation of color factors, see for instance Ref.~\cite{Dokshitzer:1995fv} or Appendix B of Ref.~\cite{Baier:1996kr}.} The diagram in the denominator of \eq{qcd-spec-pict} is simply the color factor associated to the elastic cross section. We can thus rewrite \eq{qcd-spec-pict} explicitly as
\bea
x \frac{\dd{I}}{\dd x \dd^2 \kvec} = \frac{\alpha_s}{\pi^2} \left\{  C_R \left| \frac{\kvec -  x\qvec}{(\kvec - x\qvec)^{2}} \right|^2  + C_R \left| \frac{\kvec}{\kvec^{2}} \right|^2 + N_c \left| \frac{\kvec - \qvec}{(\kvec - \qvec)^{2}} \right|^2  \right. \hskip 10 mm &&  \nn \\
\left. - 2 \, \frac{N_c}{2} \,  \frac{\kvec \cdot (\kvec -  \qvec)}{\kvec^{2}(\kvec - \qvec)^{2}}  -2 \left( C_R -\frac{N_c}{2} \right) \frac{\kvec \cdot (\kvec - x\qvec)}{\kvec^{2}(\kvec - x\qvec)^{2}}   - 2 \, \frac{N_c}{2} \,   \frac{(\kvec -  \qvec) \cdot (\kvec - x\qvec)}{(\kvec - \qvec)^{2} (\kvec - x\qvec)^{2}} \right\} \, . \hskip 0 mm  && 
\eea
Grouping together the terms $\propto C_R$ and those $\propto N_c$ we obtain
\be
\label{GBspec-exact}
x \frac{\dd{I}}{\dd x \dd^2 \kvec} = \frac{\alpha_s}{\pi^2} \left\{  C_R \, \frac{x^2 \qvec^2}{\kvec^2 (\kvec - x\qvec)^{2}} + N_c \, \frac{\kvec \cdot (\kvec - x\qvec)}{\kvec^2 (\kvec - x\qvec)^{2}} \cdot \frac{(1-x) \qvec^2}{(\kvec - \qvec)^{2}}   \right\} \, .
\ee
The latter expression for the radiation spectrum induced by a single transverse momentum exchange $\qvec$ is exact (within the soft radiation and high-energy limit defined above), and will be used in the following to review several limiting cases. 

\subsection{Abelian-like contribution}
\label{sec:abelian}

In \eq{GBspec-exact} the term $\propto C_R$ is the abelian-like contribution to the spectrum. Indeed, in QED color factors as well as the diagram (ii) of Fig.~\ref{fig:GBamplitude}b are absent, and the photon emission amplitude off an electron is of the form
\be
\label{qed-amp}
{\cal M}_{\rm rad}^{\rm QED} \propto e \left[  \frac{\kvec}{\kvec^{2}}  -  \frac{\kvec - x\qvec}{(\kvec - x\qvec)^{2}} \right] \, ,
\ee
which square reproduces the first term of \eq{GBspec-exact}. This amplitude vanishes in the limit $x q_\perp \ll k_\perp$, \ie, when the scattering angle of the incoming electron $\theta_s \simeq  2 q_\perp/p^+$ is negligible compared to the photon emission angle $\theta \simeq  2 k_\perp/k^+$. In QED, a non-zero scattering angle $\theta_s$ is necessary to perturb the electron quantum state and thus to induce radiation. As a consequence, the abelian-like contribution to the $\kvec$-integrated spectrum $x \dd{I} / {\dd x}$ arises from $k_\perp \lsim x q_\perp$. Explicitly, 
\be
\label{qed-spec}
\left. x \frac{\dd{I}}{\dd x} \right|_{\rm abelian} = \frac{C_R \, \alpha_s}{\pi} \int \frac{\dd^2 \kvec}{\pi} \,  \frac{x^2 \qvec^2}{\kvec^2 (\kvec - x\qvec)^{2}} = 2\, \frac{C_R \, \alpha_s}{\pi} \, \log{\left( \frac{x^2 \qvec^2}{\Lambda_{\rm IR}^2} \right)}  \, ,
\ee
where we introduced an infrared cutoff $\Lambda_{\rm IR}$ to regularize the singularities at $\kvec = \zerovec$ and $\kvec = x\qvec$.\footnote{\label{footreg}The integral in \eq{qed-spec} is regularized by replacing $\kvec^2 \to \kvec^2 + \Lambda_{\rm IR}^2$ and $(\kvec - x\qvec)^{2} \to (\kvec - x\qvec)^{2} + \Lambda_{\rm IR}^2$ in the denominator of the integrand. Taking then the limit $\Lambda_{\rm IR} \to 0$ yields the r.h.s. of \eq{qed-spec}.} Those singularities contribute equally to \eq{qed-spec}, the logarithm arising either from the domain $\Lambda_{\rm IR} \lsim |\kvec| \ll x|\qvec|$ (singularity at $\kvec = \zerovec$), or from the domain $\Lambda_{\rm IR} \lsim |\kvec - x\qvec| \ll x|\qvec|$ (singularity at $\kvec = x\qvec$). This corresponds to the two well-known cones,  centered at $\kvec = \zerovec$ and $\kvec = x\qvec$ and of opening $\theta_s$, of abelian-like radiation.

\subsection{Eikonal limit -- Gunion-Bertsch spectrum}

In QCD, gluon radiation can occur even in the eikonal limit where the energetic parton trajectory is approximated as a straight line ($\theta_s \to 0$), due to the parton {\em color rotation} in the elastic scattering. Indeed, in the limit $x q_\perp \ll k_\perp$ the amplitude \eq{GB} remains finite,
\be
\label{qcd-amp}
{\cal M}_{\rm rad} \propto g \com{T_R^a}{T_R^b}  \left[ \frac{\kvec}{\kvec^{2}} - \frac{\kvec-\qvec}{( \kvec-\qvec )^{2}} \right]  \, ,
\ee
where $T_R^a$ are color matrices in the representation of the parton, $T_R^a T_R^a = C_R \mathbb{1}$. The eikonal limit allows to single out the purely non-abelian contribution to the radiation amplitude, as reflected by the commutator in \eq{qcd-amp}. The resulting spectrum, obtained by neglecting $x\qvec$ compared to $\kvec$ in \eq{GBspec-exact}, is the so-called Gunion-Bertsch spectrum~\cite{Gunion:1981qs} in the same $\kvec$ range,
\be
\label{GBspec}
x \frac{\dd{I}}{\dd x \dd^2 \kvec} \simeq \frac{N_c \, \alpha_s}{\pi^2} \, \frac{\qvec^2}{\kvec^{2} (\kvec-\qvec)^{2}} \hskip 1cm  (x q_\perp \ll k_\perp) \, .
\ee

\subsection{Hard scattering limit}
\label{sec:hard-scatt-pointlike}

We now consider the `hard scattering limit', defined as $q_\perp \gg k_\perp$, but still keeping a small scattering angle, $q_\perp \ll p^+$. In this limit the second term of \eq{GB} (diagram (ii) of Fig.~\ref{fig:GBamplitude}b) can be neglected, leading to 
\be
\label{GBhard}
{\cal M}_{\rm rad} \propto  g \left[ T_R^a T_R^b \, \frac{\kvec}{\kvec^{2}}  - T_R^b T_R^a \, \frac{\kvec - x\qvec}{(\kvec - x\qvec)^{2}} \right] \, .
\ee
The associated spectrum can be read off from \eq{GBspec-exact},
\be
\label{GBspec-hard}
x \frac{\dd{I}}{\dd x \dd^2 \kvec} \simeq \frac{\alpha_s}{\pi^2} \left\{  C_R \, \frac{x^2 \qvec^2}{\kvec^2 (\kvec - x\qvec)^{2}} + N_c \, \frac{\kvec \cdot (\kvec - x\qvec)}{\kvec^2 (\kvec - x\qvec)^{2}}  \right\} \hskip 6mm  (k_\perp \ll q_\perp) \, .
\ee
In addition to the two narrow `abelian cones' of opening $\theta_s$ (term $\propto C_R$), there is a non-abelian term $\propto N_c$\footnote{In \eq{GBspec-hard} the non-abelian contribution is $\propto N_c$ due to our assumption of a single gluon exchange. In general the purely non-abelian contribution is proportional to the color charge exchanged in the $t$-channel of the elastic scattering~\cite{Dokshitzer:1995if}. The same remark applies to the Gunion-Bertsch spectrum \eq{GBspec}.} arising from emission angles {\it larger} than the scattering angle $\theta_s$~\cite{Dokshitzer:1995if}. The latter property follows from averaging the non-abelian contribution over the azimutal angle of $\avec \equiv x\qvec$ using the identity 
\be
\label{azim-integral}
\int \frac{\dd \varphi_{a}}{2\pi} \, \frac{\kvec - \avec}{(\kvec - \avec)^{2}} = \frac{\kvec}{\kvec^{2}} \, \Theta(\kvec^2 - \avec^2) \, .
\ee
Effectively, \eq{GBspec-hard} can thus be written as
\be
\label{GBspec-hard-2}
x \frac{\dd{I}}{\dd x \dd^2 \kvec} \simeq \frac{\alpha_s}{\pi^2} \left\{  C_R \, \frac{x^2 \qvec^2}{\kvec^2 (\kvec - x\qvec)^{2}} + N_c \, \frac{\Theta(\kvec^2 - x^2\qvec^2)}{\kvec^2}  \right\} \hskip 6mm  (k_\perp \ll q_\perp) \, .
\ee
When $x q_\perp \sim k_\perp$, the abelian-like and non-abelian contributions have similar magnitudes. In the region 
$x q_\perp \ll k_\perp \ll q_\perp$, the non-abelian contribution dominates, and the spectrum \eq{GBspec-hard-2} obviously coincides with the Gunion-Bertsch spectrum \eq{GBspec}, 
\be
\label{log-interval}
x \frac{\dd{I}}{\dd x \dd^2 \kvec}  \simeq 
\frac{N_c \, \alpha_s}{\pi^2} \cdot \frac{1}{\kvec^2}  \hskip 10mm  (x q_\perp \ll k_\perp \ll q_\perp) \, .
\ee

\subsection{$\kvec$-integrated radiation spectrum}
\label{sec:qdep}

In situations where only the final energetic parton is tagged, the relevant quantity is the {\it total}, $\kvec$-integrated spectrum $x \dd{I}/{\dd x}$. Strictly speaking, the integral over $\kvec$ of \eq{GBspec-exact} is ill-defined due to singularities at $\kvec = \zerovec$, $\kvec = x\qvec$ and $\kvec = \qvec$, and is not a physical observable. In particular it includes collinear radiation with respect to the initial ($\kvec = \zerovec$) and final ($\kvec = x\qvec$) parton directions. However, what usually matters in practice is the {\it variation} of the total spectrum with $\qvec$, which variation is infrared safe and physical. 

Here we discuss the total radiation spectrum associated to a single scattering $\qvec$, and the variation of the total spectrum with $\qvec$. The main goal of this discussion is to provide some insight to the medium-induced spectrum in a finite size target studied in section \ref{sec:hard-smallangle}. In particular, we will show in section \ref{sec:hard-smallangle} that the induced spectrum off a parton of color charge $C_R$ has a color factor $2C_R -N_c$, \ie\ $N_c$ for an energetic gluon, and $-1/N_c$ for a quark. A {\it negative} induced energy loss of a quark is somewhat unusual, but can be intuitively understood from the case of a pointlike target. Indeed, we show below that the factor $2C_R -N_c$ is precisely the color factor of the contribution to the total spectrum arising from the `soft' $k_\perp$-domain, which turns out to be the only domain contributing to the induced loss. 

In the following we regularize the $\kvec$-integral by introducing an infrared cutoff $\Lambda_{\rm IR}$, as was done in section \ref{sec:abelian} for the abelian-like contribution (see footnote \ref{footreg}). The same results can be obtained using dimensional regularization. 

The exact $\kvec$-differential spectrum \eq{GBspec-exact} being approximated by \eq{GBspec} for $x q_\perp \ll k_\perp$ and by \eq{GBspec-hard-2} for $k_\perp \ll q_\perp$, the total spectrum can be simply obtained to logarithmic accuracy by introducing an arbitrary separation scale $\Lambda_{\rm S}$ satisfying $x q_\perp \ll \Lambda_{\rm S} \ll q_\perp$. The contribution to the total spectrum arising from $k_\perp \leq \Lambda_{\rm S}$ is obtained from \eq{qed-spec} and \eq{GBspec-hard-2},\footnote{Let us note that when restricted to the domain $k_\perp \leq \Lambda_{\rm S}$, the abelian-like contribution \eq{qed-spec} is modified by terms $\sim \morder{x^2 \qvec^2/\Lambda_{\rm S}^2}$ which are negligible compared to the dominant logarithmic term. Similarly, since \eq{GBspec-hard-2} is an approximation at $k_\perp \ll q_\perp$, there are corrections $\sim \morder{\Lambda_{\rm S}^2/ \qvec^2}$ (which can also be dropped) to the term $\propto N_c$ in Eq.~\eq{soft-k}.}
\be
\label{soft-k}
\left. x \frac{\dd{I}}{\dd x} \right|_{k_\perp \leq \Lambda_{\rm S}} \simeq \frac{\alpha_s}{\pi} \left[ \, 2 C_R \log{\left( \frac{x^2 \qvec^2}{\Lambda_{\rm IR}^2} \right)} + N_c \log{\left( \frac{\Lambda_{\rm S}^2}{x^2 \qvec^2} \right)} \, \right] \, .
\ee
The contribution from $k_\perp \geq \Lambda_{\rm S}$ follows from \eq{GBspec}, 
\be
\label{hard-k}
\left. x \frac{\dd{I}}{\dd x} \right|_{k_\perp \geq \Lambda_{\rm S}} \simeq \frac{N_c\,\alpha_s}{\pi} \int \frac{\dd^2 \kvec}{\pi} \,\frac{\qvec^2}{\kvec^2 \left[ (\kvec - \qvec)^{2} + \Lambda_{\rm IR}^2 \right] } \, \Theta(\kvec^2 - \Lambda_{\rm S}^2) \, ,
\ee
where the singularity at $\kvec = \qvec$ is screened by the infrared cutoff. To logarithmic accuracy, the integral in \eq{hard-k} is dominated by two logarithmic intervals, namely $\Lambda_{\rm S} \ll |\kvec| \ll |\qvec|$ and $\Lambda_{\rm IR} \ll |\kvec - \qvec| \ll |\qvec|$, and thus reads\footnote{Alternatively, the integral in \eq{hard-k} can be calculated analytically for any fixed $|\qvec|$, $\Lambda_{\rm S}$ and $\Lambda_{\rm IR}$. Taking then the limit $\Lambda_{\rm IR} \to 0$, we obtain \eq{hard-k-2} up to corrections $\sim \morder{\Lambda_{\rm S}^2/ \qvec^2}$.}
\be
\label{hard-k-2}
\left. x \frac{\dd{I}}{\dd x} \right|_{k_\perp \geq \Lambda_{\rm S}} \simeq \frac{N_c\,\alpha_s}{\pi} \left[ \, \log{\left( \frac{\qvec^2}{\Lambda_{\rm S}^2} \right)} +  \log{\left( \frac{\qvec^2}{\Lambda_{\rm IR}^2} \right)}  \, \right] \, .
\ee

When adding the contributions \eq{soft-k} and \eq{hard-k-2}, the dependence on the arbitrary scale $\Lambda_{\rm S}$ cancels out, as expected. Second, as already mentioned, the total spectrum diverges in the limit $\Lambda_{\rm IR} \to 0$ but its variation when $|\qvec|$ is shifted to $|\qvec'| \sim |\qvec|$ is infrared-safe, 
\be
\label{spectrum-shift}
\left. x \frac{\dd{I}}{\dd x}  \right|_{\qvec'} - \left. x \frac{\dd{I}}{\dd x}  \right|_{\qvec}  \simeq \frac{\alpha_s}{\pi} \left[ \, (2 C_R - N_c) \log{\left( \frac{\qvec'^2}{\qvec^2} \right)} + 2 N_c \log{\left( \frac{\qvec'^2}{\qvec^2} \right)} \, \right] \, ,
\ee  
where the first and second terms arise respectively from the `soft' ($k_\perp \sim  x q_\perp \ll \Lambda_{\rm S}$) and `hard' ($k_\perp \sim q_\perp \gg \Lambda_{\rm S}$) domains. 

The modification of the soft contribution ($k_\perp \ll \Lambda_{\rm S}$) with respect to an increase of $|\qvec|$ is proportional to $2 C_R - N_c$. Indeed, in the soft part \eq{soft-k} the abelian-like radiation at angles smaller than the scattering angle ($k_\perp \leq x q_\perp$, first term of \eq{soft-k}) {\it increases} with a rate $\propto 2C_R$, but the purely non-abelian radiation arising from larger angles ($k_\perp \geq x q_\perp$, second term of \eq{soft-k}), {\it decreases} with a rate $\propto N_c$. Hence a net variation $\propto 2C_R - N_c$. In the particular case of an asymptotic quark, we have $2 C_F - N_c =-1/N_c < 0$, implying that the number of gluons (of a given energy) radiated into a cone of fixed size {\it decreases} with increasing momentum transfer from the target. At the same time, the {\it total} number of radiated gluons increases with a rate $\propto 2 C_F + N_c$, due to radiation of gluons with $|\kvec| \sim |\qvec|$.  

Although the spectrum at zeroth order in opacity receives contributions from both soft and hard $k_\perp$-domains, we will see in section~\ref{sec:hard-smallangle} that in the {\it medium-induced} spectrum, the gluon $k_\perp$ is constrained to be {\it soft}, the additional scale $\ell_\perp \sim \Delta q_\perp \ll q_\perp$ coming into play in a finite size target acting as an upper cut-off. We thus expect the induced radiation to correspond to the medium-modification of the {\it soft part} of the total spectrum, and to have the same color factor $2C_R - N_c$.

\section{Small angle scattering of an asymptotic parton off a finite size target: fully coherent medium-induced radiation}
\label{sec:hard-smallangle}

\subsection{Setup: single hard scattering plus soft rescatterings}
\label{sec:setup}

We now consider the case where the energetic parton crosses a nuclear target of thickness $L$, still assuming a small deviation angle $\theta_s \ll 1$ and a negligible longitudinal momentum transfer to the medium. The parton can undergo several {\em soft} scatterings $\ellvec_i$ in the target, leading to nuclear transverse momentum broadening $\ave{\ellvec^2}$, with $\ellvec \equiv \sum \ellvec_i$. We focus on the kinematics where the final parton is `tagged' with a transverse momentum $\pvec'$ much larger than nuclear broadening. Thus $\pvec'$ must arise dominantly from a {\em single hard} scattering $|\qvec| \simeq |\pvec'| \gg |\ellvec|$.

We want to derive the medium-induced gluon radiation spectrum associated to such a hard scattering accompanied by any number of soft rescatterings. 
We focus on soft radiation as compared to the hard process, \ie, $x\equiv k^+/p^+ \ll 1$ but also $|\kvec| \ll |\qvec|$. The latter assumption will be justified a posteriori, when verifying that the $\kvec$-integrated medium-induced spectrum arises dominantly from $|\kvec| \sim |\ellvec|$.\footnote{For the calculation at first order in the opacity expansion, see \eq{kbounds} and the discussion after \eq{elossn1}. For the calculation at all orders in opacity, see the final comments of section \ref{useful-appr}.}

The medium-induced radiation spectrum can be derived in an opacity expansion \cite{Gyulassy:2000er}. At zeroth order, \ie, in absence of soft rescattering, the spectrum associated to the hard scattering $q_\perp$ is given by \eq{GBspec-hard} and cancels in the induced radiation spectrum. At a generic order $n$, the latter reads\footnote{See Appendix \ref{app:3.1} for a simple derivation of \eq{spec-order-n} in the limit of large formation time $\tf \gg L$.}
\be
x \frac{\dd I^{(n)}}{\dd x}  =
\frac{\alpha_s}{\pi^2} \int \dd^2 \kvec \left[ \prod_{i=1}^{n} \int \frac{\dd z_i}{\lambda_R} \int \dd^2 \ellvec_i \, V(\ellvec_i) \right]  \, \frac{1}{F_{n}} \, \sum \: \mbox{\raisebox{-7mm}{\hskip 0mm \hbox{\epsfxsize=4.7cm\epsfbox{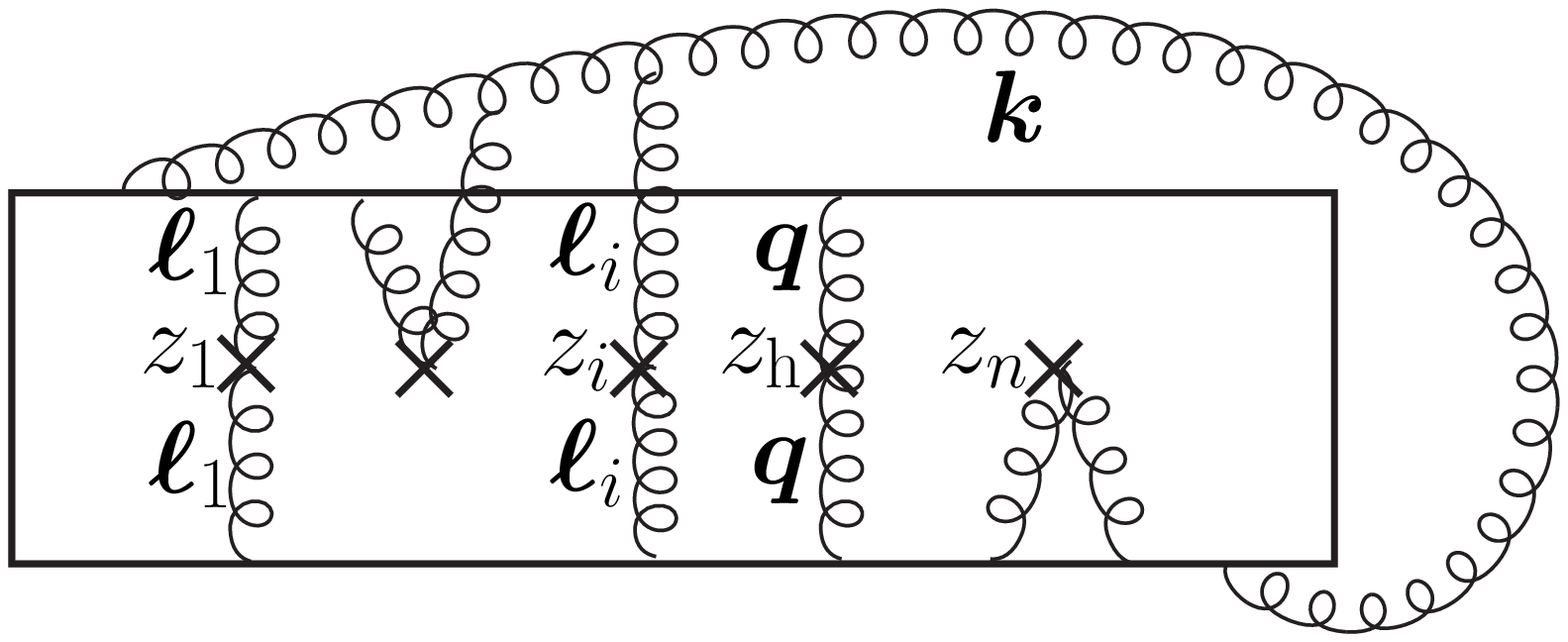}}}} 
\hskip 0mm \label{spec-order-n} 
\ee
using the pictorial rules defined in section~\ref{sec:pointlike} (see Fig.~\ref{fig:pictorial-rules}). The `elastic' color factor $F_n$ reads
\be
\label{Fn} 
F_n = \mbox{\raisebox{-4mm}{\hskip 0mm \hbox{\epsfxsize=3.7cm\epsfbox{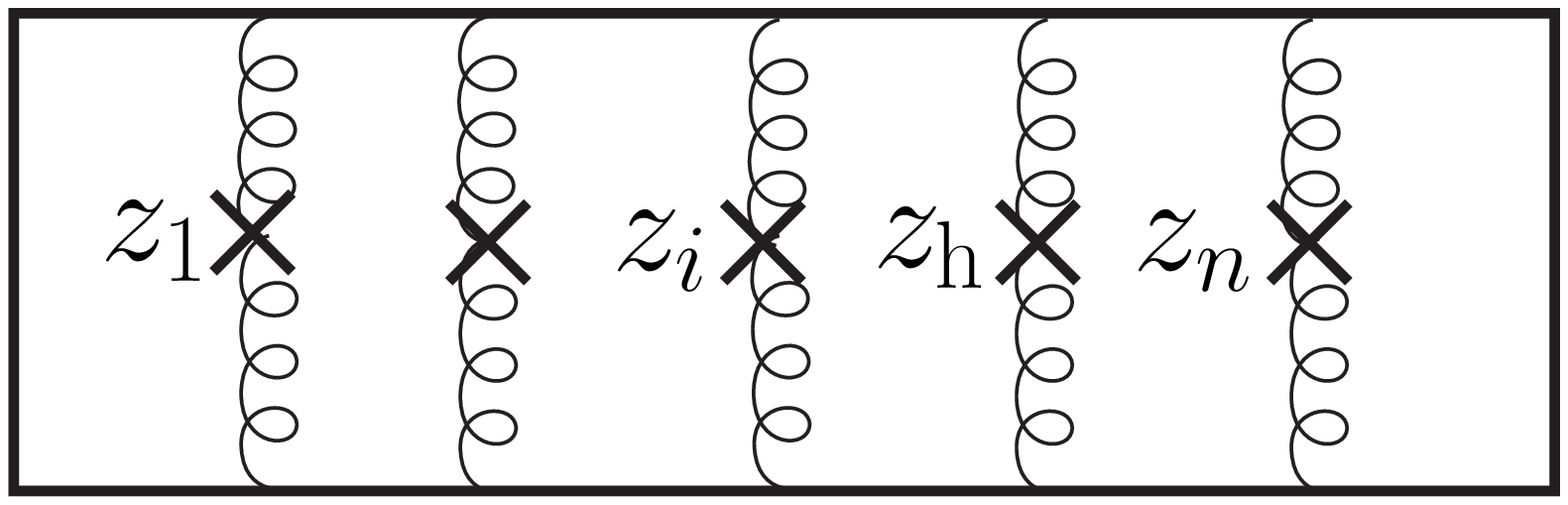}}}} = d_R \, (C_R)^{n+1} \, ,
\ee
where $d_R$ is the dimension of the parton color representation, $d_R = N_c$ for a quark and $d_R = N_c^2-1$ for a gluon. In \eq{spec-order-n} $\lambda_R$ is the parton elastic mean free path and $V(\ellvec)$ is the probability distribution for the transverse momentum transfer in the elastic scattering. Modelling each scattering center as a static center creating a screened Coulomb potential \cite{Baier:1996kr,Gyulassy:1993hr}, we have
\be
\label{coulomb-pot}
V(\ellvec) = \frac{\mu^2}{\pi (\ellvec^2 + \mu^2)^2} \; ; \ \ \int \dd^2 \ellvec \, V(\ellvec) = 1 \, .
\ee
As in Ref.~\cite{Baier:1996kr}, we choose $\lambda_R$ and the screening length $1/\mu$ of the scattering potential as the model parameters and assume $1/\mu \ll \lambda_R$, implying that successive soft scatterings can be treated as independent. They can thus be ordered in light-cone time $x^+ = t + z$, which in the high-energy limit is equivalent to ordering in usual time $t$ or longitudinal position, \ie, $0 < z_1 < z_ 2 < \ldots < z_n < L$ in \eq{spec-order-n}. The longitudinal position of the hard scattering $q_\perp$ will be denoted as $z_{\rm h}$, with $0 < z_{\rm h} < L$. 

The sum of diagrams in \eq{spec-order-n} is in general complicated to evaluate. However, anti\-ci\-pa\-ting that the medium-induced radiation spectrum $x \dd{I}/\dd x$ is dominated, in the present case of small angle scattering, by large gluon formation times $\tf \gg L$, the calculation can be greatly simplified by working in this limit from the beginning.  

When $\tf \gg L$, the dominant diagrams are those where the gluon emission times in the radiation amplitude and conjugate amplitude, denoted respectively by $t$ and $t^*$, occur either before or after the nuclear target. The diagrams where $t$ (or $t^*$) occurs within the target, $z_i < t < z_{i+1}$, are proportional to the difference $\sim (e^{i \varphi_{i+1}}- e^{i \varphi_{i}})$ between two `phase factors', where $\varphi_{i} \propto z_i/\tf$ (see Ref.~\cite{Baier:1996kr} and section \ref{sec:IS-FS} below), and vanish when $\tf \gg L$. The dominant diagrams in the sum of \eq{spec-order-n} can thus be classified into three classes, namely (i) $t, t^* > L$ (final state radiation)  (ii) $t, t^* < 0$ (initial state radiation) and (iii) $t < 0$ and $t^* >L$ (interference).\footnote{The generic diagram drawn in \eq{spec-order-n} belongs to the latter class.} 

Another simplification arises from our assumption $q_\perp \gg k_\perp$. In this limit the diagrams where the hard gluon $\qvec$ connects to the radiated gluon are negligible, similarly to the case of a pointlike target (section~\ref{sec:hard-scatt-pointlike}). 

Finally, let us stress that in the limit $t_{\mathrm{f}} \gg L$, the radiation cannot probe the target structure, but this does not imply that the radiation spectrum is medium-independent. Indeed, the spectrum may depend on the target size $L$ through the amount of soft rescattering $\ellvec^2$ (or through the {\em probability} for such rescattering) suffered by the incoming parton-gluon fluctuation across the target. This results in a non-vanishing medium-induced radiation, as confirmed in the following sections by an explicit calculation, at first and all orders in the opacity expansion.

\subsection{First order in the opacity expansion}

\subsubsection{QCD case}
\label{sec:n1qcd}

\begin{figure}[t]
\centering
\includegraphics[width=15.2cm]{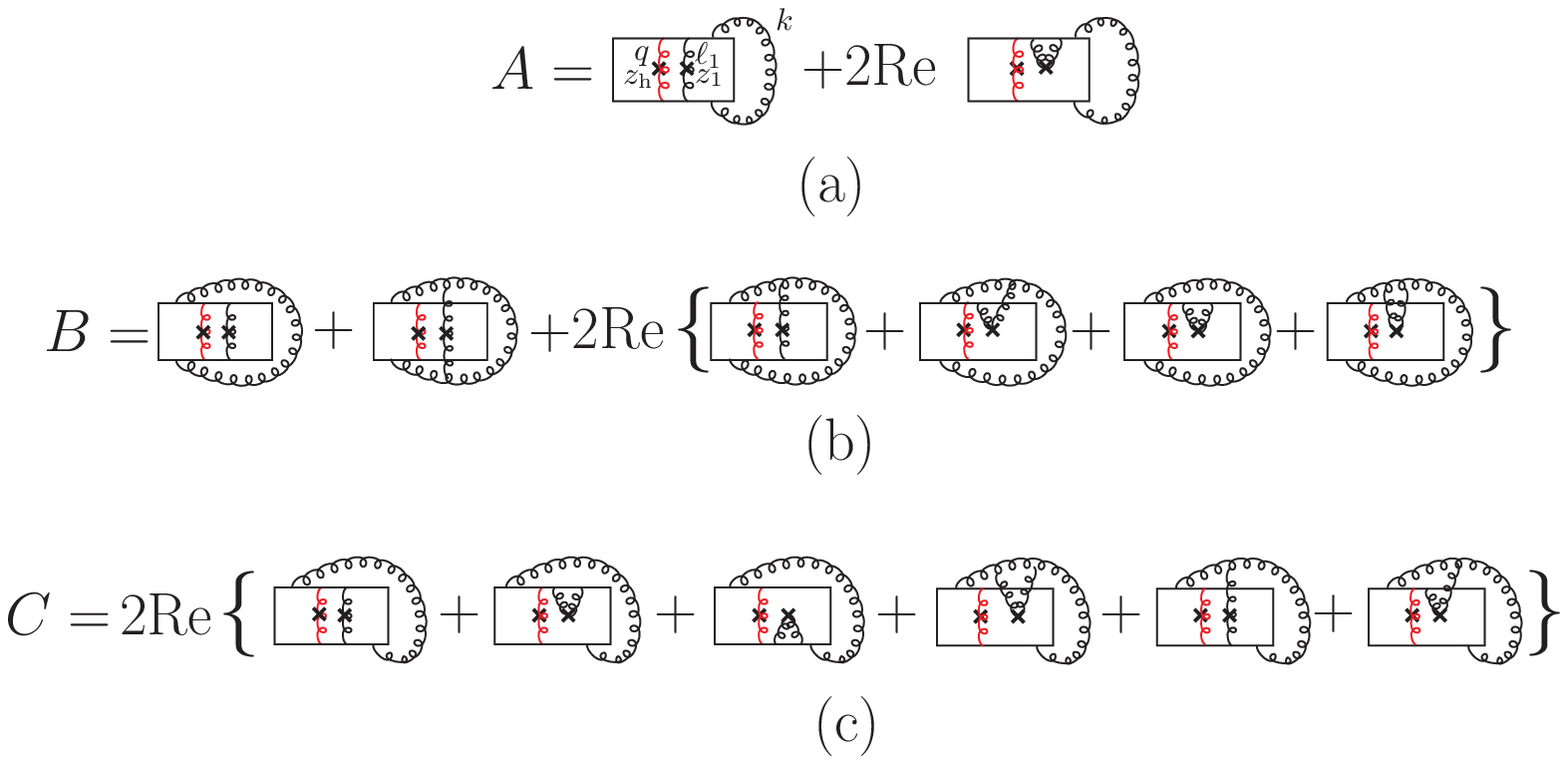}
\caption{Sets $A$, $B$, $C$ of diagrams, at first order in the opacity expansion, which may contribute to the radiation spectrum in the limit $t_{\mathrm{f}} \gg L$, see \eq{spec-order-1}. Only set $C$ turns out to be non-vanishing.}
\label{fig:ABCn1}
\end{figure}

At order $n=1$, and within the limit $\tf \gg L$ and $q_\perp \gg k_\perp$, the spectrum \eq{spec-order-n} is of the form
\be
x \frac{\dd I}{\dd x}  =
\frac{\alpha_s}{\pi^2} \int  \dd^2 \kvec \int \frac{\dd z_1}{\lambda_R} \int \dd^2 \ellvec_1 \, V(\ellvec_1)   \, \frac{1}{F_{1}} \, \left[ A + B + C \right] \, , 
\label{spec-order-1}
\ee
where the sets of diagrams $A$, $B$ and $C$ are shown in Fig.~\ref{fig:ABCn1}. When $\tf \gg L$ one may approximate $z_{\rm h} \simeq z_1 \to 0$ in those diagrams, which thus become independent of $z_{\rm h}$ and $z_1$ and in particular of their precise ordering. As a consequence we have $\int \dd z_1 =  L$ in \eq{spec-order-1}.

\bc (i) set $A$: $t, t^* > L$ \ec

The set $A$ of diagrams corresponds to the class (i) defined in section~\ref{sec:setup}, namely $t, t^* > L$ (final state radiation), and is obtained from the rules of Fig.~\ref{fig:pictorial-rules} as
\be
\label{setAn1}
\int  \dd^2 \kvec \  \frac{A}{F_{1}} = \int  \dd^2 \kvec \, \left\{ C_R \left| \frac{\kvec -  x\,(\qvec + \ellvec_1)}{(\kvec - x\,(\qvec+ \ellvec_1))^{2}} \right|^2  + 2 \cdot (-1) \cdot \halft \, C_R \left| \frac{\kvec -  x\qvec}{(\kvec - x\qvec)^{2}} \right|^2 \right\} = 0 \, .
\ee
The virtual contribution (right diagram of Fig.~\ref{fig:ABCn1}a) exactly cancels the real one (left diagram of Fig.~\ref{fig:ABCn1}a). The relative weights of virtual corrections have been discussed previously, see for instance Ref.~\cite{Baier:1998kq}. They contribute a factor $(-1)$, and a symmetry factor $\halft$ when the two gluon lines ($\ellvec_i$ and $-\ellvec_i$) connect to the same parton line (energetic parton or radiated gluon).  The associated color factor is obtained from the rules of Fig.~\ref{fig:pictorial-rules}b.

The vanishing of the set $A$ can be understood as follows. When $t, t^* > L$, the gluon is effectively produced after the nuclear medium. Gluon radiation should thus be similar to that in vacuum, up to the change in the direction of the final parton due to the rescattering $\ellvec_1$. When integrating over $\kvec$, the shift in the final parton direction becomes irrelevant, and the contribution to the {\em medium-induced} radiation spectrum vanishes.\footnote{In the limit $|\qvec| \gg |\ellvec|$ considered here, which amounts to neglect the deviation angle of the energetic parton due to soft rescatterings, the vanishing of the set $A$ occurs at fixed $\kvec$, \ie, at the integrand level in \eq{setAn1}.} In general we expect no contribution to the $\kvec$-integrated, medium-induced spectrum when $t$ and $t^*$ belong to a same time-interval where no rescattering happens.

\bc (ii) set $B$: $t, t^* < 0$ \ec

We now turn to the set $B$ of diagrams, see Fig.~\ref{fig:ABCn1}b, corresponding to the class (ii) $t, t^* < 0$ (initial state radiation). Applying the rules of Fig.~\ref{fig:pictorial-rules}, the sum of the six diagrams of the set $B$ simplifies to
\be
\label{setBn1}
\int  \dd^2 \kvec \ \frac{B}{F_{1}} = N_c \, \int \dd^2 \kvec \, \left[ \frac{1}{(\kvec-\ellvec_1)^{2}} -  \frac{1}{\kvec^{2}} \right] = 0 \, . 
\ee
The general property mentioned above is at work. When $t, t^* < 0$, the gluon is effectively radiated before the nuclear target, and the following soft rescatterings cannot affect the total ($\kvec$-integrated) radiation rate. 

\bc (iii) set $C$: $t<0$ and $t^* > L$ \ec

In the limit $\tf \gg L$, the only non-vanishing contribution to $x \dd{I}/\dd x$ arises from the set $C$ of interference diagrams with $t<0$ and $t^* > L$, see Fig.~\ref{fig:ABCn1}c. Applying the diagrammatic rules we obtain ($\ellvec \equiv \ellvec_1$)
\bea
\label{setCn1}
\int  \dd^2 \kvec \  \frac{C}{F_{1}} = \frac{2 \, C_R - N_c}{C_R}  \, \int \dd^2 \kvec \, \left\{ C_R \, \frac{\kvec}{\kvec^{2}} \cdot \left[ \frac{\kvec - x\qvec}{(\kvec - x\qvec)^{2}} - \frac{\kvec - x\,(\qvec + \ellvec)}{(\kvec - x\, (\qvec+\ellvec))^{2}} \right] \right. && \nn \\
+ \left. \frac{N_c}{2} \left[ \frac{\kvec}{\kvec^{2}}- \frac{\kvec -  \ellvec}{(\kvec - \ellvec)^{2}} \right] \cdot \left[ \frac{\kvec - x\qvec}{(\kvec - x\qvec)^{2}} + \frac{\kvec - x\,(\qvec + \ellvec)}{(\kvec - x\, (\qvec+\ellvec))^{2}} \right] \right\} \, . &&
\eea
When $|\ellvec| \ll |\qvec|$, the latter expression simplifies and the medium-induced spectrum \eq{spec-order-1} becomes (use $C_R \lambda_R = N_c \lambda_g$)
\be
\label{spectrum-n1}
x \frac{\dd I}{\dd x}  = (2 C_R - N_c) \, \frac{\alpha_s}{\pi^2} \, \frac{L}{\lambda_g}  \int \dd^2 \kvec  \int \dd^2 \ellvec \, V(\ellvec) \, \left[ \frac{\kvec -  \ellvec}{(\kvec - \ellvec)^{2}} -  \frac{\kvec}{\kvec^{2}} \right] \cdot \frac{-(\kvec - x\qvec)}{(\kvec - x\qvec)^{2}} \, .
\ee
The last factor in the integrand of \eq{spectrum-n1} is proportional to the (conjugate) wavefunction $\psi^*(x,\kvec- x\qvec)$ of the final parton-gluon fluctuation, see section~\ref{sec:exact-pointlike}. The other factor in between brackets can be interpreted as the incoming `medium-induced wavefunction'. Indeed, this factor is given by the incoming wavefunction `evolved' by rescatterings $\sim (\kvec - \ellvec)/(\kvec - \ellvec)^{2}$, from which the `vacuum' wavefunction $\sim \kvec / \kvec^2$ is subtracted. This interpretation of the radiation spectrum as the overlap between initial and final wavefunctions of the parton-gluon fluctuation holds to all orders in opacity, as we will see in section~\ref{sec:allorders}, see \eq{spec-alln-rescaled}.

The expression \eq{spectrum-n1} can be simplified by averaging over the azimutal angles of $\ellvec$ and $\qvec$. Using \eq{azim-integral} we get
\be
\label{kbounds}
\int \frac{\dd \varphi_{\ell}}{2\pi} \int \frac{\dd \varphi_{q}}{2\pi} \, \left[ \frac{\kvec}{\kvec^{2}}- \frac{\kvec -  \ellvec}{(\kvec - \ellvec)^{2}} \right] \cdot \frac{\kvec - x\qvec}{(\kvec - x\qvec)^{2}} = \frac{\Theta\left( \ellvec^2 - \kvec^2 \right) \,  \Theta\left( \kvec^2 - x^2 \qvec^2 \right) }{\kvec^2}  \, ,
\ee
leading to 
\be
\label{ipp}
x \frac{\dd I}{\dd x}  = (2 C_R - N_c) \,
\frac{\alpha_s}{\pi} \, \frac{L}{\lambda_g} \int_{x^2 \small{\qvec}^2}^{\infty} \dd \ellvec^2 \, \frac{\mu^2}{(\ellvec^2 + \mu^2)^2} \log{\frac{\ellvec^2}{x^2 \qvec^2}} \, .
\ee
The integral over $\ellvec^2$ can be performed using a simple integration by parts, giving the final expression for the medium-induced spectrum at first order in the opacity expansion,
\be
\label{specn1}
x \frac{\dd I}{\dd x}  = (2 C_R - N_c ) \, \frac{\alpha_s}{\pi} \, \frac{L}{\lambda_g} \, \log{\left( 1+ \frac{\mu^2}{x^2 \qvec^2} \right) } \, .
\ee

The associated average energy loss is proportional to $p^+$,
\be
\label{elossn1}
\Delta p^+ \equiv p^+ \int \dd x \, x \frac{\dd I}{\dd x} = \left( 2 C_R - N_c \right)  \alpha_s \frac{L}{\lambda_g} \, \frac{\mu}{q_\perp} \, p^+ \, .
\ee
Due to the fast decrease of the spectrum \eq{specn1} as $1/x^2$ when $x = k^+/p^+ \gg \mu/q_\perp$, the energy loss \eq{elossn1} is dominated by $x \sim \mu/q_\perp \ll 1$. Using \eq{kbounds} and \eq{ipp} we infer $\ellvec^2 \sim \kvec^2 \sim \mu^2$, and the energy loss is thus dominated by gluon formation times 
\be
\tf \sim \frac{k^+}{\kvec^2} \sim \frac{p^+}{q_\perp \, \mu} \, 
\ee
which scale as $p^+$ when $p^+ \to \infty$. This justifies our working assumption of large formation times $\tf \gg L$. The radiative energy loss \eq{elossn1} arises from gluon radiation which is {\it fully coherent} over the medium size $L$.

The overall color factor $2 C_R - N_c$ of the energy loss \eq{elossn1} equals $N_c$ for an energetic gluon, and $-1/N_c$ for a quark. The latter suppression in the large $N_c$ limit is simply understood, since in the quark case the relevant diagrams (Fig.~\ref{fig:ABCn1}c) are non-planar \cite{tHooft:1973jz}. The fact that the {\it quark} induced loss is {\it negative} is reminiscent from the case of a pointlike target, see the discussion in section \ref{sec:qdep}. We have seen there that the total radiation spectrum induced by a single hard scattering $\qvec$ acquires its $\qvec$-dependence from two domains of $k_\perp$, the soft $k_\perp \sim x q_\perp$ and hard $k_\perp \sim q_\perp$ domains, see \eq{spectrum-shift}, with associated color factors $2C_R - N_c$ and $2 N_c$ respectively. In the case of {\it medium-induced} radiation studied here, the rescattering $\ellvec$ effectively provides an upper cut-off on the radiated gluon transverse momentum, $\kvec^2 \leq \ellvec^2$, see \eq{kbounds}. This justifies the dominance of soft $k_\perp$ for medium-induced radiation, and elucidates the origin of the overall color factor.

\subsubsection{QED case and the Brodsky-Hoyer bound}
\label{sec:BH-bound}

It is instructive to state the result for the medium-induced radiation spectrum in QED. Replacing $N_c \to 0$, $C_R \to 1$ in \eq{setCn1} and substituting it into \eq{spec-order-1} we obtain
\bea
\label{specn1-QED}
\left. x \frac{\dd I}{\dd x} \right|_{\rm QED} =
\frac{2 \alpha}{\pi^2} \, \frac{L}{\lambda_e}   \Ave{ \int \dd^2 \kvec \  \frac{\kvec}{\kvec^{2}} \cdot \left[ \frac{\kvec - x\qvec}{(\kvec - x\qvec)^{2}} - \frac{\kvec - x\,(\qvec + \ellvec)}{(\kvec - x\, (\qvec+\ellvec))^{2}} \right] }_{\ellvec}  \hskip 10mm && \nn \\ 
= \frac{\alpha}{\pi^2} \, \frac{L}{\lambda_e}  \Ave{ \int  \dd^2 \kvec \left\{
\left[ \frac{\kvec}{\kvec^{2}} - \frac{\kvec -  x\,(\qvec+\ellvec)}{(\kvec - x\,(\qvec+\ellvec))^{2}} \right]^2 - \left[ \frac{\kvec}{\kvec^{2}} - \frac{\kvec -  x\qvec}{(\kvec - x\qvec)^{2}} \right]^2 \right\}  }_{\ellvec} \, , && 
\eea
with $\lambda_e$ the electron mean free path and $\ave{\{ \ldots \}}_{\ellvec} \equiv \int \dd^2 \ellvec \, V(\ellvec) \{ \ldots \}$. The interpretation of the latter spectrum as a {\em medium-induced} spectrum is clear from the second line of \eq{specn1-QED}, which is the difference between abelian-like spectra \eq{qed-spec} for single effective scatterings $\qvec+\ellvec$ and $\qvec$ respectively. In QED the medium-induced spectrum vanishes when the electron scattering angle is the same with or without rescattering, $|\qvec+\ellvec|/p^+ = |\qvec|/p^+$. 

In Ref.~\cite{Brodsky:1992nq}, the medium-induced spectrum in QED was derived assuming {\em both} $\tf \gg L$ {\em and} $\qvec+\ellvec = \qvec$ (in our notations), and was found to vanish. This led the authors to conclude that the assumption $\tf \gg L$ is invalid, leaving only radiation with small formation time $t_{\mathrm{f}} \lsim L$, resulting in some bound on induced energy loss, $\Delta E < {\rm cst} \cdot L^2$ when $E \to \infty$.

We stress here that it is instead the approximation $\qvec+\ellvec = \qvec$ which is too drastic. Indeed, relaxing the latter and keeping $\tf \gg L$, a non-zero medium-induced spectrum follows. Shifting $\kvec \to x\, \kvec$ in the first line of \eq{specn1-QED}, we see that the spectrum is actually independent of $x$, and thus given by the same expression with $x$ set to unity. Changing then $\kvec \to \kvec + \qvec$, the integral over $\kvec$ becomes identical to the integral in \eq{spectrum-n1} evaluated at $x=1$. We can thus read off the result from \eq{specn1},
\be
\left. x \frac{\dd I}{\dd x} \right|_{\rm QED} =  \frac{2 \alpha}{\pi} \, \frac{L}{\lambda_e}  \log{\left( 1 + \frac{\mu^2}{q_\perp^2} \right) } \, ,
\ee
leading to the medium-induced energy loss (use $\mu \ll q_\perp$) 
\be
\label{elossQED}
\left. \Delta p^+ \right|_{\rm QED} \propto  \alpha \, \frac{L}{\lambda_e} \cdot \frac{\mu^2}{q_\perp^2} \, p^+ \, .
\ee

To summarize, the Brodsky-Hoyer bound does not apply to the case under study of small angle scattering of an energetic charge, neither in QED (see \eq{elossQED}) nor in QCD (see \eq{elossn1}). In both cases the scaling  $\Delta p^+ \propto p^+$ originates from radiation with large formation times $\tf \gg L$, \ie, which is fully coherent over the medium size. In QED the medium-induced spectrum arises from the electron scattering angle being affected by in-medium rescatterings ($|\qvec+\ellvec| \neq  |\qvec|$), whereas in QCD a non-vanishing spectrum is found even in the limit $\qvec+\ellvec \simeq \qvec$, due to the parton color rotation. 

Finally, let us note that the Brodsky-Hoyer argument does apply to purely initial or purely final state radiation, both in QED and QCD. Indeed, in those cases radiation with large formation time cancels out, see \eq{setAn1} and \eq{setBn1}. As we will review in section \ref{sec:IS-FS}, this leads to a medium-induced energy loss satisfying (up to logarithms, see \eq{FS-eloss}) the Brodsky-Hoyer bound.

\subsection{All orders in the opacity expansion}
\label{sec:allorders}

\subsubsection{Exact derivation of the medium-induced spectrum}
\label{sec:exact-der}

\begin{figure}[t]
\centering
\includegraphics[width=14cm]{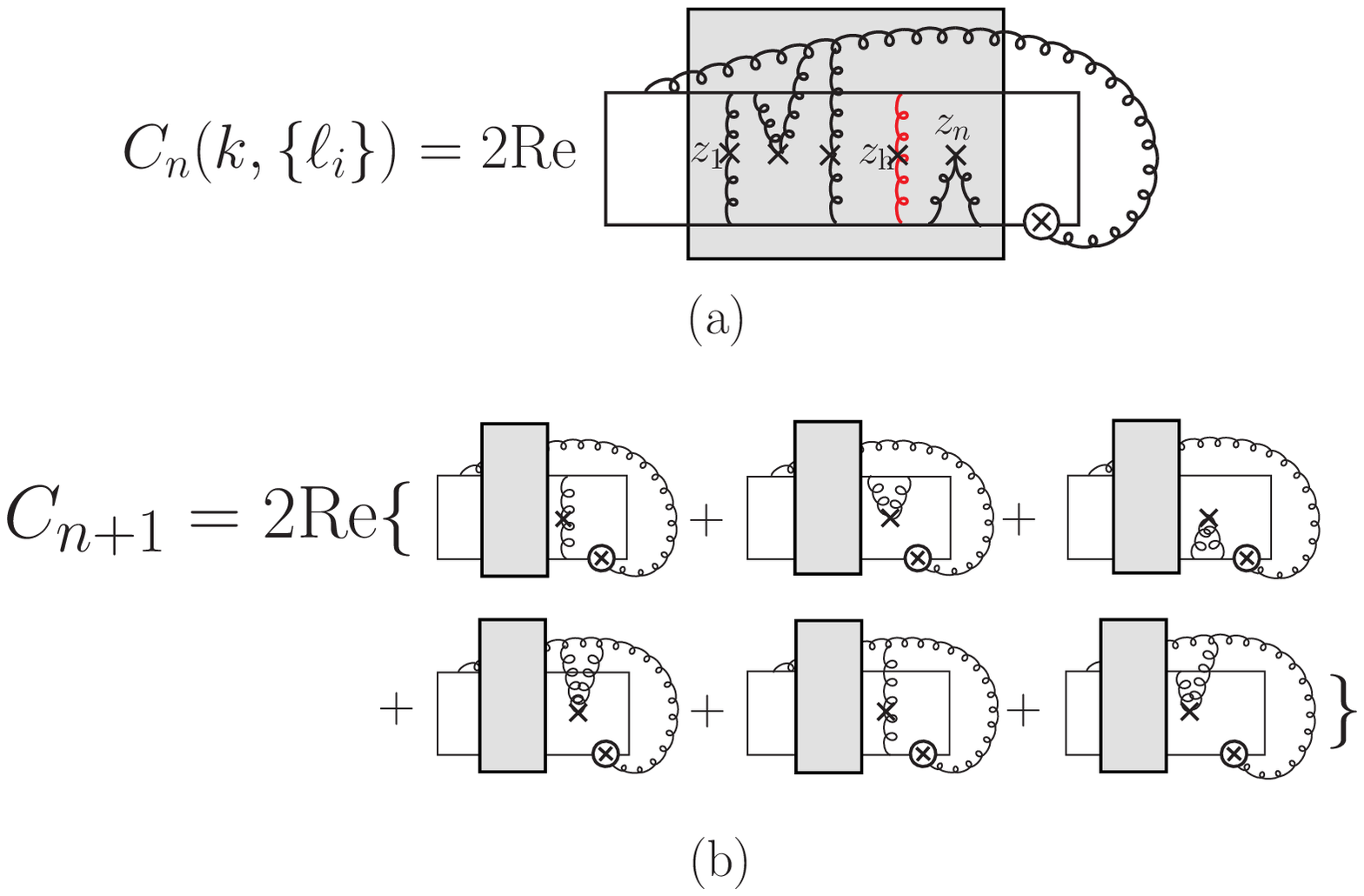}
\caption{(a) Set of diagrams $\Cvec_{n}$ contributing to \eq{fn}. The shaded area denotes all possible attachments of the lines $\ellvec_1, \ldots, \ellvec_n$, and the cross in the conjugate amplitude means that the emission vertex $-(\kvec - x\qvec)/(\kvec - x\qvec)^{2}$ is not included in the definition of $\Cvec_{n}$. (b) Diagrammatic expression of $\Cvec_{n+1}$ in terms of $\Cvec_{n}$, leading to \eq{C-rec}.}
\label{fig:Cnplus1}
\end{figure}

Here we derive the radiation spectrum to all orders in the opacity expansion. At any order $n$, it is clear that in the limit $\tf \gg L$ the dominant diagrams are the interference diagrams corresponding to $t<0$ and $t^* > L$, as in the case $n=1$, see Fig.~\ref{fig:ABCn1}c. It is convenient to rewrite \eq{spec-order-n} as 
\be
x \frac{\dd I^{(n)}}{\dd x}  =
\frac{\alpha_s}{\pi^2} \int \dd^2 \kvec \; \fvec_n(\kvec,L) \cdot \frac{-(\kvec - x\qvec)}{(\kvec - x\qvec)^{2}}  
\label{spec-order-n-bis} 
\ee
\be
\fvec_n(\kvec,L)  \equiv \frac{1}{n!}\left( \frac{L}{\lambda_R} \right)^n  \left[ \prod_{i=1}^{n} \int \dd^2 \ellvec_i \, V(\ellvec_i) \right] \, \frac{{\Cvec_n(\kvec; \left\{ \ellvec_1 \ldots  \ellvec_n \right \})}}{F_n}  \, ,
\label{fn}
\ee
where the quantity $\Cvec_{n}$ corresponds to the set of diagrams of Fig.~\ref{fig:Cnplus1}a, defined by the rules of Fig.~\ref{fig:pictorial-rules} but with the emission factor in the conjugate amplitude $-(\kvec - x\qvec)/(\kvec - x\qvec)^{2}$ {\it removed}.  

The medium-induced spectrum to all orders in the opacity expansion is obtained by summing \eq{spec-order-n-bis} over $n$, 
\bea
&& \hskip 2mm x \frac{\dd I}{\dd x}  = \sum_{n=1}^{\infty} x \frac{\dd I^{(n)}}{\dd x}  =
\frac{\alpha_s}{\pi^2} \int \dd^2 \kvec \; \left[ \fvec(\kvec,L) - \fvec_0(\kvec) \right] \cdot  \frac{-(\kvec - x\qvec)}{(\kvec - x\qvec)^{2}} \, ,  
\label{spec-alln-0}  \\
&& \fvec(\kvec,L)  \equiv \sum_{n=0}^{\infty}  \fvec_n(\kvec,L)  \;  ; \ \  \fvec_0(\kvec,L)  \equiv \fvec_0(\kvec) = (2C_R-N_c) \,  \frac{\kvec}{\kvec^2}  \, .
\label{f-def} 
\eea
The function $\fvec(\kvec,L)$ can be derived by noticing that at order $n+1$, the set of diagrams $\Cvec_{n+1}$ is given by the recurrency relation (see Fig.~\ref{fig:Cnplus1}b) 
\be
\Cvec_{n+1}(\kvec; \left\{ \ellvec_1 \ldots  \ellvec_{n+1} \right \}) = -N_c \,  \Cvec_n(\kvec; \left\{ \ellvec_1 \ldots  \ellvec_n \right \}) +N_c \, \Cvec_n(\kvec - \ellvec_{n+1} ; \left\{ \ellvec_1 \ldots  \ellvec_n \right \}) \, ,
\label{C-rec}
\ee
where the first term arises from the first four diagrams, and the second term from the last two diagrams of Fig.~\ref{fig:Cnplus1}b. From \eq{fn} and \eq{C-rec} we obtain
\be
\frac{\partial \fvec_{n+1}(\kvec,L)}{\partial L} = - \frac{N_c}{C_R \lambda_R} \int \dd^2 \ellvec \, V(\ellvec) \left[ \fvec_n(\kvec,L) - \fvec_n(\kvec - \ellvec,L) \right] \, .
\label{rec-rel}
\ee 
Using $C_R \lambda_R = N_c \lambda_g$ and defining $r \equiv L/\lambda_g$, we easily show that $\fvec(\kvec,L)$ defined in \eq{f-def} satisfies the equation 
\be
\frac{\partial \fvec(\kvec,r)}{\partial r} = -  \fvec(\kvec,r) + \int \dd^2 \ellvec \, V(\ellvec) \fvec(\kvec - \ellvec,r)   \; ; \ \ \fvec(\kvec,r=0) = \fvec_0(\kvec) \, .
\label{equation-f}
\ee
Rescaling $\fvec(\kvec,r)$ by the factor $2C_R-N_c$, the spectrum \eq{spec-alln-0} becomes
\be
x \frac{\dd I}{\dd x}  = (2C_R-N_c) \, \frac{\alpha_s}{\pi^2} \int \dd^2 \kvec \; \left[ \fvec(\kvec,r) - \fvec_0(\kvec) \right] \cdot (- \fvec_0(\kvec - x\qvec)) \, ,
\label{spec-alln-rescaled}
\ee
where $\fvec(\kvec,r)$ is solution of \eq{equation-f} with the rescaled initial condition $\fvec_0(\kvec) = \kvec/\kvec^2$. We easily check that keeping in \eq{spec-alln-rescaled} only the $n=1$ term of the opacity expansion, we recover the result \eq{spectrum-n1} obtained in section \ref{sec:n1qcd}.\footnote{To do this, replace $\fvec(\kvec,r) - \fvec_0(\kvec) \to \fvec_1(\kvec,r)$ in the bracket of \eq{spec-alln-rescaled}, and get $\fvec_1(\kvec,r)$ from \eq{rec-rel}.}

As is explicit from \eq{spec-alln-rescaled}, the spectrum is given by the overlap between the (conjugate) wavefunction $\sim - \fvec_0(\kvec - x\qvec)$ of the final state parton-gluon fluctuation and the `medium-induced wavefunction' $\fvec(\kvec,r) - \fvec_0(\kvec)$, where $\fvec(\kvec,r)$ is obtained by evolving the incoming wavefunction $\fvec_0(\kvec)$ according to \eq{equation-f}. The latter equation accounts for the modification of the radiated gluon transverse momentum $\kvec$ due to multiple soft rescatterings, and is identical to the equation satisfied by the gluon transverse momentum probability distribution $\rho(\kvec^2, t)$ after a `time' $t = z/\lambda_g$ travelled in the medium, see Ref.~\cite{Baier:1996sk}. 

By going to the impact parameter space, 
\be
\fvect(\bvec,r) = \int \dd^2 \kvec \,  \fvec(\kvec,r)  \, e^{- i  {\kvec \cdot \bvec}}  \, ,
\label{impact-space}
\ee
the set of equations \eq{equation-f} and \eq{spec-alln-rescaled} becomes
\bea
&& \frac{\partial \fvect(\bvec,r)}{\partial r} = - \left(1 - \Vt(\bvec) \right) \fvect(\bvec,r)  \; ; \ \ \fvect(\bvec,r=0) = \fvect_0(\bvec) = - 2i\pi \frac{\bvec}{\bvec^2} \, , \label{equation-f-bspace} \\ 
&& x \frac{\dd I}{\dd x}  = (2C_R-N_c) \, \frac{\alpha_s}{\pi^2} \int \! \! \frac{\dd^2 \bvec}{(2\pi)^2} \, [ \fvect(\bvec,r) - \fvect_0(\bvec) ] \cdot  (- \fvect_0(-\bvec))  \,  e^{i  x  \qvec \cdot \bvec} \, .
\label{spec-alln-bspace}
\eea 
The solution of \eq{equation-f-bspace} is
\be
\fvect(\bvec,r) = \fvect_0(\bvec) \, \exp{\left[- \left(1 - \Vt(\bvec) \right)  r \right]} \, ,
\ee
and \eq{spec-alln-bspace} can be rewritten as
\be
x \frac{\dd I}{\dd x}  = (2C_R-N_c) \, \frac{\alpha_s}{\pi^2} \int  \frac{\dd^2 \bvec}{\bvec^2} \, \left\{ 1 - \exp{\left[ - \left(1 - \Vt(\bvec) \right)  r \right]} \right\} \,  e^{i  x \qvec \cdot \bvec}  \, .
\label{spec-5}
\ee
Finally, using $\Vt(\bvec) = b \mu \, {\rm K}_1(b \mu)$ for the Coulomb potential \eq{coulomb-pot}, performing the azimutal integral in \eq{spec-5} and changing variable $B =b \mu$, we obtain
\bea
x \frac{\dd I}{\dd x} &=&  (2C_R-N_c) \, \frac{\alpha_s}{\pi} \, S[\Omega; r] \; ; \ \ \Omega \equiv \frac{x |\qvec|}{\mu} \; ; \ \ r \equiv \frac{L}{\lambda_g}  \, , \label{spec-alln} \\
S[\Omega;r] &\equiv& 2 \int_0^{\infty} \frac{\dd B}{B} \, {\rm J}_0(\Omega B)  \left\{ 1 -\exp{\left[- r \, (1-B \, {\rm K}_1(B) ) \right]}  \right\} \, . \label{sar}
\eea
The medium-induced spectrum to all orders in the opacity expansion given by Eqs.~\eq{spec-alln}--\eq{sar} is
one of the main results of our study. 

Let us emphasize that the factor $1 - \Vt(\bvec) = 1-b \mu \, {\rm K}_1(b \mu)$ in the exponential's argument is (up to a normalization factor) the {\it dipole} scattering cross section of a color singlet $q \bar{q}$ pair on a color center, see for instance Ref.~\cite{Zakharov:2000iz}.

\subsubsection{Limiting behaviors of the coherent spectrum} 
\label{useful-appr}

Here we focus on the parametric dependence of the exact spectrum \eq{spec-alln} (\ie\ of the function $S[\Omega;r]$ defined in \eq{sar}) in various limits. For simplicity we however assume a medium of large size, $r \equiv L/\lambda_g \gg 1$, and derive the limits {\it at fixed $r \gg 1$} of $S[\Omega;r]$ at small and large $\Omega$. 

\vskip 3mm
{\bf small $x$ limit}
\vskip 3mm

First, we note that at large $r \equiv L/\lambda_g \gg 1$, the integral over $B$ in \eq{sar} is dominated by small $B \ll 1$, and we can thus approximate $1-B \, {\rm K}_1(B)$ at small $B$, giving
\be
\label{sar-appr}
S[\Omega;r]  \mathop{\simeq}_{r \gg 1}  \int_0^{1} \frac{\dd B^2}{B^2} \, {\rm J}_0(\Omega B)  \left\{ 1 -\exp{\left[- r \frac{B^2}{4} \log{\frac{1}{B^2}} \, \right]}  \right\} \, .
\ee

\begin{figure}[t]
\centering
\includegraphics[width=7.1cm]{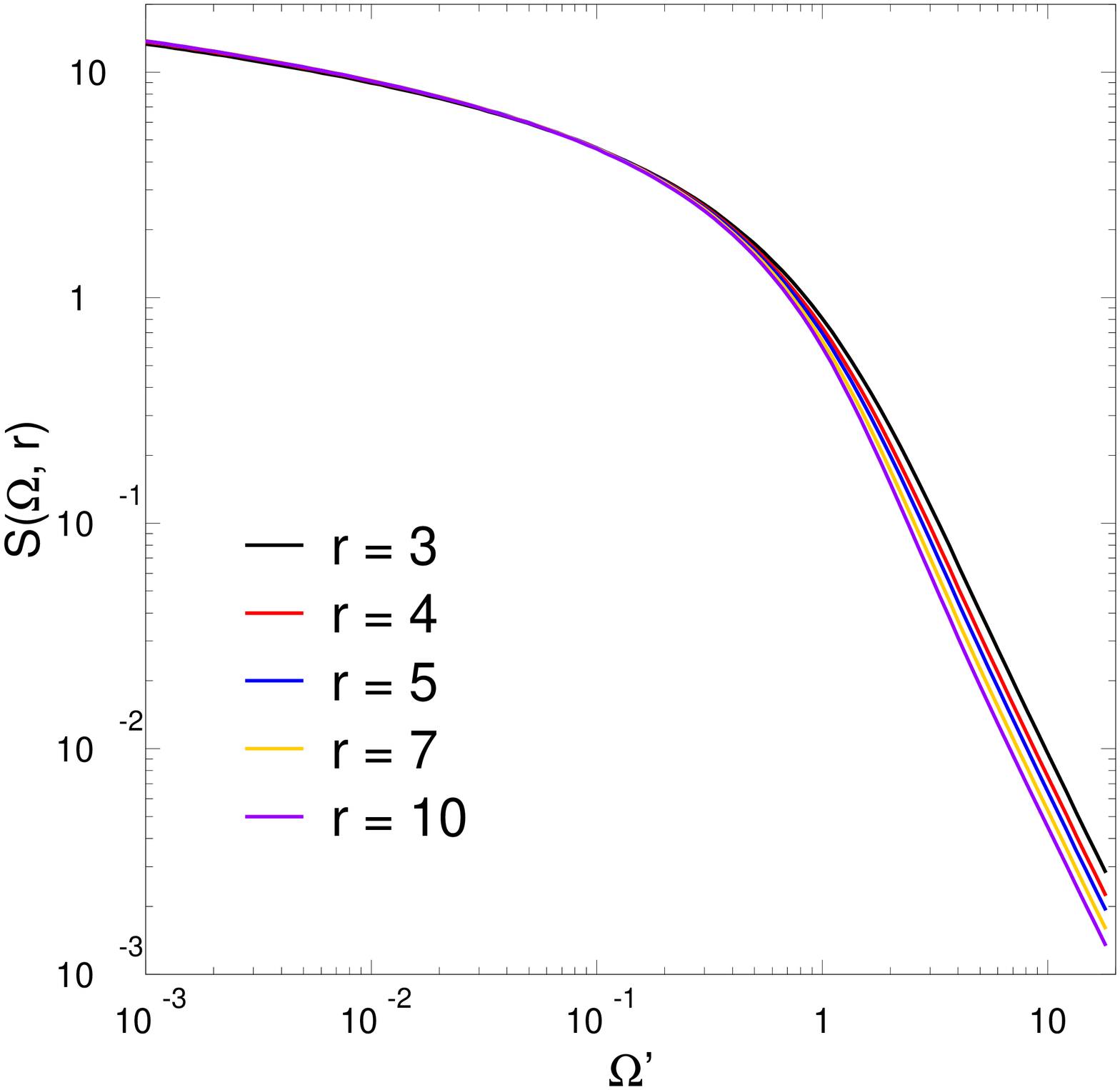} \hskip 8mm \includegraphics[width=7.1cm]{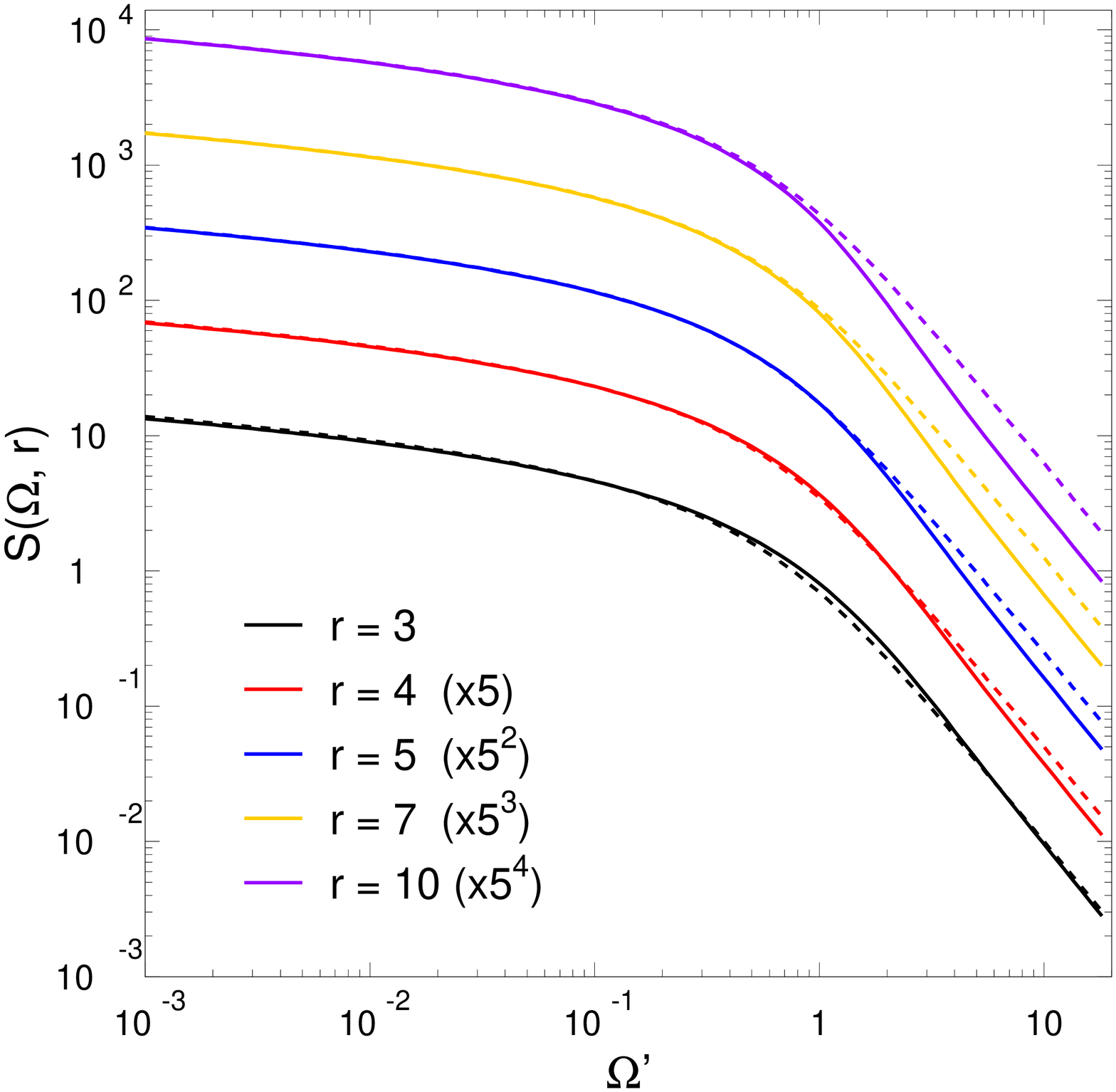}
\caption{{Left}:  approximate scaling of the function $S[\Omega; r]$ defined in \eq{sar} with $\Omega' = \Omega /\sqrt{r \log{r}}$, for various values of $r$. {Right}: $S[\Omega; r]$ (solid lines) compared to $S_{\rm appr}[\Omega; r]$ (dashed lines) for various values of $r$. For clarity the curves at a given $r$ are multiplied by a factor $5^n$.}
\label{fig:appr-scaling}
\end{figure}

In order to extract the small $\Omega$ limit of the latter expression (the precise meaning of `small $\Omega$' to be defined shortly), we first note that when $\Omega \to 0$, the integral \eq{sar-appr} is dominated by $B^2 \sim 1/r$. To {\it logarithmic accuracy} ($\log{r} \gg 1$), we can thus replace $\log{(1/B^2)} \to \log{r}$ in the exponential of \eq{sar-appr}, and after the change of variable $u = (r \log{r})  B^2$ we obtain
\be
S[\Omega;r] \simeq \int_0^{\infty} \frac{\dd u}{u}  \, {\rm J}_0 \left( \frac{\Omega}{\sqrt{r \log{r}}} \, \sqrt{u} \right) \left( 1 - e^{-u/4} \right)  
= -{\rm Ei} \left( - \frac{\Omega^2}{r \log{r}} \right)
\, ,
\label{u-int}
\ee
where we have set $r \log{r} \to \infty$ in the upper bound of the $u$-integral, and ${\rm Ei}(x) \equiv -\int_{-x}^{\infty} \frac{e^{-t}}{t} \dd t$. Thus, at `small $\Omega$', $S[\Omega; r]$ has an approximate scaling with the variable $\Omega' \equiv \Omega /\sqrt{r \log{r}}$, as illustrated in Fig.~\ref{fig:appr-scaling} (left). This very fact implies that the `small $\Omega$' limit should be defined as $\Omega' \ll 1$. Since \eq{u-int} strictly holds only in this limit, we can use, without loss of generality,  $-{\rm Ei}(-\Omega'^{\: \! 2}) \simeq \log{(1/\Omega'^{\: \! 2})}$ in the r.h.s. of \eq{u-int},  
leading to 
\be
\label{S-log-int}
S[\Omega;r]  \simeq \log{\left( \frac{r \log{r}}{\Omega^2} \right) } \ \ {\rm when} \ \ \Omega \ll \sqrt{r \log{r}} \, .
\ee
This yields the limiting behavior of the spectrum \eq{spec-alln} at small $x$,
\be  
\label{small-x}
x \frac{\dd I}{\dd x} \simeq (2C_R-N_c) \, \frac{\alpha_s}{\pi} \, \log{\left( \frac{\mu^2 r \log{r}}{x^2 \qvec^2 } \right) } \ \ {\rm when} \ \  x \ll \frac{\sqrt{\mu^2 r \log{r}}}{|\qvec|}  \, .
\ee

We stress that in order to obtain the small $x$ limit, we have replaced $\log{(1/B^2)} \to \log{r}$ in  
the exponential of \eq{sar-appr}, which amounts to neglect the logarithmic dependence on $b$ of the dipole scattering cross section at small $b$, and thus to approximate the latter as an exactly quadratic function of $b$, namely $1 - \Vt(\bvec) = 1- b \mu \, {\rm K}_1(b \mu) \propto b^2$. In other words, the so-called {\it harmonic oscillator} approximation (see \eg\ \cite{Zakharov:2000iz}) allows one to access the correct small $x$ limit of the coherent spectrum.  

We also note that our initial working assumption $|\kvec| \ll |\qvec|$ was le\-gi\-ti\-mate in order to derive \eq{small-x}. Indeed, from the above discussion (following \eq{sar-appr}), the behavior \eq{small-x} arises from impact parameters $b^2 = B^2/\mu^2 \sim (\mu^2 r \log{r})^{-1}$, \ie, from gluon transverse momenta $\kvec^2 \sim 1/b^2 \sim \mu^2 r \log{r} \sim \ellvec^2$, with $|\ellvec| \ll |\qvec|$ within the setup defined in section \ref{sec:setup}.

\vskip 3mm
{\bf large $x$ limit}
\vskip 3mm

When $\Omega' \gsim 1$ visible deviations to the scaling of $S[\Omega; r]$ in $\Omega'$ appear, see Fig.~\ref{fig:appr-scaling} (left). Indeed, at large $\Omega$ the above derivation does not hold,
due to the rapid oscillations of ${\rm J}_0(\Omega B)$ in \eq{sar-appr}.  
At large $\Omega$, we show in Appendix \ref{app:limits} that $S[\Omega; r]$ scales with $\Omega /\sqrt{r}$ (rather than with $\Omega' = \Omega /\sqrt{r \log{r}}$) and behaves as
\be 
S[\Omega;r] \simeq \frac{r}{\Omega^2} \ \ {\rm when} \ \ \Omega \gg \sqrt{r}  \, .
\label{large-Omega}
\ee
This gives the limiting behavior of the coherent spectrum \eq{spec-alln} at large $x$,
\be  
\label{large-x}
x \frac{\dd I}{\dd x} \simeq (2C_R-N_c) \, \frac{\alpha_s}{\pi} \, \frac{\mu^2 r}{x^2 \qvec^2 } \ \ {\rm when} \ \  x \gg \frac{\sqrt{\mu^2 r}}{|\qvec|}  \, .
\ee

It can be easily verified (see Appendix \ref{app:limits}) that obtaining the large $x$ behavior \eq{large-x} requires keeping the exact $b$-dependence of the dipole scattering cross section at small $b$, namely $1 - \Vt(\bvec) = 1- b \mu \, {\rm K}_1(b \mu) \propto b^2 \log{({1}/{b^2})}$. This can be simply understood as follows. At large $\Omega \gg \sqrt{r}$, the integral \eq{sar-appr} is not dominated by $B^2 \sim 1/r$ any longer, but by smaller values $B^2 \lsim 1/\Omega^2 \ll 1/r$, hence the importance of keeping the $\log{({1}/{B^2})}$ factor in the exponential when $\Omega \to \infty$.
The harmonic oscillator approximation would not be adequate to derive the parametric behavior of the coherent spectrum in the `large' $x$ domain $x \gg \sqrt{\mu^2 r}/{|\qvec|}$.

Although \eq{large-x} arises from smaller impact parameters than those contributing to \eq{small-x}, we can  still verify the consistency of the assumption $|\kvec| \ll |\qvec|$ used throughout our study. Indeed, here we have $b^2 = B^2/\mu^2 \sim 1/(\mu^2 \Omega^2) \sim 1/(x^2 \qvec^2)$, implying $\kvec^2 \sim x^2 \qvec^2 \ll \qvec^2$.

\vskip 3mm
{\bf a simple approximation}
\vskip 3mm

Although $S[\Omega; r]$ is not exactly a scaling function of $\Omega'$ (in particular when $\Omega' \gsim 1$), we consider the function 
\be
\label{sar-interpolation}
S_{\rm appr}[\Omega'] \equiv \log{\left( 1 + \frac{1}{\Omega'^{\: \! 2}} \right) } = \log{\left( 1 + \frac{r \log{r}}{\Omega^2} \right) } 
\ee
as a simple approximation to $S[\Omega; r]$. In the small $\Omega$ limit, \eq{sar-interpolation} has the same parametric behavior as $S[\Omega; r]$ (see \eq{S-log-int}). At large $\Omega$, both $S_{\rm appr}[\Omega']$ and $S[\Omega;r]$ behave as $1/\Omega^2$, with however the normalization of $S_{\rm appr}[\Omega']$ being enhanced (by a factor $\log{r}$) when compared to $S[\Omega;r]$ (see \eq{large-Omega}).

The numerical accuracy of \eq{sar-interpolation} is illustrated in Fig.~\ref{fig:appr-scaling} (right). The deviations of $S_{\rm appr}[\Omega']$ (dashed lines) with respect to $S[\Omega;r]$ (solid lines) are limited to roughly a percent when $\Omega' \leq 10^{-1}$ and to $\sim 10\%$ at $\Omega' = 1$. 
When $\Omega' \gg 1$, $S_{\rm appr}[\Omega']$ has a correct shape but
not the correct normalization, and the exact expression \eq{sar} should be preferred.
 
In summary, when $r = L/\lambda_g$ is large enough, $S_{\rm appr}[\Omega']$ is a good approximation to $S[\Omega;r]$ up to values $\Omega' \lsim \morder{1}$, and the spectrum \eq{spec-alln} can thus be approximated by
\be
\label{spec-alln-appr}
x \lsim \frac{\sqrt{\mu^2 r \log{r}}}{|\qvec|} \ \ \Rightarrow \ \ x \frac{\dd I}{\dd x} \simeq  (2C_R-N_c) \, \frac{\alpha_s}{\pi} \, \log{\left( 1 + \frac{\mu^2 r \log{r}}{x^2 \qvec^2} \right) }    \, .
\ee

Let us remark that the logarithm in the expression \eq{spec-alln-appr} at all orders in opacity ($r \gg 1$) can be formally obtained from the logarithm in the spectrum \eq{specn1} at first order in opacity ($r \ll 1$)  by replacing the typical transverse exchange in a single scattering $\mu^2$ by the typical exchange acquired in multiple {\it Coulomb} rescattering, given by $\mu^2 r \log{r}$ when $r \gg 1$~\cite{Baier:1996sk}.\footnote{Note also that when $r \ll 1$ the overall factor $L/\lambda_g$ in \eq{specn1} can be interpreted as the rescattering probability, $L/\lambda_g = r  \simeq 1-e^{-r}$, which becomes unity when $r \gg 1$.}

As already mentioned in Section~\ref{sec:intro}, in Refs.~\cite{Arleo:2010rb,Arleo:2012rs} the result \eq{spec-APS-0} was obtained (in the case $C_R = N_c$) from a calculation at first order in opacity, by a similar formal replacement. The above calculation shows that this procedure actually yields a correct approximation to the exact coherent spectrum \eq{spec-alln} at moderate $x$, provided $\Delta q_\perp^2(L)$ in \eq{spec-APS-0} is understood as the {\it typical} exchange in multiple Coulomb scattering $\Delta q_\perp^2(L) = \hat{q} L \log{(L/\lambda_g)}$. At large $x$ however, the result \eq{spec-APS-0} (with $\Delta q_\perp^2(L) = \hat{q} L \log{(L/\lambda_g)}$) overestimates by a factor $\log{(L/\lambda_g)}$ the limiting behavior \eq{large-x} of the exact spectrum \eq{spec-alln}. 

As already mentioned, deriving the large $x$ limit \eq{large-x} requires working {\it beyond} the harmonic oscillator approximation. In particular, the spectrum \eq{spec-armesto2} (corresponding to the case of $q \to q$ scattering mediated by color singlet $t$-channel exchange), which can be shown to ensue from Ref.~\cite{Armesto:2013fca} where the harmonic oscillator approximation is used, does not capture either the proper large $x$ limit of the spectrum. Note that \eq{spec-armesto2} coincides\footnote{Putting aside the overall color factor, and identifying $Q_{s {\rm A}}^2$ in \eq{spec-armesto2} as the typical transverse exchange in the target nucleus A.} with \eq{u-int}, which as we discussed holds only in the small $x$ region (see the discussion after \eq{u-int}). 

\subsubsection{Average energy loss}

The average coherent energy loss associated to the exact spectrum \eq{spec-alln} reads
\be
\label{eloss-all-n}
\Delta p^+ = p^+ \int_0^{\infty} \dd x \, x \frac{\dd I}{\dd x} = \left( 2 C_R - N_c \right)  \frac{\alpha_s}{\pi} \frac{\mu}{q_\perp} \, p^+  \int_0^{\infty} \! \dd \Omega \, S[\Omega;r] \, .
\ee
When $r \gg 1$, the first moment of $S[\Omega;r]$ is obtained starting from \eq{sar-appr} and using the identities $\int_0^{\infty} \dd \Omega \, {\rm J}_0(\Omega B) = 1/B$ and $\int_0^{\infty} \! \! \frac{\dd u}{u\sqrt{u}} (1 - e^{-u/4}) = \sqrt{\pi}$,
\be
\label{firstmoment}
\int_0^{\infty} \! \dd \Omega \, S[\Omega;r]  \simeq  \sqrt{\pi \, r \log{r}} \, .
\ee
We thus obtain
\be
\label{final-eq}
\Delta p^+  \mathop{\simeq}_{r \gg 1} \left( 2 C_R - N_c \right)  \frac{\alpha_s}{\sqrt{\pi}} \frac{\sqrt{\mu^2 r \log{r}}}{q_\perp}  \, p^+   \, .
\ee

Similarly to the result \eq{elossn1} obtained at first order in opacity, the average energy loss is proportional to $p^+$. The integral in \eq{eloss-all-n} is dominated by $\Omega \sim \morder{\sqrt{r \log{r}}}$, \ie, $x \sim \sqrt{\mu^2 r \log{r}}/q_\perp$. This stresses that our calculation assuming soft radiation ($x \ll 1$) is consistent provided the accumulated transfer $|\ellvec| = |\sum \ellvec_i | \sim  \sqrt{\mu^2 r \log{r}}$ is smaller than $q_\perp$. Note that the calculation of the average loss using the approximation \eq{sar-interpolation} instead of $S[\Omega; r]$ would overestimate the exact result \eq{final-eq} by a factor $\sqrt{\pi}$. This is because in the region $\Omega \sim \morder{\sqrt{r \log{r}}}$ (equivalently $\Omega' \sim \morder{1}$), deviations of $S_{\rm appr}[\Omega']$ with respect to $S[\Omega; r]$ are formally of order $\morder{1}$.  

\section{Fully coherent medium-induced radiation in other processes} 
\label{sec:other-processes}

Up to now the hard process was chosen as the small angle (but large enough $q_\perp$) scattering of a fast parton, either quark or gluon, see Fig.~\ref{fig:GBamplitude}a. In this case the fully coherent induced radiation derived above can be interpreted as the radiative energy loss of an asymptotic quark or gluon. The coherent radiation originates dominantly from the interference between initial and final state radiation (see Fig.~\ref{fig:ABCn1}c), and should arise independently of the hard process as long as the energetic incoming and outgoing particles are colored.

\begin{figure}[t]
\centering
\includegraphics[width=10cm]{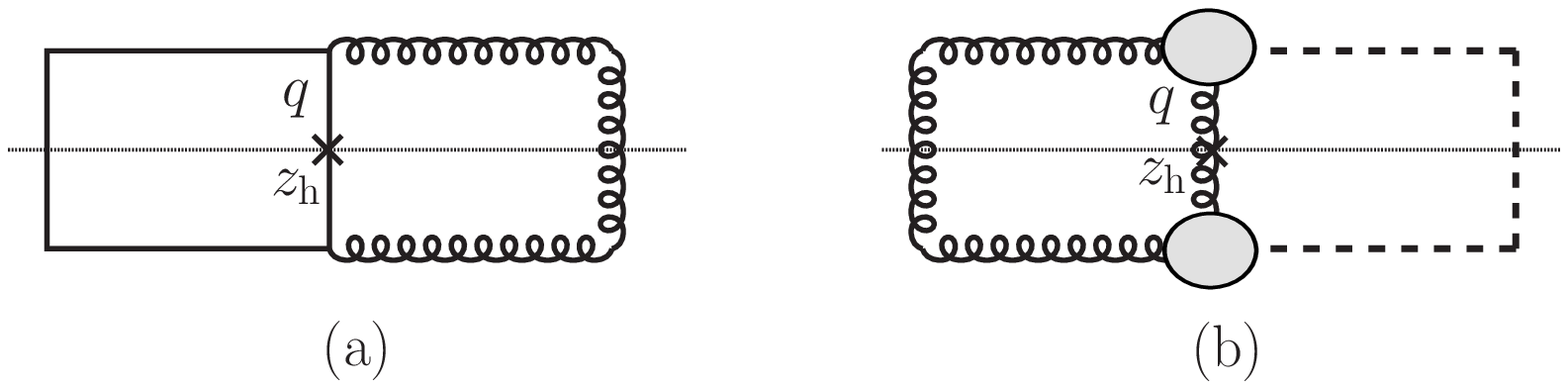}
\caption{(a) Scattering of an energetic quark to an energetic gluon via color-triplet $t$-channel exchange. (b) Scattering of an energetic gluon to an energetic massive pointlike color octet state (denoted by the dashed line) via single gluon exchange.}
\label{fig:other-hard}
\end{figure}

To illustrate this, we study in this section two examples where the nature of the charged particle is modified by the hard process. First, we consider the case of an incoming fast quark being scattered at small angle  (in the target rest frame) to an outgoing fast gluon through color-triplet $t$-channel exchange, see Fig.~\ref{fig:other-hard}a. Second, we study the scattering of a fast gluon to a fast pointlike color octet state of mass $M$ via single gluon exchange, see Fig.~\ref{fig:other-hard}b. This case is suited to the production of an octet heavy quark pair in gluon-gluon fusion, $gg \to [ Q \bar Q ]_{\rm 8}$, in the approximation where the $Q \bar Q$ pair is compact. In those examples the coherent radiation arises from the interference between emission amplitudes off different objects, and cannot be interpreted as the energy loss of a well-defined particle. It might be called `parton energy loss' by abuse of language, but it is simply the medium-induced radiation associated to a given hard process.

\subsection{Fast quark scattered to a fast gluon}
\label{sec:qtog}

Focussing as before on radiation with large formation time $\tf \gg L$, the medium-induced radiation spectrum associated to the hard process of Fig.~\ref{fig:other-hard}a is given, at first order in opacity, by some sets $A$, $B$, $C$ of diagrams analogous to those of Fig.~\ref{fig:ABCn1}. When $\tf \gg L$, the sets $A$ and $B$ corresponding to purely final and initial state radiation vanish, and what remains is the set $C$ of interference diagrams. 

\begin{figure}[t]
\centering
\includegraphics[width=15.2cm]{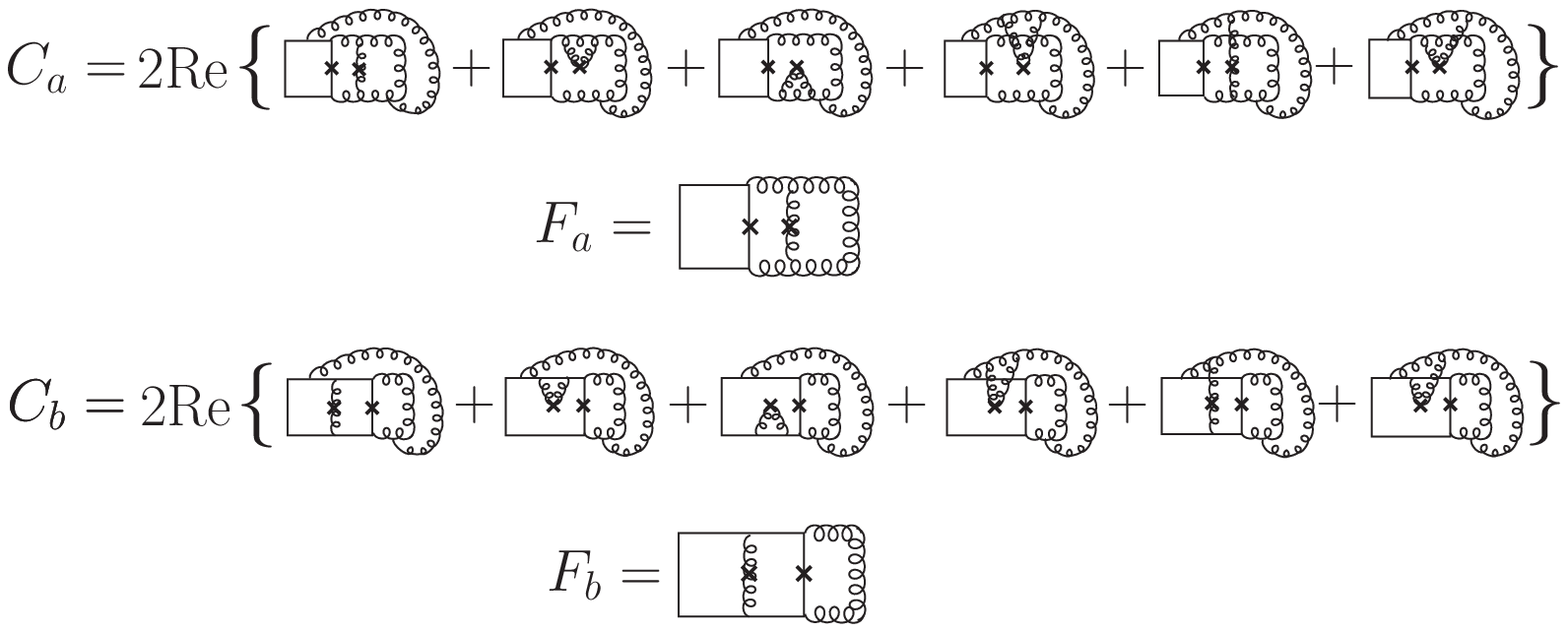}
\caption{Sets $C_a$ and $C_b$ of diagrams contributing to the radiation spectrum when $z_1 > z_{\rm h}$ and $z_1 < z_{\rm h}$ respectively (see \eq{z1>zh} and \eq{z1<zh}), as well as color factors $F_a$ and $F_b$.}
\label{fig:CaCb}
\end{figure}

In the present situation we must distinguish the cases where the rescattering center located at $z_1$ is before or after the hard scattering vertex. For $z_1 > z_{\rm h}$, what rescatters is a color octet (see Fig.~\ref{fig:other-hard}a), giving a contribution to the spectrum similar to \eq{spec-order-1}, 
\be
\left. x \frac{\dd I}{\dd x} \right|_{z_1 > z_{\rm h}}  =
\frac{\alpha_s}{\pi^2} \int  \dd^2 \kvec \int_{z_{\rm h}}^L \frac{\dd z_1}{\lambda_g} \int \dd^2 \ellvec \, V(\ellvec)   \, \frac{C_a}{F_a} \, , 
\label{z1>zh}
\ee
where the set $C_a$ is given in Fig.~\ref{fig:CaCb}a and $F_a = N_c^2 C_F$. For $z_1 < z_{\rm h}$, what rescatters is a color triplet, yielding the contribution
\be
\left. x \frac{\dd I}{\dd x} \right|_{z_1 < z_{\rm h}}  =
\frac{\alpha_s}{\pi^2} \int  \dd^2 \kvec \int_0^{z_{\rm h}} \frac{\dd z_1}{\lambda_q} \int \dd^2 \ellvec \, V(\ellvec)   \, \frac{C_b}{F_b}  \, , 
\label{z1<zh}
\ee
where the set $C_b$ is given in Fig.~\ref{fig:CaCb}b and $F_b = N_c C_F^2$. Note the proper normalization with respect to the quark mean free path $\lambda_q$ in \eq{z1<zh}.

Applying the pictorial rules of Fig.~\ref{fig:pictorial-rules} we easily obtain
\be
\frac{1}{N_c} \frac{C_a}{F_a} = \frac{C_F}{N_c^2} \frac{C_b}{F_b} =  \left[ \frac{\kvec}{\kvec^{2}}- \frac{\kvec -  \ellvec}{(\kvec - \ellvec)^{2}} \right] \cdot \frac{\kvec - x\qvec}{(\kvec - x\qvec)^{2}} \, .
\ee
Using $C_F \lambda_q = N_c \lambda_g$, the contributions from $z_1 > z_{\rm h}$ and $z_1 < z_{\rm h}$ add up to
\be
x \frac{\dd I}{\dd x}  = N_c \, \frac{\alpha_s}{\pi^2} \, \frac{L}{\lambda_g}  \int \dd^2 \kvec  \int \dd^2 \ellvec \, V(\ellvec) \, \left[ \frac{\kvec}{\kvec^{2}} - \frac{\kvec -  \ellvec}{(\kvec - \ellvec)^{2}} \right] \cdot \frac{\kvec - x\qvec}{(\kvec - x\qvec)^{2}} \, .
\label{spec-qtog}
\ee

The medium-induced spectrum associated to the hard process of Fig.~\ref{fig:other-hard}a is thus identical to the spectrum \eq{spectrum-n1} associated to the hard scattering of an asymptotic gluon.\footnote{\label{foot:colorfact}The fact that those spectra have an identical color factor $N_c$ is simply understood as follows. In general, the overall color factor of the fully coherent radiation, due to the structure of the interference term, is of the form $\sim 2 T_R^a T_{R'}^a$, where the incoming and outgoing particles are in the color representations $R$ and $R'$ respectively. Using $ 2 T_R^a T_{R'}^a = (T_R^a)^2 + (T_{R'}^a)^2  - (T_{R}^a - T_{R'}^a)^2 = C_R + C_{R'} - C_{t}$, where $C_{t}$ is the color charge of the $t$-channel exchange, we recover the factor $N_c+N_c-N_c= N_c$ in the case of asymptotic gluon scattering considered in section \ref{sec:hard-smallangle}, and the factor $C_F + N_c -C_F =N_c$ in the present case.} The spectrum \eq{spec-qtog} and associated radiative loss were evaluated analytically in section \ref{sec:n1qcd}, see \eq{specn1} and \eq{elossn1}. As expected, a fully coherent radiative loss proportional to $p^+$ arises despite different (but non-zero) initial and final color charges in the process of Fig.~\ref{fig:other-hard}a. The above results trivially extend to all orders in opacity, as well as to the process of a fast gluon scattered to a fast quark.

\subsection{Fast gluon scattered to a compact massive color octet}

We now consider a hard process where an octet state of mass $M$ is produced in gluon-gluon fusion, see Fig.~\ref{fig:other-hard}b. For soft gluon radiation with $\tf \gg L$, the precise partonic content of the octet state ($Q \bar Q$, $Q Q Q$,...) as well as the internal structure of the hard process producing it (the blob in Fig.~\ref{fig:other-hard}b) is irrelevant as long as the octet state is compact enough. Within this assumption the amplitude for soft gluon emission off the final state is effectively the same as off a pointlike, massive gluon (the dashed line in Fig.~\ref{fig:other-hard}b). 

The calculation of the fully coherent radiation spectrum is thus identical to the case of an asymptotic gluon, up to the replacement
\be
\label{insert-mass}
\frac{\kvec - x\qvec}{(\kvec - x\qvec)^{2}} \to \frac{\kvec - x\qvec}{(\kvec - x\qvec)^{2}+x^2 M^2}
\ee
to account for the mass dependence of the gluon emission vertex off the final state. 

\subsubsection{First order in opacity}

At first order in the opacity expansion, we start from the expression \eq{spectrum-n1} modified accordingly, 
\be
x \frac{\dd I}{\dd x}  = N_c \, \frac{\alpha_s}{\pi^2} \, \frac{L}{\lambda_g}  \int \dd^2 \kvec  \int \dd^2 \ellvec \, V(\ellvec) \, \left[ \frac{\kvec}{\kvec^{2}}- \frac{\kvec -  \ellvec}{(\kvec - \ellvec)^{2}} \right] \cdot \frac{\kvec - x\qvec}{(\kvec - x\qvec)^{2}+x^2 M^2} \, .
\label{spec-massive-octet}
\ee
The explicit calculation is performed along the same lines as in section \ref{sec:n1qcd}. We first average over the azimutal angles of $\ellvec$ and $\qvec$, 
\be
\int \frac{\dd \varphi_{\ell}}{2\pi} \, \left[ \frac{\kvec}{\kvec^{2}}- \frac{\kvec -  \ellvec}{(\kvec - \ellvec)^{2}} \right] = \Theta\left( \ellvec^2 - \kvec^2 \right) \, \frac{\kvec}{\kvec^{2}}  \, ,  
\ee
\be
\int \frac{\dd \varphi_{q}}{2\pi} \, \frac{\kvec - x\qvec}{(\kvec - x\qvec)^{2}+x^2 M^2} = \frac{1}{2}  \left[ 1 + \frac{\kvec^2 - x^2 M_\perp^2 }{\sqrt{(\kvec^2 - x^2 M_\perp^2)^2 + 4 x^2 M^2 \kvec^2}} \right] \frac{\kvec}{\kvec^{2}} \equiv \Phi(\kvec^2) \, \frac{\kvec}{\kvec^{2}} \, , 
\ee
where $M_\perp \equiv \sqrt{M^2 + \qvec^2}$, to obtain
\be
x \frac{\dd I}{\dd x}  = N_c \, \frac{\alpha_s}{\pi} \, \frac{L}{\lambda_g} \int_0^{\infty} \! \dd \ellvec^2 \, \pi V(\ellvec) \, \int_0^{\ellvec^2} \! \frac{\dd \kvec^2}{\kvec^2} \, \Phi(\kvec^2) \, .
\label{spec-massive-octet-2}
\ee
Integrating by parts yields
\be
x \frac{\dd I}{\dd x}  = N_c \, \frac{\alpha_s}{\pi} \, \frac{L}{\lambda_g} \int_0^{\infty} \! \dd \ellvec^2 \, \frac{\mu^2}{\ellvec^2 (\ellvec^2 + \mu^2)} \, \Phi(\ellvec^2) \, .
\label{spec-massive-octet-3}
\ee
With the change of variable $u=\ellvec^2/(x^2 M_\perp^2)$ we arrive at 
\be
x \frac{\dd I}{\dd x}  = N_c \, \frac{\alpha_s}{\pi} \, \frac{L}{\lambda_g} \, \Sigma(\Omega_{\perp}, \delta) \; ; \ \ \Omega_{\perp} \equiv \frac{x M_\perp}{\mu} \; ; \ \ \delta \equiv \frac{M}{M_\perp}  \, ,
\label{spec-massive-octet-n1}
\ee
where the function $\Sigma$ is defined as
\be
\Sigma(\Omega_{\perp}, \delta) \equiv \int_0^{\infty} \! \frac{\dd u}{2 u \, (\Omega_{\perp}^2 u +1)} \left[ 1 + \frac{u-1}{\sqrt{(u-1)^2 + 4 u \,\delta^2}} \right] \, .
\ee
The integral over $u$ can be evaluated analytically, providing an exact expression of the spectrum 
\eq{spec-massive-octet-n1}. However, a simple approximation to the spectrum can be obtained by examining the limiting behaviors of $\Sigma(\Omega_{\perp}, \delta)$. 
\bi
\item[(i)]  For $\Omega_{\perp} \ll 1$ and fixed $\delta$ (note that $0 \leq \delta \leq 1$) , the integral over $u$  defining $\Sigma(\Omega_{\perp}, \delta)$ is dominated by the logarithmic interval $1 \ll u \ll 1/\Omega_{\perp}^2$, leading to
\be
\label{Sigma-limit1}
\Sigma(\Omega_{\perp}, \delta) \, \mathop{\simeq}_{\Omega_{\perp} \ll 1} \, \log{\left( \frac{1}{\Omega_{\perp}^2} \right) } \, .
\ee
\item[(ii)]  When $\Omega_{\perp} \gg 1$, the $u$-integral can be shown to be dominated by $ 1/\Omega_{\perp}^2 \ll u \ll 1$, 
\be
\label{Sigma-limit2}
\Sigma(\Omega_{\perp}, \delta) \, \mathop{\simeq}_{\Omega_{\perp} \gg 1} \, \frac{1}{\Omega_{\perp}^2} \int_{1/\Omega_{\perp}^2}^1 \frac{\dd u}{2 u^2}  \left[ 2 \delta^2 u \right] \simeq \frac{\delta^2 \log{\left( \Omega_{\perp}^2 \right)}}{\Omega_{\perp}^2} \, .
\ee
\ei
Those limiting behaviors suggest the simple approximation
\be
\Sigma(\Omega_{\perp}, \delta) \simeq \log{\left( 1+ \frac{1}{\Omega_{\perp}^2} \right) } \, ,
\ee
which reproduces the leading parametric behaviors of $\Sigma(\Omega_{\perp}, \delta)$ in the limits $\Omega_{\perp} \ll 1$ and $\Omega_{\perp} \gg 1$. The spectrum \eq{spec-massive-octet-n1} can thus be approximated as 
\be
\label{spec-massive-octet-n1-appr}
x \frac{\dd I}{\dd x}  \simeq N_c \, \frac{\alpha_s}{\pi} \, \frac{L}{\lambda_g} \, \log{\left( 1+ \frac{\mu^2}{x^2 M_\perp^2} \right) } \, .
\ee
This result is simply obtained from the spectrum \eq{specn1} by replacing the hard scale $q_\perp$ (in the $M=0$ case of section \ref{sec:n1qcd}) by $M_\perp$. The spectrum \eq{spec-massive-octet-n1-appr} is not exact (except when $M=0$), but as mentioned above exhibits a correct parametric dependence. In particular it yields the average radiative loss
\be
\Delta p^+ \equiv p^+ \int \dd x \, x \frac{\dd{I}\ }{\dd x} = N_c \, \alpha_s \frac{L}{\lambda_g} \, \frac{\mu}{M_\perp} \, p^+ \, ,
\label{eloss-massive-appr}
\ee
to be compared to the exact loss derived from \eq{spec-massive-octet-n1}, 
\be
\Delta p^+ =  N_c \, \alpha_s \frac{L}{\lambda_g} \, \frac{\mu}{M_\perp} \, p^+ \, \Xi(\delta) \; ; \ \ \Xi(\delta) \equiv \int_0^{\infty} \frac{\dd y}{\pi} \, \Sigma(y, \delta) \, .
\label{eloss-massive-octet-n1-exact}
\ee
Eqs.~\eq{eloss-massive-appr} and \eq{eloss-massive-octet-n1-exact} differ only by the factor $\Xi(\delta)$, which can be checked to be a monotonous function of $\delta = M/M_\perp$, with values $1$ and $\pi/2$ at $\delta = 0$ and $\delta = 1$ respectively. Thus, the factor $\Xi(\delta)$ does not affect the parametric dependence of \eq{eloss-massive-appr}.

At first order in opacity, the medium-induced radiation associated to the production of a heavy color octet state (Fig.~\ref{fig:other-hard}b) is parametrically similar to that associated to the scattering of an energetic (massless) gluon studied in section \ref{sec:n1qcd}, up to the change in the hard scale, $q_\perp \to M_\perp$. Below we verify that this statement still holds when resumming all orders in opacity. 

\subsubsection{All orders in opacity}
\label{sec:massive-all-orders}

At all orders in opacity, the spectrum is given by the expression \eq{spec-alln-rescaled} modified according to \eq{insert-mass},
\be
x \frac{\dd I}{\dd x}  = N_c \, \frac{\alpha_s}{\pi^2} \int \dd^2 \kvec \; \left[ \fvec(\kvec,r) - \fvec_0(\kvec)  \right] \cdot \frac{- (\kvec - x\qvec)}{(\kvec - x\qvec)^{2}+x^2 M^2}  \, .
\label{spec-alln-mass}
\ee
The spectrum can be derived as in section \ref{sec:exact-der}. Only one additional identity is needed,
\be 
\int \dd^2 \kvec \, \frac{\kvec}{\kvec^{2}+x^2 M^2}  \, e^{- i  {\kvec \cdot \bvec}} = \fvect_0(\bvec) \, xMb \, {\rm K}_1(xMb)\, .
\ee
The final result is
\bea
x \frac{\dd I}{\dd x} =  N_c \, \frac{\alpha_s}{\pi} \, S[\Omega; r ; \Omega_{_M}] \; ; \ \ \Omega \equiv \frac{x |\qvec|}{\mu} \; ; \ \ r \equiv \frac{L}{\lambda_g} \; ; \ \ \Omega_{_M} \equiv \frac{x M}{\mu}  \, , \hskip 1.5cm && \label{spec-alln-mass-2} \\
S[\Omega; r; \Omega_{_M}] = 2 \int_0^{\infty} \frac{\dd B}{B} \, {\rm J}_0(\Omega B)  \left\{ 1 -\exp{\left[- r \, (1-B \, {\rm K}_1(B) ) \right]}  \right\} \, \Omega_{_M}B \, {\rm K}_1(\Omega_{_M} B) \, . &&
\label{SOmegaM}
\eea
Comparing \eq{spec-alln-mass-2} and \eq{SOmegaM} to \eq{spec-alln} and \eq{sar}, we see that the presence of the mass parameter $M$ introduces the factor $\Omega_{_M}B \, {\rm K}_1(\Omega_{_M} B)$, which as expected tends to unity when $M \to 0$. 

Using the same procedure as in section \ref{useful-appr}, we find that at fixed $r \gg 1$, the function $S[\Omega; r; \Omega_{_M}]$ can be approximated by
\be
\label{massive-S-appr}
S[\Omega; r; \Omega_{_M}] \simeq S_{\rm appr}[\Omega; r; \Omega_{_M}] \equiv \log{\left( 1 + \frac{r \log{r}}{\Omega_\perp^2} \right) }  \, ,
\ee
where $\Omega_\perp \equiv x M_\perp /\mu = (\Omega^2 + \Omega_{_M}^2)^{1/2}$ was introduced before. Thus the spectrum \eq{spec-alln-mass-2} takes the simpler form
\be
\label{massive-spec-appr}
x \frac{\dd I}{\dd x} \simeq  N_c \, \frac{\alpha_s}{\pi} \, \log{\left( 1 + \frac{\mu^2 r \log{r}}{x^2 M_\perp^2} \right)} \, .
\ee
As was the case at first order in opacity (see previous section), at all orders in opacity the spectrum \eq{massive-spec-appr} is obtained from that of an energetic massless gluon \eq{spec-alln-appr} by replacing $q_\perp \to M_\perp$. The result \eq{massive-spec-appr} corresponds to the spectrum derived semi-heu\-ris\-ti\-cally in \cite{Arleo:2010rb} and \cite{Arleo:2012rs}, and used for phenomenology in \cite{Arleo:2012hn,Arleo:2012rs,Arleo:2013zua}. 

\begin{figure}[t]
\centering
\includegraphics[width=4.9cm]{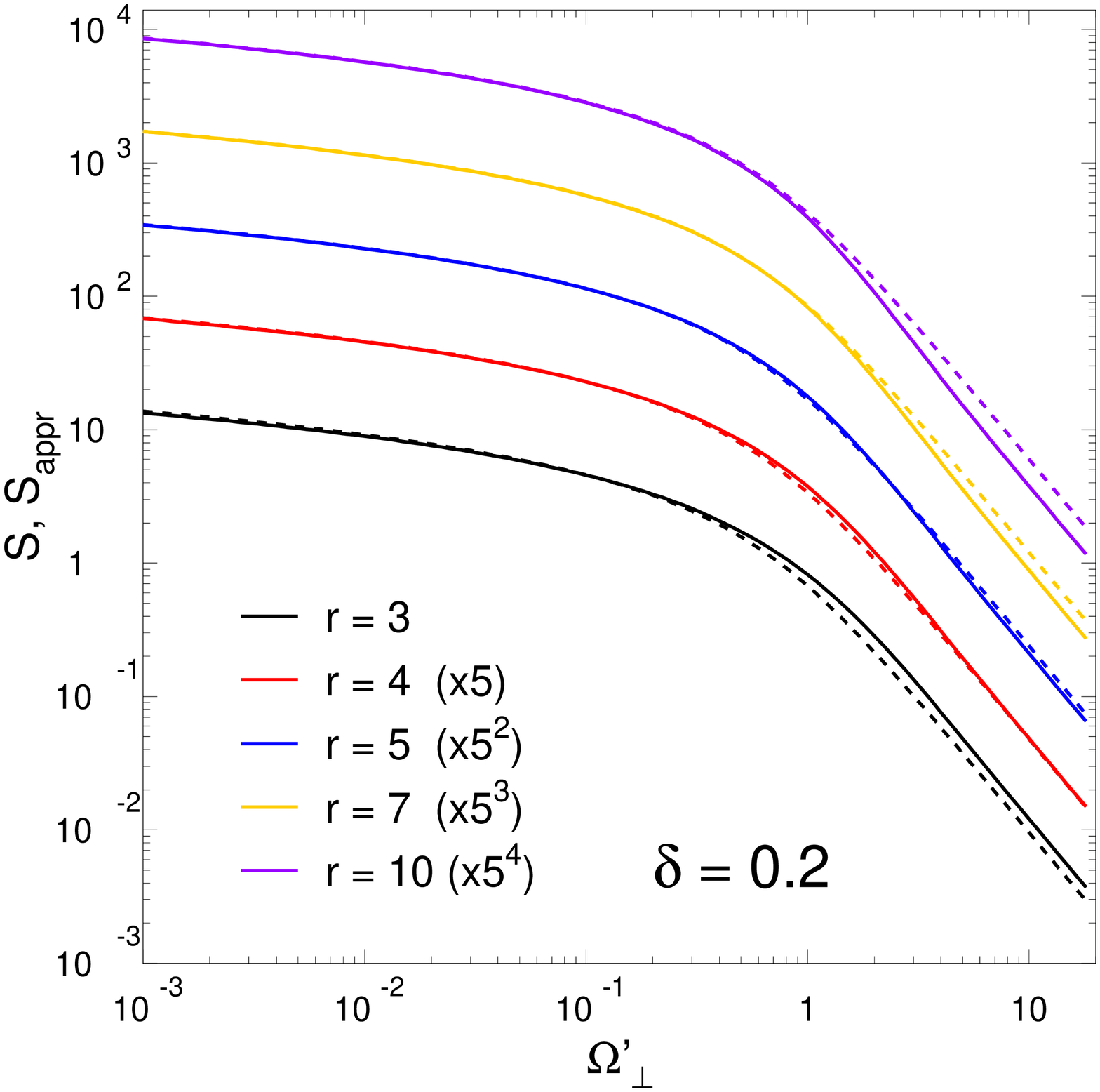}
\includegraphics[width=4.9cm]{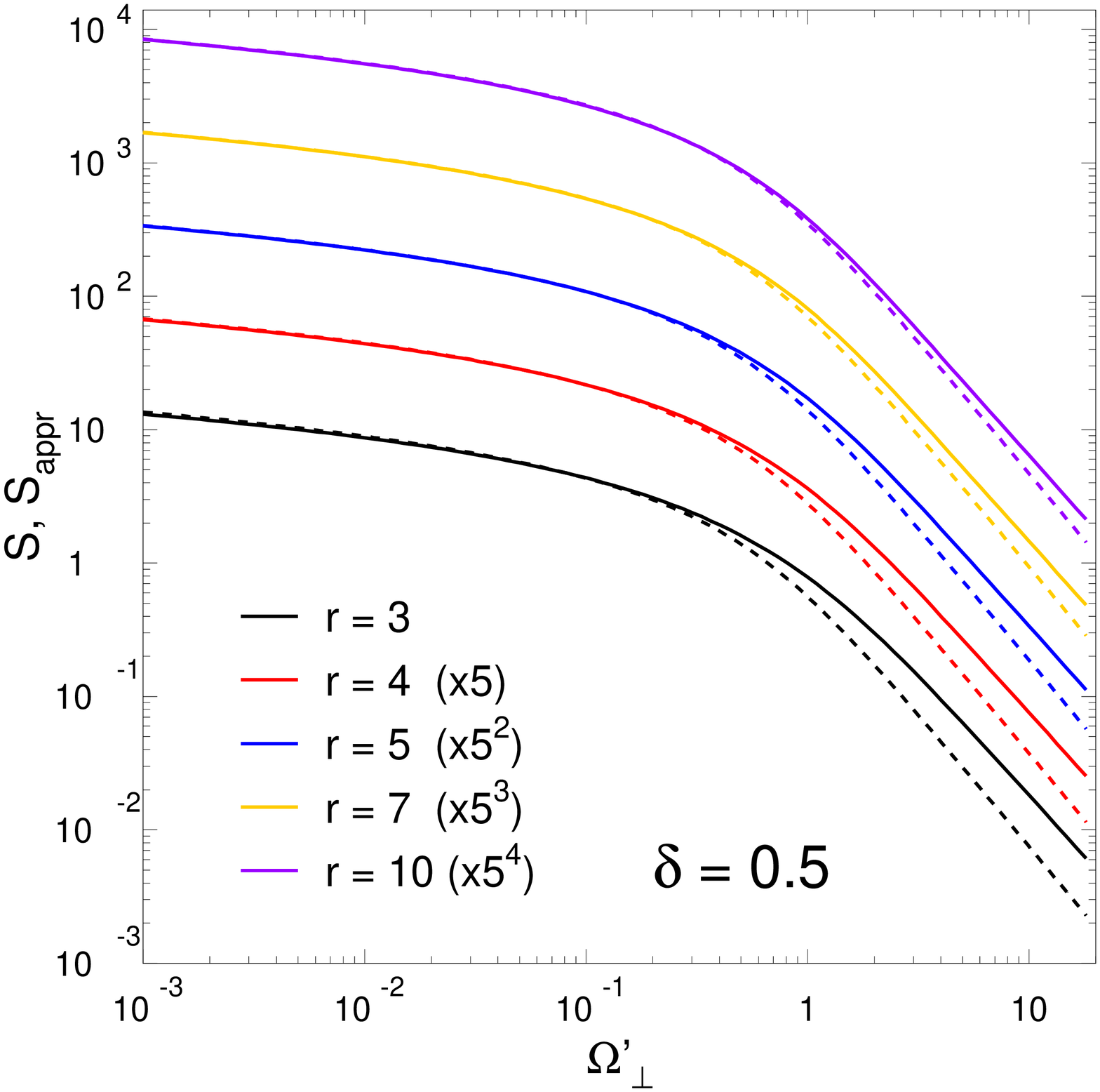}
\includegraphics[width=4.9cm]{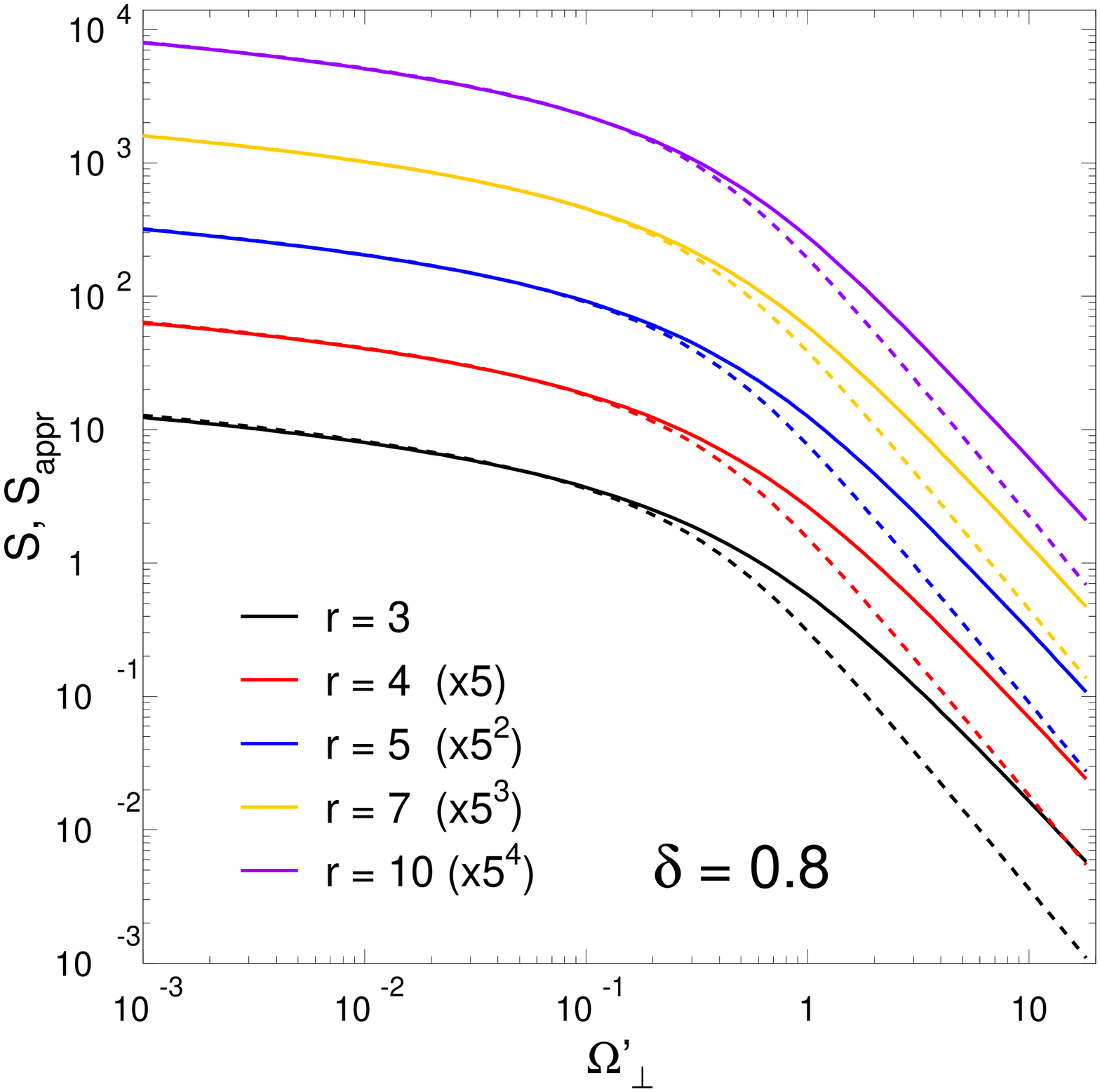}
\caption{Function $S[\Omega; r; \Omega_{_M}]$ defined in \eq{SOmegaM} (solid lines) compared to the approximation $S_{\rm appr}$ defined in \eq {massive-S-appr} (dashed lines), as a function of the variable $\Omega'_{\perp} \equiv \Omega_\perp /\sqrt{r \log{r}}$, for different values of $r$ ($r=3, 4, 5, 7, 10$) and of $\delta \equiv M/M_\perp$: $\delta= 0.2$ (left),  $\delta= 0.5$ (middle), and $\delta= 0.8$ (right).}
\label{fig:massive-S-appr}
\end{figure}

The numerical accuracy of \eq{massive-S-appr} is shown in Fig.~\ref{fig:massive-S-appr} for various values of $r$ and of the variable $\delta \equiv M/M_\perp$. The case $\delta = 0$ ($M=0$) was studied in section \ref{useful-appr}, see Fig.~\ref{fig:appr-scaling}. 

\section{Purely initial/final state medium-induced radiation}
\label{sec:IS-FS}

\begin{figure}[t]
\centering
\includegraphics[width=10cm]{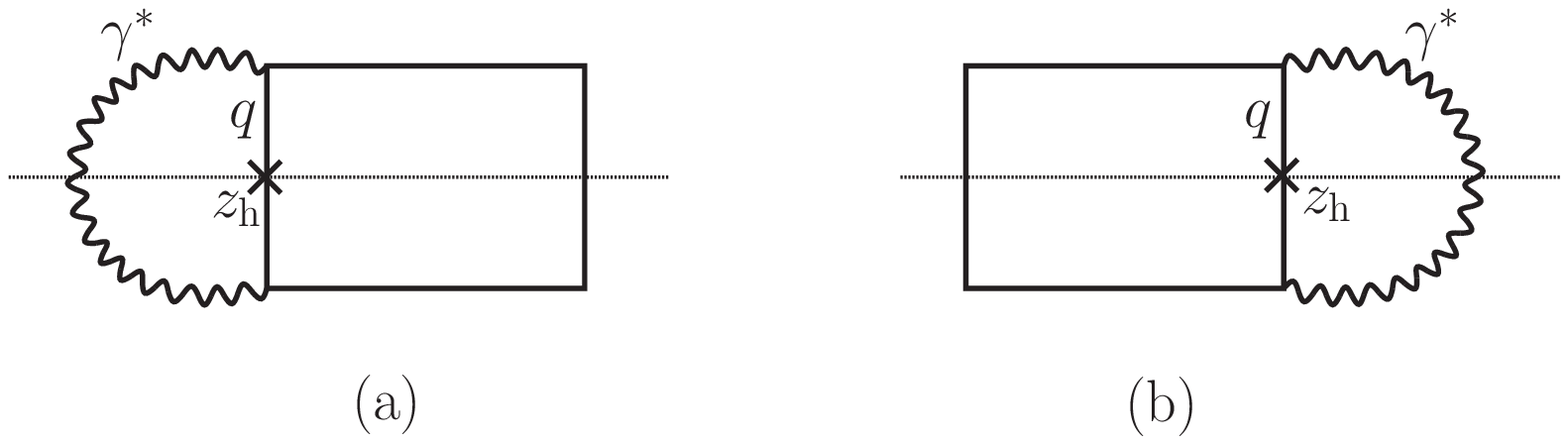}
\caption{(a) Sudden acceleration of a quark in the $\gamma^* q \to q$ hard process. (b)  Fast quark annihilation in the $q \bar q \to \gamma^*$ hard process.}
\label{fig:no-interference}
\end{figure}

In hard processes where either the incoming or the outgoing energetic particle is colorless (see Fig.~\ref{fig:no-interference}), the interference responsible for the fully coherent radiation is absent, and the induced radiation reduces to purely final or purely initial state radiation.\footnote{In Ref.~\cite{Vitev:2007ve} where both incoming and outgoing energetic particles carry color, but the interference term is simply neglected, the medium-induced radiation reduces to the incoherent sum of initial and final state radiation.} In the present section we briefly review the latter contributions and recover known results. For simplicity we work at first order in the opacity expansion. The calculations at all orders in opacity \cite{Baier:1996kr,Zakharov:1997uu,Gyulassy:2000er} do not bring any important qualitative change. 

As we have seen in section \ref{sec:n1qcd}, initial and final state radiation cancels out in the ($\kvec$-integrated) radiation spectrum in the limit $\tf \gg L$, see \eq{setAn1} and \eq{setBn1}. This means that assuming $\tf \gg L$ is too drastic to derive initial/final state radiation. The purely initial/final average energy loss turns out to be dominated by $\tf \sim L$, scaling as $\Delta p^+ \propto L^2$ and independent of $p^+$ (up to logarithms), see \eq{FS-eloss}. This type of energy loss, $\Delta p^+ \propto L^2$, received much attention in the last two decades. We emphasize that in situations where the interference between initial and final state radiation cannot be neglected (as studied in the previous sections), $\Delta p^+ \propto L^2$ is actually negligible at large enough $p^+$, compared to the fully coherent loss $\Delta p^+ \propto p^+$.

\subsection{Final state radiation}

\begin{figure}[t]
\centering
\includegraphics[width=13cm]{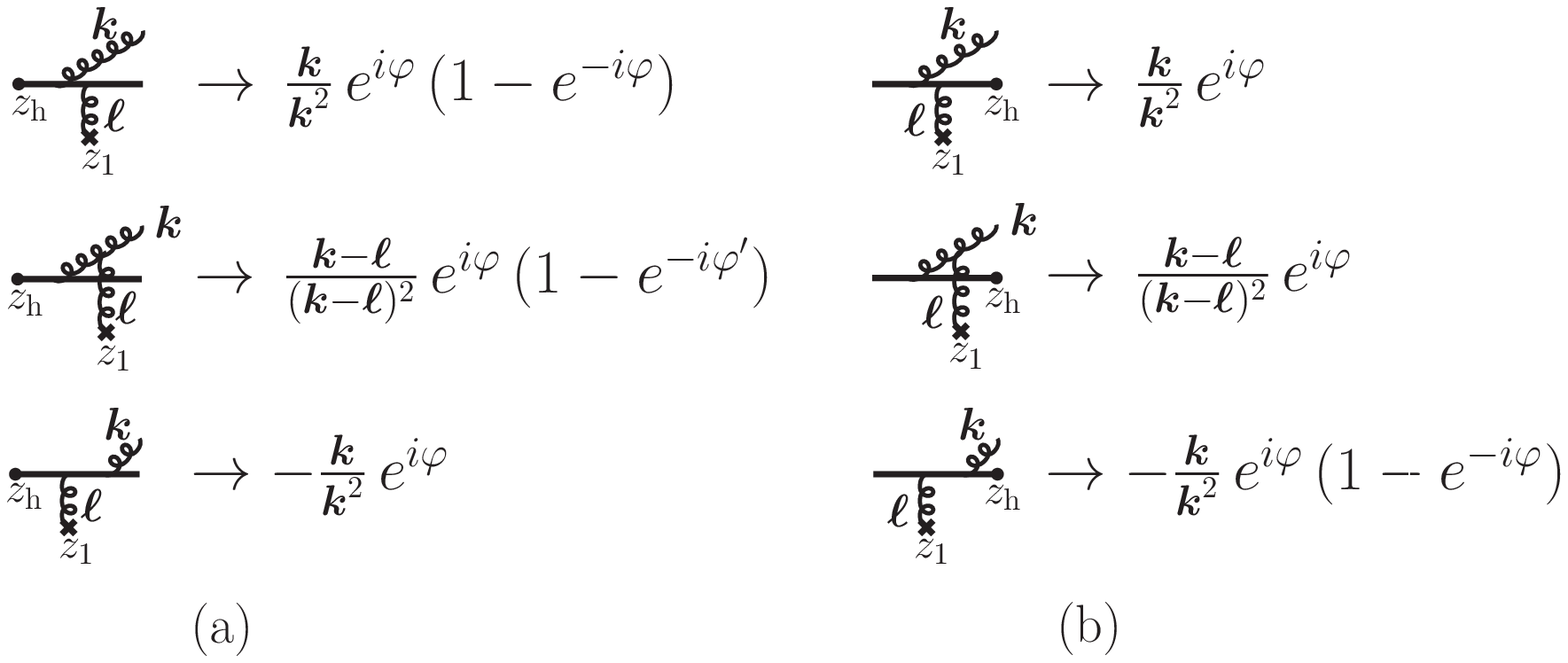}
\caption{Pictorial rules for (a) final state ($z_1 > z_{\rm h}$) and (b) initial state ($z_1 < z_{\rm h}$) emission vertices supplemented by phase factors.  For convenience we set $z_{\rm h}=0$, and we define $\varphi \equiv z_1 \kvec^2/k^+$, $\varphi' \equiv z_1 (\kvec - \ellvec)^2/k^+$.}
\label{fig:pictorial-rules-phases}
\end{figure}

A toy model for a process where only final state radiation contributes is shown in Fig.~\ref{fig:no-interference}a. Here the hard transfer $\qvec$ is irrelevant and we can choose a frame where the final energetic quark has zero transverse momentum. At order $n=1$ in the opacity expansion, the spectrum for final state radiation off an energetic quark is given by (compare to \eq{spec-order-n})
\be
\left. x \frac{\dd I}{\dd x} \right|_{\rm FS}  =
\frac{\alpha_s}{\pi^2} \int \dd^2 \kvec \int \frac{\dd z_1}{\lambda_q} \int \dd^2 \ellvec \, V(\ellvec)  \, \frac{1}{N_c C_F} \, \sum \: \mbox{\raisebox{-8mm}{\hskip 0mm \hbox{\epsfxsize=3.8cm\epsfbox{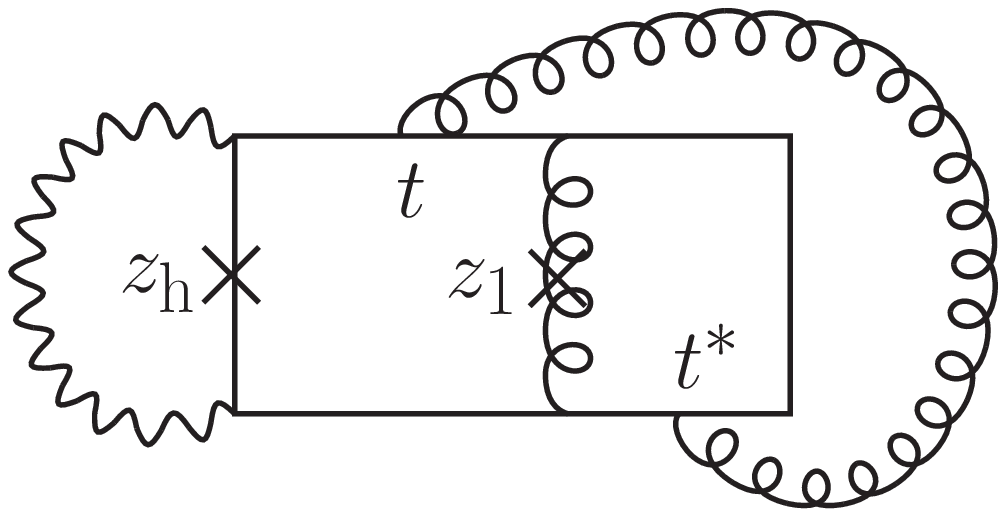}}}} \ \ ,
\label{FS-spec-order-1} 
\ee
where only one generic contribution is shown is the sum of diagrams. Since we do not assume $\tf \gg L$ any longer, diagrams where the gluon emission time $t$ in the amplitude (or $t^*$ in the conjugate amplitude) occurs within the target, $z_{\rm h} < t < z_1$, cannot be neglected. Such diagrams are proportional to a difference of phase factors $\sim (e^{i \varphi(z_1)}- e^{i \varphi(z_{\rm h})})$ which can be derived in time-ordered perturbation theory \cite{Baier:1996kr}. Taking into account those phase factors, the rules of Fig.~\ref{fig:pictorial-rules}a for gluon emission vertices must be modified to the rules defined in Fig.~\ref{fig:pictorial-rules-phases}. 

As can be easily verified, among the diagrams contributing to \eq{FS-spec-order-1}, those where $t$ and $t^*$ belong to the same time-interval (\ie, either $t, t^* > z_1$ or $0 < t, t^* < z_1$) give a vanishing contribution to the $\kvec$-integrated radiation spectrum. This property was already mentioned in section \ref{sec:n1qcd}. The remaining diagrams sum up to\footnote{The virtual photon participating to the hard process (see Fig.~\ref{fig:no-interference}a) is not drawn in Eq.~\eq{fig:FS-diagrams}.}
\bea
&& \hskip -3mm \mbox{\raisebox{-5.5mm}{\hbox{\epsfxsize=14.4cm\epsfbox{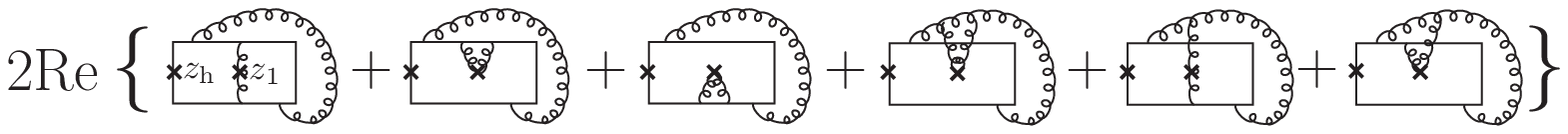}}}} \! \! \! = \nn  \\
&& 2 N_c C_F \left\{\left[-\frac{1}{2 N_c} - 2 \frac{C_F}{2} - \frac{N_c}{2} \right] \frac{-1}{\kvec^2} (1-\cos{\varphi}) \right. + \left. \left[ \frac{N_c}{2} + \frac{N_c}{2} \right] \frac{-\kvec \cdot (\kvec -\ellvec)}{\kvec^2 (\kvec - \ellvec)^2} (1-\cos{\varphi'}) \right\} \nn  \\
&&  \hskip 25mm = 2 N_c^2 C_F \left\{ \frac{1}{\kvec^2} (1-\cos{\varphi}) - \frac{\kvec \cdot (\kvec - \ellvec)}{\kvec^2 (\kvec - \ellvec)^2} (1-\cos{\varphi'})\right\} \, . \label{fig:FS-diagrams}  
\eea
Inserting the latter result in \eq{FS-spec-order-1} and shifting variable $\kvec \to \ellvec - \kvec$ in the first term we get
\be
\left. x \frac{\dd I}{\dd x} \right|_{\rm FS} = \frac{\alpha_s}{\pi^2} \int \dd^2 \kvec \int_{z_{\rm h}}^L \frac{\dd z_1}{\lambda_q} \int \dd^2 \ellvec \, V(\ellvec)  \, 2 N_c (1-\cos{\varphi'}) \left[ \frac{1}{(\kvec - \ellvec)^2} - \frac{\kvec \cdot (\kvec - \ellvec)}{\kvec^2 (\kvec - \ellvec)^2} \right] \, .
\ee
Finally, integrating over $z_1$ and using $C_F \lambda_q = N_c \lambda_g$ yields
\be
\left. x \frac{\dd I}{\dd x} \right|_{\rm FS} = \frac{C_F \alpha_s}{\pi^2 \lambda_g} \int \dd^2 \kvec \int \dd^2 \ellvec \, V(\ellvec)  \, \frac{2 \kvec \cdot \ellvec}{\kvec^2 (\kvec - \ellvec)^2} \left[ L- \frac{k^+}{(\kvec - \ellvec)^2 }\sin{L \frac{(\kvec - \ellvec)^2}{k^+}} \right]  \, ,
\label{FS-spec-order-1-Vitev} 
\ee
where $L$ is the distance crossed by the quark from the hard production point $z_{\rm h}=0$ to the boundary of the medium. Clearly, the spectrum for purely final state radiation off a color charge $C_R$ is obtained by replacing $C_F \to C_R$ in the expression \eq{FS-spec-order-1-Vitev}, which then coincides with the final state radiation spectrum derived in Ref.~\cite{Vitev:2007ve}. 

The $\kvec$-integrated spectrum \eq{FS-spec-order-1-Vitev} is mathematically well-defined, in particular there is no singularity at $\kvec = \zerovec$ or $\kvec = \ellvec$. Changing variable $\kvec \to \ellvec - \kvec$, \eq{FS-spec-order-1-Vitev} becomes
\be
\left. x \frac{\dd I}{\dd x} \right|_{\rm FS} = \frac{C_F \alpha_s}{\pi^2 \lambda_g} \int \frac{\dd^2 \kvec}{\kvec^2} \left[ L- \frac{k^+}{\kvec^2}\sin{L \frac{\kvec^2}{k^+}} \right]  \int \dd^2 \ellvec \, V(\ellvec)  \, \frac{2 \ellvec \cdot (\ellvec - \kvec)}{(\kvec - \ellvec)^2}   \, .
\label{FS-spec-order-1-Vitev-2} 
\ee
The integral over $\ellvec$ can be performed using
\be
\int \frac{\dd \varphi_{\ell}}{2\pi} \, \frac{2 \ellvec \cdot (\ellvec - \kvec)}{(\kvec - \ellvec)^2} = 
\int \frac{\dd \varphi_{\ell}}{2\pi} \, \left[ 1 + \frac{\ellvec^2 - \kvec^2}{(\kvec - \ellvec)^2} \right] = 2 \Theta\left( \ellvec^2 - \kvec^2 \right) \, ,
\ee
and introducing the variable $u = k^+/(\kvec^2 L)$ we obtain
\be 
\left. x \frac{\dd I}{\dd x} \right|_{\rm FS} = \frac{2 C_F \alpha_s}{\pi \lambda_g} \, \frac{L^2 \mu^2}{k^+} \int_0^{\infty} \! \dd u \,  \frac{1 - u \sin{\frac{1}{u}}}{1 + \frac{\mu^2 L}{k^+} \, u} \, .
\ee
When $\mu^2 L/k^+ \ll 1$, the $u$-integral simplifies to $\int_0^{\infty} \! \dd u \, (1 - u \sin{\frac{1}{u}}) = \frac{\pi}{4}$, giving 
\be
\left. x \frac{\dd I}{\dd x} \right|_{\rm FS} = \frac{C_F \alpha_s}{2 \lambda_g} \, \frac{L^2 \mu^2}{x p^+} \hskip 1cm (x p^+ \gg \mu^2 L) \, .
\ee
The spectrum arises from an integration domain where $u \sim \morder{1}$, \ie, from formation times $\tf \sim k^+/\kvec^2 \sim L$, as expected. 

To logarithmic accuracy the associated average energy loss reads
\be
\left. \Delta p^+ \right|_{\rm FS} \equiv p^+ \int \dd x \, \left. x \frac{\dd{I}}{\dd x} \right|_{\rm FS} = \frac{C_F \alpha_s}{2} \, \frac{L^2 \mu^2}{\lambda_g} \log{\frac{p^+}{\mu^2 L}} \, ,
\label{FS-eloss}
\ee
arising from the logarithmic interval $\mu^2 L \ll k^+ \ll p^+$. The result \eq{FS-eloss} was derived previously, see for instance Ref.~\cite{Zakharov:2000iz}. When $p^+ \to \infty$ the purely final state energy loss is independent of $p^+$ (up to logarithms) and is proportional to $L^2$. 

\subsection{Initial state radiation}

In the case where the outgoing energetic particle participating to the hard process carries no color charge, as in Fig.~\ref{fig:no-interference}b, the medium-induced radiation arises solely from the initial state. The derivation of the spectrum is the same as for final state radiation, up to few modifications. The diagrams contributing to the initial state radiation spectrum 
\be
\left. x \frac{\dd I}{\dd x} \right|_{\rm IS} =
\frac{\alpha_s}{\pi^2} \int \dd^2 \kvec \int \frac{\dd z_1}{\lambda_q} \int \dd^2 \ellvec \, V(\ellvec)  \, \frac{1}{N_c C_F} \, \sum \: \mbox{\raisebox{-9mm}{\hskip 0mm \hbox{\epsfxsize=3.6cm\epsfbox{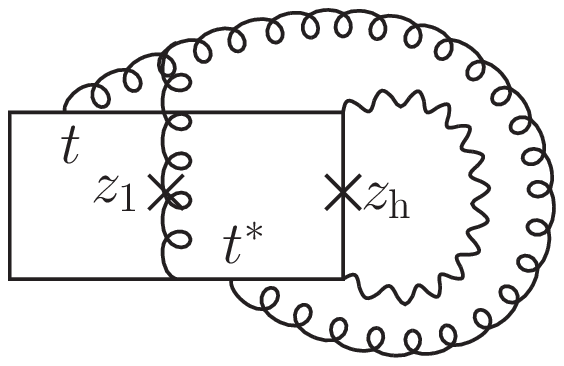}}}} 
\label{IS-spec-order-1} 
\ee
are evaluated using the pictorial rules defined in Fig.~\ref{fig:pictorial-rules-phases}b, which differ from those of Fig.~\ref{fig:pictorial-rules-phases}a only in the phase factors. The only non-vanishing contribution to the $\kvec$-integrated spectrum \eq{IS-spec-order-1} arises from the set of diagrams 
\bea
&& \hskip -3mm \mbox{\raisebox{-5.5mm}{\hbox{\epsfxsize=14.4cm\epsfbox{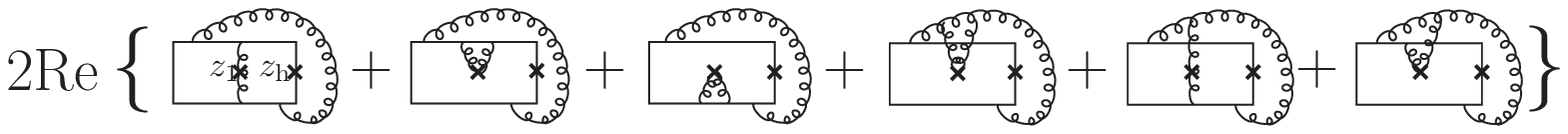}}}} \! \! \! = \nn  \\
&& 2 N_c C_F \left\{\left[-\frac{1}{2 N_c} - 2 \frac{C_F}{2} - \frac{N_c}{2} \right] \frac{-1}{\kvec^2} (1-\cos{\varphi}) \right. + \left. \left[ \frac{N_c}{2} + \frac{N_c}{2} \right] \frac{-\kvec \cdot (\kvec -\ellvec)}{\kvec^2 (\kvec - \ellvec)^2} (1-\cos{\varphi}) \right\} \nn  \\
&&  \hskip 45mm = 2 N_c^2 C_F \, \frac{\ellvec \cdot (\ellvec - \kvec)}{\kvec^2 (\kvec - \ellvec)^2} \, (1-\cos{\varphi}) \, . \label{fig:IS-diagrams}  
\eea
As a result,
\be
\left. x \frac{\dd I}{\dd x} \right|_{\rm IS} = \frac{C_F \alpha_s}{\pi^2 \lambda_g} \int \dd^2 \kvec \int \dd^2 \ellvec \, V(\ellvec)  \, \frac{2 \ellvec \cdot (\ellvec - \kvec)}{\kvec^2 (\kvec - \ellvec)^2} \left[ L- \frac{k^+}{\kvec^2 }\sin{L \frac{\kvec^2}{k^+}} \right]  \, ,
\label{IS-spec-order-1-end} 
\ee
where $L$ is the distance travelled by the quark in the medium up to the point $z_{\rm h}=0$.

Thus, provided the radiated gluon transverse momentum $\kvec$ is integrated over, the purely initial and purely final state radiation spectra are identical, see \eq{FS-spec-order-1-Vitev-2} and \eq{IS-spec-order-1-end}. The associated average energy loss is given by \eq{FS-eloss}.

\section{Summary and discussion}
\label{sec:discussion}

In this study we have derived, to all orders in the opacity expansion, and for the case of a Coulomb rescattering potential, the {\it medium-induced} gluon radiation spectrum associated to the hard forward scattering of an energetic parton. The latter spectrum arises from gluon formation times scaling as the parton energy, $\tf \propto E$, corresponding to {\it fully coherent} radiation (\ie\ $\tf \gg L$) over the medium size. In the light-cone gauge $A^+=0$, the induced coherent spectrum is dominated by the interference between initial and final state emission amplitudes. 

Using the result \eq{spec-alln-appr} for a parton of color charge $C_R$ (process $q \to q$ or $g \to g$), the demonstration in Section~\ref{sec:qtog} that \eq{spec-alln-appr} also holds (up to the color factor) for processes where the nature of the fast parton is changed in the hard scattering ($q \to g$ and $g \to q$), and finally the result \eq{massive-spec-appr} for the scattering of a gluon to a {\it massive} compact color octet, we infer the following general expression for the induced coherent spectrum,
\be
\label{spec-general}
x \frac{\dd I}{\dd x} =  (C_R + C_{R'} - C_{t}) \, \frac{\alpha_s}{\pi} \, \log{\left(1+ \frac{\Delta q_\perp^2(L)}{x^2 M_\perp^2} \right) } \, .
\ee

The color factor trivially follows from the structure of the interference term. The latter is indeed of the form $\sim 2 T_R^a T_{R'}^a$ (where $T_R^a$ and $T_{R'}^a$ are the color generators of the incoming and outgoing parton color representations $R$ and $R'$), which can be written as
\be
\label{color-rule}
2 T_R^a T_{R'}^a = (T_R^a)^2 + (T_{R'}^a)^2  - (T_{R}^a - T_{R'}^a)^2 = C_R + C_{R'} - C_{t} \, ,
\ee
with $C_R$, $C_{R'}$ the incoming and outgoing color charges, and $C_t$ the color charge of the $t$-channel exchange. We emphasize that the factor $C_R + C_{R'} - C_{t}$ holds at any finite $N_c$, and is negative in the particular $q \to q$ case, where $C_R + C_{R'} - C_{t} = 2C_F -N_c = -1/N_c$. We argued in section \ref{sec:qdep} that a medium-induced energy {\it gain} for the $q \to q$ process, although surprising, could have been anticipated from the features of the {\it total} radiation spectrum associated to $q \to q$. Note however that the negativity of the induced spectrum associated to $q \to q$ is in general not a specificity of the quark, but depends on the $t$-channel color exchange $C_t$. In particular, the induced spectrum associated to $g \to g$ would also be negative in an academic situation where $C_t$ would be constrained to satisfy $C_t > 2 N_c$. 

A natural question to ask is whether this negative color factor could have important phenomenological consequences. Although phenomenology is not the purpose of this paper, let us mention a couple of observables which could be sensitive to an \emph{energy gain} arising from a negative medium-induced gluon spectrum. The production of photons in proton--nucleus (\pA) collisions is likely dominated by the Compton scattering channel, $q \to q \gamma$, at forward rapidity. Although this process is formally $1 \to 2$, it was shown in~\cite{Peigne:2014rka} that $C_{R^\prime}$ should be taken as the \emph{global} {color charge} of the 2-particle final state; consequently, the color factor of the medium-induced gluon spectrum in $q\to q$ and $q \to q \gamma$ should be identical and equal to $-1/N_c$. Similarly, forward light-hadron production should be sensitive to the scattering channel $q \to q g$, and therefore be sensitive to negative medium-induced gluon spectrum when the final $qg$-state is in the color triplet representation (unlike $q \gamma$, the $qg$-state can be in other, higher-dimension, color representations). In both channels, $q\to q\gamma$ and ${q\to (q g)}_3$, one could legitimately expect some enhancement in \pA\  with respect to \pp\  collisions, although this remains to be quantified. In particular, the smallness of the color factor $-1/N_c$ may somehow tame such an enhancement.

The induced coherent spectrum explicitly depends on the hard process (through $C_t$ and the hard scale $M_\perp = \sqrt{M^2 + \qvec^2}$). When the incoming particle is colorless ($C_R =0$), the interference between initial and final state emission amplitudes is absent and fully coherent radiation vanishes (which formally follows by noting that when $C_R =0$, color conservation implies $C_{R'} = C_{t}$). In this case, we verified in Section~\ref{sec:IS-FS} (at first order in the opacity expansion) that the spectrum found in our setup arises from formation times $\tf \lsim L$ and coincides with previously known results~\cite{Vitev:2007ve,Zakharov:2000iz}.

Strictly speaking, \eq{spec-general} is an approximation to the exact spectrum, which however has the correct parametric behavior when $x \ll \Delta q_\perp(L)/M_\perp$ (see Sections~\ref{useful-appr} and \ref{sec:massive-all-orders}). In the case of a Coulomb scattering potential considered in our study, the {\it typical} transverse broadening entering \eq{spec-general} reads $\Delta q_\perp^2(L) \simeq \hat{q} L \log{L/\lambda_g}$ at large $L$ (specifically when $\log{L/\lambda_g} \gg 1$, see~\cite{Baier:1996sk}). We however found that \eq{spec-general} is {\it numerically} a very accurate appro\-xi\-mation to the exact coherent spectrum as soon as $\log{L/\lambda_g} \geq 1$, \ie\ already at moderate values of $L$, and up to values of $x \sim \morder{\Delta q_\perp(L)/M_\perp}$. We have seen that at larger $x$, $\Delta q_\perp(L)/M_\perp \ll x \ll 1$, the parametric behavior $x\, {\dd I}/{\dd x} \propto \Delta q_\perp^2(L)/(x^2 M_\perp^2)$ of \eq{spec-general} does not reproduce the proper normalization of the exact spectrum. In the massless case, it overestimates (by a factor $\log{(L/\lambda_g)}$) the correct behavior  $x\, {\dd I}/{\dd x} \propto \hat{q} L /(x^2 q_\perp^2)$, see \eq{large-x} and Fig.~\ref{fig:appr-scaling} (right). In the massive case, the approximation \eq{spec-general} is also inaccurate at large $x$, see Fig.~\ref{fig:massive-S-appr}. In the large $x$ domain the exact expression of the spectrum (Eqs.~\eq{spec-alln-mass-2}--\eq{SOmegaM} in the massive case) should be preferred.

We stress that the spectrum \eq{spec-general} corresponds to the induced coherent radiation in a target of finite size $L$ with respect to an ideal target of zero size. For the sake of the following discussion, we introduce the induced spectrum in \pA\ collisions with respect to \pp\ collisions, defined as the difference of the spectrum \eq{spec-general} for a target nucleus A and a proton target, 
\be
\label{spec-general-pAvspp}
\left. x \frac{\dd I}{\dd x} \right|_{{\rm pA}-{\rm pp}} =  (C_R + C_{R'} - C_{t}) \, \frac{\alpha_s}{\pi} \, \log{\left(\frac{x^2 M_\perp^2 + \Delta q_\perp^2(L_{\rm A})}{x^2 M_\perp^2 + \Delta q_\perp^2(L_{\rm p})} \right) } \, .
\ee

In some aspects, our study of coherent gluon radiation resembles some studies of gluon radiation in the saturation framework, however with a crucial difference: in the saturation formalism, the focus is usually on that part of gluon production which is {\it process-independent} (and related to the {\it unintegrated} or {\it $k_t$-dependent} gluon distribution in the target), whereas we concentrated here on the {\it process-dependent} part of gluon radiation.

To illustrate this point, it is instructive to compare our study to the study of gluon production in nuclear DIS and \pA\ collisions of Ref.~\cite{Kovchegov:1998bi}. In this work, the focus is on the contributions to the {\it total}, $k_\perp$-differential radiation spectrum which exhibit a {\it collinear} (logarithmic) singularity when $k_\perp \to 0$. As a consequence, gluon production in \pA\ collisions is dominated (in light-cone gauge) by purely initial and purely final state radiation. The interference term,\footnote{In our notations, \eq{MK60} corresponds to equation (60) of Ref.~\cite{Kovchegov:1998bi} and represents the contribution of the interference to the total gluon radiation spectrum off a fast massless quark in \pA\ collisions.}
\be
\label{MK60}
\left. x \frac{\dd I}{\dd x \, \dd^2 \vec{k}_\perp} \right|_{\rm pA, \, int} = -2 C_F \frac{\alpha_s}{\pi^2} \frac{1-e^{-{k_\perp^2}/{\Delta q_\perp^2(L_{\rm A})}}}{k_\perp^2} \, ,
\ee
is non-logarithmic when $k_\perp \to 0$ and is thus neglected in~\cite{Kovchegov:1998bi}. 

However, considering {\it medium-induced} instead of {\it total} radiation in the setup of Ref.~\cite{Kovchegov:1998bi}, it is straightforward to check that purely initial and final state radiation (integrated over $\vec{k}_\perp$) cancels out (as in the present study), leaving only a contribution from the interference term. Thus, in Ref.~\cite{Kovchegov:1998bi} the medium-induced spectrum can be simply obtained by subtracting from \eq{MK60} a similar contribution in \pp \ collisions, 
\be
\label{MK60-induced}
\left. x \frac{\dd I}{\dd x \, \dd^2 \vec{k}_\perp} \right|_{{\rm pA}-{\rm pp}} = -2 C_F \frac{\alpha_s}{\pi^2} \, \frac{e^{-{k_\perp^2}/{\Delta q_\perp^2(L_{\rm p})}} -e^{-{k_\perp^2}/{\Delta q_\perp^2(L_{\rm A})}}}{k_\perp^2} \, .
\ee
Integrating \eq{MK60-induced} over $\vec{k}_\perp$ yields
\be
\label{MK60-induced-2}
\left. x \frac{\dd I}{\dd x} \right|_{{\rm pA}-{\rm pp}}  = 2 C_F \frac{\alpha_s}{\pi} \log{\left(\frac{\Delta q_\perp^2(L_{\rm A})}{\Delta q_\perp^2(L_{\rm p})} \right)} \, ,
\ee
arising from the domain $\Delta q_\perp^2(L_{\rm p}) \ll k_\perp^2 \ll \Delta q_\perp^2(L_{\rm A})$. This  emphasizes that the induced contribution \eq{MK60-induced-2} lies beyond the scope of the $k_\perp \to 0$ limit (which formally implies $k_\perp^2 \ll \Delta q_\perp^2(L_{\rm p})$) used in Ref.~\cite{Kovchegov:1998bi}.

Quite remarkably, \eq{MK60-induced-2} can be obtained from the general expression \eq{spec-general-pAvspp} by the formal replacements $C_t \to 0$, $M_\perp \to 0$ (and $C_R, C_{R'} \to C_F$, since \eq{MK60-induced-2} corresponds to the scattering of a fast quark).\footnote{Note that \eq{MK60-induced-2} can also be obtained (including the color factor) {\it directly} from the $q_\perp \to 0$ limit of the expression \eq{spec-armesto2} (which expression follows from the setup of Ref.~\cite{Armesto:2013fca}). Indeed, as mentioned in \cite{Armesto:2013fca}, when $q_\perp \to 0$ the setups of \cite{Kovchegov:1998bi} and \cite{Armesto:2013fca} coincide. This is because both studies consider the $q \to q$ process, and $C_t = 0$ in the model of Ref.~\cite{Armesto:2013fca}.} This can be understood from the fact that there is no `hard process' in the setup of Ref.~\cite{Kovchegov:1998bi}, the fast quark undergoing only soft rescatterings in the nuclear target.

\acknowledgments

We are most grateful to Al Mueller and St\'ephane Munier for useful exchanges, and to Tseh Liou and Al Mueller for a very rich and instructive correspondence on the similarities between their recent work \cite{Liou:2014rha} and our study. Despite different theoretical formalisms and setups used in \cite{Liou:2014rha} and here, our studies agree on important points, in particular on the parametric dependence of the medium-induced soft gluon radiation spectrum.

Feynman diagrams have been drawn with the JaxoDraw software~\cite{Binosi:2008ig}. This work is funded by ``Agence Nationale de la Recherche'' under grant ANR-PARTONPROP.

\appendix

\section{Derivation of Eq.~\eq{spec-order-n}}
\label{app:3.1}

In this appendix we briefly outline the derivation of Eq.~\eqref{spec-order-n}. 

The soft radiation  {\it intensity} accompanying the (hard) production of a parton of momentum $p'$ is given by the ratio
\be
\dd I = \cfrac{\dd \sigma_{\rm rad}(p',k)}{\dd \sigma_{\rm prod}(p')} \, ,
\label{app:dI-general}
\ee
where the numerator is the radiative cross section for producing a two-parton final state (the leading parton $p'$ and a soft radiated gluon of momentum $k$) while the denominator represents the inclusive cross section for producing the leading parton. Formally the latter inclusive cross section includes the two-parton production cross section entering the numerator, integrated over the radiated gluon phase space. However, to leading order in $\alpha_s$, the denominator can be approximated by the production cross section of the leading parton {\it without} soft gluon emission. Moreover, assuming that soft rescatterings of the leading parton do not affect $\dd \sigma_{\rm prod}(p')$, the latter can be calculated as if there were no scattering center in the medium, \ie, at zeroth order in the opacity expansion \cite{Gyulassy:2000er}.

On the contrary, the numerator of \eqref{app:dI-general} is modified by soft rescatterings and must be
represented as an expansion in the number $n$ of scattering centers encountered by the leading parton. The $n=0$ term in this expansion coincides with the radiation cross section in the absence of a medium, and thus cancels out in the {\it medium-induced} radiation intensity defined as
\be
\dd I^{\rm induced} \equiv \dd I - \dd I^{\rm vacuum} = \sum_{n=1}^{\infty} \frac{\dd\sigma_{\rm rad}^{(n)}(p',k)}{\dd \sigma_{\rm prod}(p')} \equiv \sum_{n=1}^{\infty} \dd I^{(n)} \, .
\label{app:dI-induced}
\ee

Each term in the sum over $n$ contains an implicit average over the positions of the scattering centers on which the parton-gluon system rescatters. Assume there are $N\gg 1$ scattering centers overall, which are randomly but, on the average, uniformly distributed in a layer of thickness $L$ and transverse area $S \gg 1/\mu^2$. (Recall that $1/\mu$ is the screening length of the Coulomb scattering potential \eq{coulomb-pot}.) In the calculation we consider the interaction of each scattering center to the lowest order in $\alpha_s$. This implies that each of the $n$ scattering centers either contributes a single Born interaction in both the amplitude and the conjugate, or it can contribute a double Born interaction in the amplitude with no interaction in the conjugate amplitude (or vice versa), thus providing a so-called virtual correction to the aforementioned single-Born interaction.

Averaging over the position of each scattering center introduces a factor $(\int \! \dd z_i \! \int \! \dd^2{\xvec_i})/(LS)$. Accounting the different possibilities to choose $n_1$ scattering centers contributing via single Born interaction together with $n_2$ and $n_{\bar 2}$ centers contributing double Born interactions in the amplitude and conjugate amplitude respectively, brings a factor $\binom{N}{n_1} \binom{N-n_1}{n_2} \binom{N-n_1 -n_2}{n_{\bar 2}}$. Integrating over $\dd^2{\xvec_i}$ ensures the same transverse momentum transfer from the scattering center $i$ in the amplitude and its conjugate for a single Born scattering, or a zero net momentum transfer in 
a double Born interaction~\cite{Gyulassy:2000er}. For $N \to \infty$ one can use $\binom{N}{n} \simeq N^n/n!$ to obtain
\be
\dd I^{(n)} = \frac{1}{\dd \sigma_{\rm prod}(p')}  \left[ \prod_{i=1}^n \int_0^L \! \! \dd z_i \int \! \dd^2 \ellvec_i \, \dd^2 \ellvec'_i \, \delta^{(2)}(\ellvec_i-\ellvec_i') \right] \sum_{n_1, n_2, n_{\bar 2}} \frac{\nu^n}{n_1! n_2! n_{\bar 2}!}  \cfrac{\dd \sigma_{\rm rad}(p',k,\{\ellvec_i,\ellvec'_i\})}{\dd^2 \ellvec_1 \dd^2 \ellvec'_1 \ldots \dd^2 \ellvec_n\dd^2 \ellvec'_n}\, ,
\label{app:dsigma-n-averaged}
\ee
where $\nu\equiv N/(LS)$ is the number of scattering centers per unit volume in the target, $n = n_1+n_2 + n_{\bar 2}$, and $n_1 \ge 1$.

The next observation concerns the square of the matrix element entering the cross section 
$
{\dd \sigma_{\rm rad}(p',k,\{\ellvec_i,\ellvec'_i\})}/{\dd^2 \ellvec_1 \dd^2 \ellvec'_1\ldots \dd^2 \ellvec_n\dd^2 \ellvec'_n}
$
in \eqref{app:dsigma-n-averaged}. In our setup of small angle scattering and gluon radiation with large formation times, each rescattering contributes to the Lorentz structure a factor equal to the elastic scattering cross section $\dd \sigma_{\rm el}^{\rm Abelian}(\ellvec_i)/\dd^2\ellvec_i$ of an `abelian' parton with charge $g$. This factor is parametrized as 
\be
\frac{\dd \sigma_{\rm el}^{\rm Abelian}(\ellvec_i)}{\dd^2\ellvec_i} = \sigma_{\rm el}^{\rm Abelian} \, V(\ellvec_i) \, ,
\ee
where $V(\ellvec)$ is the normalized Coulomb potential defined in \eq{coulomb-pot}. The elastic cross section which corresponds to the hard scattering cancels the denominator except for the color factor $d_R C_R$. Therefore
\be
\dd I^{(n)} = \frac{(\nu\sigma_{\rm el}^{\rm Abelian})^n}{d_R C_R} \left[ \prod_{i=1}^{n}\int_0^L  \! \! \dd z_i \int \! \dd^2\ellvec_i \, V(\ellvec_i) \right] 
\sum\limits_{n_1, n_2, n_{\bar 2}} 
\frac{\tilde \sigma_{\rm rad}(p',k,\{\ellvec_i\})}{n_1! n_2! n_{\bar 2}!} \,(2 g)^2 \frac{\dd k^+ \dd^2 \kvec}{2 k^+  (2\pi)^3} \, ,
\ee
where the phase space factor for the radiated gluon (see \eq{phase-space-factor}) and a factor $(2 g)^2$ (see \eq{GB}) have been made explicit, and the `reduced cross section' $ \tilde \sigma_{\rm rad}$ contains only the color factors and emission vertices computed using the pictorial rules defined in Fig.~\ref{fig:pictorial-rules}. Introducing light-cone time (or longitudinal position) ordering of the scattering centers, $z_i < z_{i+1}$ for $i=1,\ldots,n-1$, 
the permutations of the scattering centers within the subsets of those contributing via single Born or double Born interaction (in either the amplitude or its conjugate) give $n_1! n_2! n_{\bar 2}!$ identical contributions. Noting that $C_R \nu \sigma_{el}^{\rm Abelian} = 1/\lambda_R$ completes the proof of \eq{spec-order-n}.

\section{Large $\Omega$ limit of $S[\Omega;r]$}
\label{app:limits}

Here we derive the behavior of the function $S[\Omega;r]$ defined in \eq{sar} in the limit $\Omega \to \infty$ at fixed $r$. The function $S[\Omega;r]$ can be written as 
\be
\label{sar-alpha-1}
S[\Omega;r] = \int_0^{\infty} \dd B \, B {\rm J}_0(\Omega B) \, \alpha(B) \, , 
\ee
where 
\be
\label{alphaB}
\alpha(B) \equiv \frac{2}{B^2} \left\{ 1 -\exp{\left[-r \left(1-B \, {\rm K}_1(B) \right)\right]}  \right\}  \, .
\ee
Using 
\be
B {\rm J}_0(\Omega B) = \frac{1}{\Omega} \frac{\dd}{\dd B} \, B {\rm J}_1(\Omega B)  \, 
\ee
a simple integration by parts of \eq{sar-alpha-1} yields
\be
\label{sar-alpha-2}
S[\Omega;r] = - \frac{1}{\Omega} \int_0^{\infty} \dd B \, {\rm J}_1(\Omega B) \, B\alpha' \, ,
\ee
where $\alpha' \equiv \dd \alpha(B)/\dd B$. Now use
\be
{\rm J}_1(\Omega B) = \frac{1}{\Omega} \frac{\dd}{\dd B} \left( 1- {\rm J}_0(\Omega B) \right) \, 
\ee
to integrate \eq{sar-alpha-2} by parts, leading to
\bea
\label{sar-alpha-3}
S[\Omega;r] &=& \frac{1}{\Omega^2} \int_0^{\infty} \dd B \,  \left[ 1- {\rm J}_0(\Omega B) \right] (B\alpha')' \nn \\
&=& \frac{r}{\Omega^2} - \frac{1}{\Omega^2} \int_0^{\infty} \dd B \,  {\rm J}_0(\Omega B) \, (B\alpha')' \, .
\eea
We used $\left. B\alpha' \right|_{B\to 0} = -r$, which can be verified from \eq{alphaB}. When $\Omega \to \infty$, the integral in the second term of \eq{sar-alpha-3} vanishes due to the rapid oscillation of ${\rm J}_0(\Omega B)$. We thus obtain
\be
S[\Omega;r] \mathop{\simeq}_{\Omega \to \infty} \frac{r}{\Omega^2}  \, .
\label{app:large-Omega}
\ee


\begin{thebibliography}{31}%
\makeatletter
\providecommand \@ifxundefined [1]{%
 \@ifx{#1\undefined}
}%
\providecommand \@ifnum [1]{%
 \ifnum #1\expandafter \@firstoftwo
 \else \expandafter \@secondoftwo
 \fi
}%
\providecommand \@ifx [1]{%
 \ifx #1\expandafter \@firstoftwo
 \else \expandafter \@secondoftwo
 \fi
}%
\providecommand \natexlab [1]{#1}%
\providecommand \enquote  [1]{``#1''}%
\providecommand \bibnamefont  [1]{#1}%
\providecommand \bibfnamefont [1]{#1}%
\providecommand \citenamefont [1]{#1}%
\providecommand \href@noop [0]{\@secondoftwo}%
\providecommand \href [0]{\begingroup \@sanitize@url \@href}%
\providecommand \@href[1]{\@@startlink{#1}\@@href}%
\providecommand \@@href[1]{\endgroup#1\@@endlink}%
\providecommand \@sanitize@url [0]{\catcode `\\12\catcode `\$12\catcode
  `\&12\catcode `\#12\catcode `\^12\catcode `\_12\catcode `\%12\relax}%
\providecommand \@@startlink[1]{}%
\providecommand \@@endlink[0]{}%
\providecommand \url  [0]{\begingroup\@sanitize@url \@url }%
\providecommand \@url [1]{\endgroup\@href {#1}{\urlprefix }}%
\providecommand \urlprefix  [0]{URL }%
\providecommand \Eprint [0]{\href }%
\providecommand \doibase [0]{http://dx.doi.org/}%
\providecommand \selectlanguage [0]{\@gobble}%
\providecommand \bibinfo  [0]{\@secondoftwo}%
\providecommand \bibfield  [0]{\@secondoftwo}%
\providecommand \translation [1]{[#1]}%
\providecommand \BibitemOpen [0]{}%
\providecommand \bibitemStop [0]{}%
\providecommand \bibitemNoStop [0]{.\EOS\space}%
\providecommand \EOS [0]{\spacefactor3000\relax}%
\providecommand \BibitemShut  [1]{\csname bibitem#1\endcsname}%
\let\auto@bib@innerbib\@empty
\bibitem [{\citenamefont {Adler}\ \emph {et~al.}(2003)\citenamefont {Adler}
  \emph {et~al.}}]{Adler:2003qi}%
  \BibitemOpen
  \bibfield  {author} {\bibinfo {author} {\bibfnamefont {S.~S.}\ \bibnamefont
  {Adler}} \emph {et~al.} (\bibinfo {collaboration} {PHENIX}),\ }\href@noop {}
  {\bibfield  {journal} {\bibinfo  {journal} {Phys. Rev. Lett.}\ }\textbf
  {\bibinfo {volume} {91}},\ \bibinfo {pages} {072301} (\bibinfo {year}
  {2003})},\ \Eprint {http://arxiv.org/abs/nucl-ex/0304022} {nucl-ex/0304022}
  \BibitemShut {NoStop}%
\bibitem [{\citenamefont {Adams}\ \emph {et~al.}(2003)\citenamefont {Adams}
  \emph {et~al.}}]{Adams:2003kv}%
  \BibitemOpen
  \bibfield  {author} {\bibinfo {author} {\bibfnamefont {J.}~\bibnamefont
  {Adams}} \emph {et~al.} (\bibinfo {collaboration} {STAR}),\ }\href {\doibase
  10.1103/PhysRevLett.91.172302} {\bibfield  {journal} {\bibinfo  {journal}
  {Phys. Rev. Lett.}\ }\textbf {\bibinfo {volume} {91}},\ \bibinfo {pages}
  {172302} (\bibinfo {year} {2003})},\ \Eprint
  {http://arxiv.org/abs/nucl-ex/0305015} {arXiv:nucl-ex/0305015} \BibitemShut
  {NoStop}%
\bibitem [{\citenamefont {Aad}\ \emph {et~al.}(2013)\citenamefont {Aad} \emph
  {et~al.}}]{Aad:2012vca}%
  \BibitemOpen
  \bibfield  {author} {\bibinfo {author} {\bibfnamefont {G.}~\bibnamefont
  {Aad}} \emph {et~al.} (\bibinfo {collaboration} {ATLAS}),\ }\href {\doibase
  10.1016/j.physletb.2013.01.024} {\bibfield  {journal} {\bibinfo  {journal}
  {Phys.Lett.}\ }\textbf {\bibinfo {volume} {B719}},\ \bibinfo {pages} {220}
  (\bibinfo {year} {2013})},\ \Eprint {http://arxiv.org/abs/1208.1967}
  {arXiv:1208.1967 [hep-ex]} \BibitemShut {NoStop}%
\bibitem [{\citenamefont {Aamodt}\ \emph {et~al.}(2011)\citenamefont {Aamodt}
  \emph {et~al.}}]{Aamodt:2010jd}%
  \BibitemOpen
  \bibfield  {author} {\bibinfo {author} {\bibfnamefont {K.}~\bibnamefont
  {Aamodt}} \emph {et~al.} (\bibinfo {collaboration} {ALICE}),\ }\href
  {\doibase 10.1016/j.physletb.2010.12.020} {\bibfield  {journal} {\bibinfo
  {journal} {Phys. Lett.}\ }\textbf {\bibinfo {volume} {B696}},\ \bibinfo
  {pages} {30} (\bibinfo {year} {2011})},\ \Eprint
  {http://arxiv.org/abs/1012.1004} {arXiv:1012.1004 [nucl-ex]} \BibitemShut
  {NoStop}%
\bibitem [{\citenamefont {Chatrchyan}\ \emph {et~al.}(2012)\citenamefont
  {Chatrchyan} \emph {et~al.}}]{CMS:2012aa}%
  \BibitemOpen
  \bibfield  {author} {\bibinfo {author} {\bibfnamefont {S.}~\bibnamefont
  {Chatrchyan}} \emph {et~al.} (\bibinfo {collaboration} {CMS}),\ }\href
  {\doibase 10.1140/epjc/s10052-012-1945-x} {\bibfield  {journal} {\bibinfo
  {journal} {Eur.\ Phys.\ J.}\ }\textbf {\bibinfo {volume} {C72}},\ \bibinfo
  {pages} {1945} (\bibinfo {year} {2012})},\ \Eprint
  {http://arxiv.org/abs/1202.2554} {arXiv:1202.2554 [nucl-ex]} \BibitemShut
  {NoStop}%
\bibitem [{\citenamefont {Baier}\ \emph
  {et~al.}(1997{\natexlab{a}})\citenamefont {Baier}, \citenamefont
  {Dokshitzer}, \citenamefont {Mueller}, \citenamefont {Peign{\'e}},\ and\
  \citenamefont {Schiff}}]{Baier:1996kr}%
  \BibitemOpen
  \bibfield  {author} {\bibinfo {author} {\bibfnamefont {R.}~\bibnamefont
  {Baier}}, \bibinfo {author} {\bibfnamefont {Y.~L.}\ \bibnamefont
  {Dokshitzer}}, \bibinfo {author} {\bibfnamefont {A.~H.}\ \bibnamefont
  {Mueller}}, \bibinfo {author} {\bibfnamefont {S.}~\bibnamefont {Peign{\'e}}},
  \ and\ \bibinfo {author} {\bibfnamefont {D.}~\bibnamefont {Schiff}},\ }
  \\ %
  \href
  {\doibase 10.1016/S0550-3213(96)00553-6} {\bibfield  {journal} {\bibinfo
  {journal} {Nucl. Phys.}\ }\textbf {\bibinfo {volume} {B483}},\ \bibinfo
  {pages} {291} (\bibinfo {year} {1997}{\natexlab{a}})},\ \Eprint
  {http://arxiv.org/abs/hep-ph/9607355} {arXiv:hep-ph/9607355} \BibitemShut
  {NoStop}%
\bibitem [{\citenamefont {Peign{\'e}}\ and\ \citenamefont
  {Smilga}(2009)}]{Peigne:2008wu}%
  \BibitemOpen
  \bibfield  {author} {\bibinfo {author} {\bibfnamefont {S.}~\bibnamefont
  {Peign{\'e}}}\ and\ \bibinfo {author} {\bibfnamefont {A.}~\bibnamefont
  {Smilga}},\ }\href {\doibase 10.3367/UFNe.0179.200907a.0697} {\bibfield
  {journal} {\bibinfo  {journal} {Phys.Usp.}\ }\textbf {\bibinfo {volume}
  {52}},\ \bibinfo {pages} {659} (\bibinfo {year} {2009})},\ \Eprint
  {http://arxiv.org/abs/0810.5702} {arXiv:0810.5702 [hep-ph]} \BibitemShut
  {NoStop}%
\bibitem [{\citenamefont {Arleo}\ \emph {et~al.}(2011)\citenamefont {Arleo},
  \citenamefont {Peign{\'e}},\ and\ \citenamefont {Sami}}]{Arleo:2010rb}%
  \BibitemOpen
  \bibfield  {author} {\bibinfo {author} {\bibfnamefont {F.}~\bibnamefont
  {Arleo}}, \bibinfo {author} {\bibfnamefont {S.}~\bibnamefont {Peign{\'e}}}, \
  and\ \bibinfo {author} {\bibfnamefont {T.}~\bibnamefont {Sami}},\ }\href
  {\doibase 10.1103/PhysRevD.83.114036} {\bibfield  {journal} {\bibinfo
  {journal} {Phys. Rev.}\ }\textbf {\bibinfo {volume} {D83}},\ \bibinfo {pages}
  {114036} (\bibinfo {year} {2011})},\ \Eprint {http://arxiv.org/abs/1006.0818}
  {arXiv:1006.0818 [hep-ph]} \BibitemShut {NoStop}%
\bibitem [{\citenamefont {Arleo}\ and\ \citenamefont
  {Peign{\'e}}(2012)}]{Arleo:2012hn}%
  \BibitemOpen
  \bibfield  {author} {\bibinfo {author} {\bibfnamefont {F.}~\bibnamefont
  {Arleo}}\ and\ \bibinfo {author} {\bibfnamefont {S.}~\bibnamefont
  {Peign{\'e}}},\ }\href@noop {} {\bibfield  {journal} {\bibinfo  {journal}
  {Phys. Rev. Lett.}\ }\textbf {\bibinfo {volume} {109}},\ \bibinfo {pages}
  {122301} (\bibinfo {year} {2012})},\ \Eprint {http://arxiv.org/abs/1204.4609}
  {arXiv:1204.4609 [hep-ph]} \BibitemShut {NoStop}%
\bibitem [{\citenamefont {Arleo}\ and\ \citenamefont
  {Peign{\'e}}(2013)}]{Arleo:2012rs}%
  \BibitemOpen
  \bibfield  {author} {\bibinfo {author} {\bibfnamefont {F.}~\bibnamefont
  {Arleo}}\ and\ \bibinfo {author} {\bibfnamefont {S.}~\bibnamefont
  {Peign{\'e}}},\ }\href@noop {} {\bibfield  {journal} {\bibinfo  {journal}
  {JHEP}\ }\textbf {\bibinfo {volume} {03}},\ \bibinfo {pages} {122} (\bibinfo
  {year} {2013})},\ \Eprint {http://arxiv.org/abs/1212.0434} {arXiv:1212.0434
  [hep-ph]} \BibitemShut {NoStop}%
\bibitem [{\citenamefont {Arleo}\ \emph {et~al.}(2013)\citenamefont {Arleo},
  \citenamefont {Kolevatov}, \citenamefont {Peign{\'e}},\ and\ \citenamefont
  {Rustamova}}]{Arleo:2013zua}%
  \BibitemOpen
  \bibfield  {author} {\bibinfo {author} {\bibfnamefont {F.}~\bibnamefont
  {Arleo}}, \bibinfo {author} {\bibfnamefont {R.}~\bibnamefont {Kolevatov}},
  \bibinfo {author} {\bibfnamefont {S.}~\bibnamefont {Peign{\'e}}}, \ and\
  \bibinfo {author} {\bibfnamefont {M.}~\bibnamefont {Rustamova}},\ }\href
  {\doibase 10.1007/JHEP05(2013)155} {\bibfield  {journal} {\bibinfo  {journal}
  {JHEP}\ }\textbf {\bibinfo {volume} {1305}},\ \bibinfo {pages} {155}
  (\bibinfo {year} {2013})},\ 
  \\ %
  \Eprint {http://arxiv.org/abs/1304.0901}
  {arXiv:1304.0901 [hep-ph]} \BibitemShut {NoStop}%
\bibitem [{\citenamefont {Armesto}\ \emph {et~al.}(2012)\citenamefont
  {Armesto}, \citenamefont {Ma}, \citenamefont {Martinez}, \citenamefont
  {Mehtar-Tani},\ and\ \citenamefont {Salgado}}]{Armesto:2012qa}%
  \BibitemOpen
  \bibfield  {author} {\bibinfo {author} {\bibfnamefont {N.}~\bibnamefont
  {Armesto}}, \bibinfo {author} {\bibfnamefont {H.}~\bibnamefont {Ma}},
  \bibinfo {author} {\bibfnamefont {M.}~\bibnamefont {Martinez}}, \bibinfo
  {author} {\bibfnamefont {Y.}~\bibnamefont {Mehtar-Tani}}, \ and\ \bibinfo
  {author} {\bibfnamefont {C.~A.}\ \bibnamefont {Salgado}},\ } 
  \\ %
  \href {\doibase
  10.1016/j.physletb.2012.09.039} {\bibfield  {journal} {\bibinfo  {journal}
  {Phys. Lett.}\ }\textbf {\bibinfo {volume} {B717}},\ \bibinfo {pages} {280}
  (\bibinfo {year} {2012})},\ \Eprint {http://arxiv.org/abs/1207.0984}
  {arXiv:1207.0984 [hep-ph]} \BibitemShut {NoStop}%
\bibitem [{\citenamefont {Armesto}\ \emph {et~al.}(2013)\citenamefont
  {Armesto}, \citenamefont {Ma}, \citenamefont {Martinez}, \citenamefont
  {Mehtar-Tani},\ and\ \citenamefont {Salgado}}]{Armesto:2013fca}%
  \BibitemOpen
  \bibfield  {author} {\bibinfo {author} {\bibfnamefont {N.}~\bibnamefont
  {Armesto}}, \bibinfo {author} {\bibfnamefont {H.}~\bibnamefont {Ma}},
  \bibinfo {author} {\bibfnamefont {M.}~\bibnamefont {Martinez}}, \bibinfo
  {author} {\bibfnamefont {Y.}~\bibnamefont {Mehtar-Tani}}, \ and\ \bibinfo
  {author} {\bibfnamefont {C.~A.}\ \bibnamefont {Salgado}},\ }\href {\doibase
  10.1007/JHEP12(2013)052} {\bibfield  {journal} {\bibinfo  {journal} {JHEP}\
  }\textbf {\bibinfo {volume} {12}},\ \bibinfo {pages} {052} (\bibinfo {year}
  {2013})},\ \Eprint {http://arxiv.org/abs/1308.2186} {arXiv:1308.2186
  [hep-ph]} \BibitemShut {NoStop}%
\bibitem [{\citenamefont {Kovchegov}\ and\ \citenamefont
  {Mueller}(1998)}]{Kovchegov:1998bi}%
  \BibitemOpen
  \bibfield  {author} {\bibinfo {author} {\bibfnamefont {Y.~V.}\ \bibnamefont
  {Kovchegov}}\ and\ \bibinfo {author} {\bibfnamefont {A.~H.}\ \bibnamefont
  {Mueller}},\ }\href {\doibase 10.1016/S0550-3213(98)00384-8} {\bibfield
  {journal} {\bibinfo  {journal} {Nucl. Phys.}\ }\textbf {\bibinfo {volume}
  {B529}},\ \bibinfo {pages} {451} (\bibinfo {year} {1998})},
  \\ %
  \ \Eprint
  {http://arxiv.org/abs/hep-ph/9802440} {arXiv:hep-ph/9802440 [hep-ph]}
  \BibitemShut {NoStop}%
\bibitem [{\citenamefont {Gunion}\ and\ \citenamefont
  {Bertsch}(1982)}]{Gunion:1981qs}%
  \BibitemOpen
  \bibfield  {author} {\bibinfo {author} {\bibfnamefont {J.~F.}\ \bibnamefont
  {Gunion}}\ and\ \bibinfo {author} {\bibfnamefont {G.}~\bibnamefont
  {Bertsch}},\ }\href {\doibase 10.1103/PhysRevD.25.746} {\bibfield  {journal}
  {\bibinfo  {journal} {Phys. Rev.}\ }\textbf {\bibinfo {volume} {D25}},\
  \bibinfo {pages} {746} (\bibinfo {year} {1982})}\BibitemShut {NoStop}%
\bibitem [{\citenamefont {Mueller}(2001)}]{Mueller:2001fv}%
  \BibitemOpen
  \bibfield  {author} {\bibinfo {author} {\bibfnamefont {A.~H.}\ \bibnamefont
  {Mueller}},\ }in\ \href@noop {} {\emph {\bibinfo {booktitle} {{QCD
  perspectives on hot and dense matter. Proceedings, NATO Advanced Study
  Institute, Summer School, Cargese, France, August 6-18, 2001}}}}\ (\bibinfo
  {year} {2001})\ pp.\ \bibinfo {pages} {45--72},\ \Eprint
  {http://arxiv.org/abs/hep-ph/0111244} {arXiv:hep-ph/0111244 [hep-ph]}
  \BibitemShut {NoStop}%
\bibitem [{\citenamefont {Lepage}\ and\ \citenamefont
  {Brodsky}(1980)}]{Lepage:1980fj}%
  \BibitemOpen
  \bibfield  {author} {\bibinfo {author} {\bibfnamefont {G.~P.}\ \bibnamefont
  {Lepage}}\ and\ \bibinfo {author} {\bibfnamefont {S.~J.}\ \bibnamefont
  {Brodsky}},\ }\href {\doibase 10.1103/PhysRevD.22.2157} {\bibfield  {journal}
  {\bibinfo  {journal} {Phys.\ Rev.}\ }\textbf {\bibinfo {volume} {D22}},\
  \bibinfo {pages} {2157} (\bibinfo {year} {1980})}\BibitemShut {NoStop}%
\bibitem [{\citenamefont {Dokshitzer}(1995{\natexlab{a}})}]{Dokshitzer:1995fv}%
  \BibitemOpen
  \bibfield  {author} {\bibinfo {author} {\bibfnamefont {{\relax Yu}.~L.}\
  \bibnamefont {Dokshitzer}},\ }in\ \href@noop {} {\emph {\bibinfo {booktitle}
  {{Strong Interactions Study Days Kloster Banz, Germany, October 10-12,
  1995}}}}\ (\bibinfo {year} {1995})\BibitemShut {NoStop}%
\bibitem [{\citenamefont {Dokshitzer}(1995{\natexlab{b}})}]{Dokshitzer:1995if}%
  \BibitemOpen
  \bibfield  {author} {\bibinfo {author} {\bibfnamefont {Y.~L.}\ \bibnamefont
  {Dokshitzer}},\ }in\ \href@noop {} {\emph {\bibinfo {booktitle} {{3rd
  European School of High-energy Physics Dubna, Russia, August 27-September 9,
  1995}}}}\ (\bibinfo {year} {1995})\BibitemShut {NoStop}%
\bibitem [{\citenamefont {Gyulassy}\ \emph {et~al.}(2001)\citenamefont
  {Gyulassy}, \citenamefont {L{\'e}vai},\ and\ \citenamefont
  {Vitev}}]{Gyulassy:2000er}%
  \BibitemOpen
  \bibfield  {author} {\bibinfo {author} {\bibfnamefont {M.}~\bibnamefont
  {Gyulassy}}, \bibinfo {author} {\bibfnamefont {P.}~\bibnamefont {L{\'e}vai}},
  \ and\ \bibinfo {author} {\bibfnamefont {I.}~\bibnamefont {Vitev}},\
  }\href@noop {} {\bibfield  {journal} {\bibinfo  {journal} {Nucl. Phys.}\
  }\textbf {\bibinfo {volume} {B594}},\ \bibinfo {pages} {371} (\bibinfo {year}
  {2001})},\ \Eprint {http://arxiv.org/abs/nucl-th/0006010} {nucl-th/0006010}
  \BibitemShut {NoStop}%
\bibitem [{\citenamefont {Gyulassy}\ and\ \citenamefont
  {Wang}(1994)}]{Gyulassy:1993hr}%
  \BibitemOpen
  \bibfield  {author} {\bibinfo {author} {\bibfnamefont {M.}~\bibnamefont
  {Gyulassy}}\ and\ \bibinfo {author} {\bibfnamefont {X.-N.}\ \bibnamefont
  {Wang}},\ }\href {\doibase 10.1016/0550-3213(94)90079-5} {\bibfield
  {journal} {\bibinfo  {journal} {Nucl.\ Phys.}\ }\textbf {\bibinfo {volume}
  {B420}},\ \bibinfo {pages} {583} (\bibinfo {year} {1994})},
  \\ %
  \ \Eprint
  {http://arxiv.org/abs/nucl-th/9306003} {arXiv:nucl-th/9306003 [nucl-th]}
  \BibitemShut {NoStop}%
\bibitem [{\citenamefont {Baier}\ \emph {et~al.}(1998)\citenamefont {Baier},
  \citenamefont {Dokshitzer}, \citenamefont {Mueller},\ and\ \citenamefont
  {Schiff}}]{Baier:1998kq}%
  \BibitemOpen
  \bibfield  {author} {\bibinfo {author} {\bibfnamefont {R.}~\bibnamefont
  {Baier}}, \bibinfo {author} {\bibfnamefont {Y.~L.}\ \bibnamefont
  {Dokshitzer}}, \bibinfo {author} {\bibfnamefont {A.~H.}\ \bibnamefont
  {Mueller}}, \ and\ \bibinfo {author} {\bibfnamefont {D.}~\bibnamefont
  {Schiff}},\ }  \href@noop {} {\bibfield  {journal} {\bibinfo  {journal} {Nucl.
  Phys.}\ }\textbf {\bibinfo {volume} {B531}},\ \bibinfo {pages} {403}
  (\bibinfo {year} {1998})},
  \\ %
  \ \Eprint {http://arxiv.org/abs/hep-ph/9804212}
  {hep-ph/9804212} \BibitemShut {NoStop}%
\bibitem [{\citenamefont {'t~Hooft}(1974)}]{tHooft:1973jz}%
  \BibitemOpen
  \bibfield  {author} {\bibinfo {author} {\bibfnamefont {G.}~\bibnamefont
  {'t~Hooft}},\ }\href {\doibase 10.1016/0550-3213(74)90154-0} {\bibfield
  {journal} {\bibinfo  {journal} {Nucl.\ Phys.}\ }\textbf {\bibinfo {volume}
  {B72}},\ \bibinfo {pages} {461} (\bibinfo {year} {1974})}\BibitemShut
  {NoStop}%
\bibitem [{\citenamefont {Brodsky}\ and\ \citenamefont
  {Hoyer}(1993)}]{Brodsky:1992nq}%
  \BibitemOpen
  \bibfield  {author} {\bibinfo {author} {\bibfnamefont {S.~J.}\ \bibnamefont
  {Brodsky}}\ and\ \bibinfo {author} {\bibfnamefont {P.}~\bibnamefont
  {Hoyer}},\ }\href {\doibase 10.1016/0370-2693(93)91724-2} {\bibfield
  {journal} {\bibinfo  {journal} {Phys. Lett.}\ }\textbf {\bibinfo {volume}
  {B298}},\ \bibinfo {pages} {165} (\bibinfo {year} {1993})},\ \Eprint
  {http://arxiv.org/abs/hep-ph/9210262} {arXiv:hep-ph/9210262} \BibitemShut
  {NoStop}%
\bibitem [{\citenamefont {Baier}\ \emph
  {et~al.}(1997{\natexlab{b}})\citenamefont {Baier}, \citenamefont
  {Dokshitzer}, \citenamefont {Mueller}, \citenamefont {Peign{\'e}},\ and\
  \citenamefont {Schiff}}]{Baier:1996sk}%
  \BibitemOpen
  \bibfield  {author} {\bibinfo {author} {\bibfnamefont {R.}~\bibnamefont
  {Baier}}, \bibinfo {author} {\bibfnamefont {Y.~L.}\ \bibnamefont
  {Dokshitzer}}, \bibinfo {author} {\bibfnamefont {A.~H.}\ \bibnamefont
  {Mueller}}, \bibinfo {author} {\bibfnamefont {S.}~\bibnamefont {Peign{\'e}}},
  \ and\ \bibinfo {author} {\bibfnamefont {D.}~\bibnamefont {Schiff}},\ } 
  \\ %
  \href
  {\doibase 10.1016/S0550-3213(96)00581-0} {\bibfield  {journal} {\bibinfo
  {journal} {Nucl. Phys.}\ }\textbf {\bibinfo {volume} {B484}},\ \bibinfo
  {pages} {265} (\bibinfo {year} {1997}{\natexlab{b}})},\ \Eprint
  {http://arxiv.org/abs/hep-ph/9608322} {arXiv:hep-ph/9608322} \BibitemShut
  {NoStop}%
\bibitem [{\citenamefont {Zakharov}(2001)}]{Zakharov:2000iz}%
  \BibitemOpen
  \bibfield  {author} {\bibinfo {author} {\bibfnamefont {B.}~\bibnamefont
  {Zakharov}},\ }\href {\doibase 10.1134/1.1358417} {\bibfield  {journal}
  {\bibinfo  {journal} {JETP Lett.}\ }\textbf {\bibinfo {volume} {73}},\
  \bibinfo {pages} {49} (\bibinfo {year} {2001})},\ \Eprint
  {http://arxiv.org/abs/hep-ph/0012360} {arXiv:hep-ph/0012360 [hep-ph]}
  \BibitemShut {NoStop}%
\bibitem [{\citenamefont {Vitev}(2007)}]{Vitev:2007ve}%
  \BibitemOpen
  \bibfield  {author} {\bibinfo {author} {\bibfnamefont {I.}~\bibnamefont
  {Vitev}},\ }\href {\doibase 10.1103/PhysRevC.75.064906} {\bibfield  {journal}
  {\bibinfo  {journal} {Phys. Rev.}\ }\textbf {\bibinfo {volume} {C75}},\
  \bibinfo {pages} {064906} (\bibinfo {year} {2007})},\ \Eprint
  {http://arxiv.org/abs/hep-ph/0703002} {arXiv:hep-ph/0703002} \BibitemShut
  {NoStop}%
\bibitem [{\citenamefont {Zakharov}(1997)}]{Zakharov:1997uu}%
  \BibitemOpen
  \bibfield  {author} {\bibinfo {author} {\bibfnamefont {B.~G.}\ \bibnamefont
  {Zakharov}},\ }\href@noop {} {\bibfield  {journal} {\bibinfo  {journal} {JETP
  Lett.}\ }\textbf {\bibinfo {volume} {65}},\ \bibinfo {pages} {615} (\bibinfo
  {year} {1997})},\ \Eprint {http://arxiv.org/abs/hep-ph/9704255}
  {hep-ph/9704255} \BibitemShut {NoStop}%
\bibitem [{\citenamefont {Peign{\'e}}\ and\ \citenamefont
  {Kolevatov}(2015)}]{Peigne:2014rka}%
  \BibitemOpen
  \bibfield  {author} {\bibinfo {author} {\bibfnamefont {S.}~\bibnamefont
  {Peign{\'e}}}\ and\ \bibinfo {author} {\bibfnamefont {R.}~\bibnamefont
  {Kolevatov}},\ }\href {\doibase 10.1007/JHEP01(2015)141} {\bibfield
  {journal} {\bibinfo  {journal} {JHEP}\ }\textbf {\bibinfo {volume} {01}},\
  \bibinfo {pages} {141} (\bibinfo {year} {2015})},\ \Eprint
  {http://arxiv.org/abs/1405.4241} {arXiv:1405.4241 [hep-ph]} \BibitemShut
  {NoStop}%
\bibitem [{\citenamefont {Liou}\ and\ \citenamefont
  {Mueller}(2014)}]{Liou:2014rha}%
  \BibitemOpen
  \bibfield  {author} {\bibinfo {author} {\bibfnamefont {T.}~\bibnamefont
  {Liou}}\ and\ \bibinfo {author} {\bibfnamefont {A.~H.}\ \bibnamefont
  {Mueller}},\ }\href {\doibase 10.1103/PhysRevD.89.074026} {\bibfield
  {journal} {\bibinfo  {journal} {Phys.\ Rev.}\ }\textbf {\bibinfo {volume}
  {D89}},\ \bibinfo {pages} {074026} (\bibinfo {year} {2014})},\ \Eprint
  {http://arxiv.org/abs/1402.1647} {arXiv:1402.1647 [hep-ph]} \BibitemShut
  {NoStop}%
\bibitem [{\citenamefont {Binosi}\ \emph {et~al.}(2009)\citenamefont {Binosi},
  \citenamefont {Collins}, \citenamefont {Kaufhold},\ and\ \citenamefont
  {Theussl}}]{Binosi:2008ig}%
  \BibitemOpen
  \bibfield  {author} {\bibinfo {author} {\bibfnamefont {D.}~\bibnamefont
  {Binosi}}, \bibinfo {author} {\bibfnamefont {J.}~\bibnamefont {Collins}},
  \bibinfo {author} {\bibfnamefont {C.}~\bibnamefont {Kaufhold}}, \ and\
  \bibinfo {author} {\bibfnamefont {L.}~\bibnamefont {Theussl}},\ } 
  \\ %
  \href
  {\doibase 10.1016/j.cpc.2009.02.020} {\bibfield  {journal} {\bibinfo
  {journal} {Comput. Phys. Commun.}\ }\textbf {\bibinfo {volume} {180}},\
  \bibinfo {pages} {1709} (\bibinfo {year} {2009})},\ \Eprint
  {http://arxiv.org/abs/0811.4113} {arXiv:0811.4113 [hep-ph]} \BibitemShut
  {NoStop}%
\end{thebibliography}
\end{document}